\documentclass[11pt]{article}
\usepackage{amsmath,amssymb,graphicx,float}
\usepackage[colorlinks=true, urlcolor=blue, citecolor=blue]{hyperref}
\usepackage[a4paper,margin=0.7in]{geometry}
\usepackage{braket}
\usepackage{slashed}
\usepackage{tensor,tcolorbox}
\usepackage{subcaption}
\usepackage[sorting=none , backend = biber, maxnames=5]{biblatex}
\numberwithin{equation}{section}

\newcommand{\mc}[1]{{\mathcal{#1}}}
\newcommand{\mb}[1]{{\mathbf{#1}}}

\let\del\partial
\let\tns\tensor
\newcommand{\dimd}{\text{d}}

\newcommand{\intm}[1]{\int \frac{d^\dimd#1}{(2\pi)^\dimd}}
\newcommand{\expp}[2]{e^{i\vec{#1}\cdot\vec{#2}}}
\newcommand{\expn}[2]{e^{-i\vec{#1}\cdot\vec{#2}}}

\renewcommand{\vec}{\mathbf}

\title{\textbf{Fermionic S-matrix and cosmological correlators: \\ T-violation at $O(H)$}}
\author{
    \parbox{\textwidth}{\centering
        Aman Goyal$^{1}$\thanks{amangoyal@iisc.ac.in}, Aneek Jana$^{1}$\thanks{aneekjana@iisc.ac.in}, Swapnanil Mandal$^{1}$\thanks{swapnanilm@iisc.ac.in}, and Aninda Sinha$^{1,2}$\thanks{asinha@iisc.ac.in} \\
        \vspace{0.3cm}
        \textit{ $^{1}$Centre for High Energy Physics, Indian Institute of Science, C. V. Raman Avenue, Bangalore 560012, India} \\ \vspace{0.3cm}
        \textit{$^{2}$Department of Physics and Astronomy, University of Calgary, Alberta T2N IN4, Canada}
    }
}

\date{\today}

\begin{document}

\maketitle

\renewcommand{\abstractname}{\large Abstract}
\begin{abstract}
    We study the Bunch-Davies (BD) and Unruh-de Witt (UdW) de Sitter S-matrices in the presence of spin-$1/2$ fermions. Building on recent work, this enables us to correlate the de Sitter S-matrix with cosmological correlators. We consider a finite-time version of the UdW S-matrix to study $O(H)$ corrections to some typical particle physics processes such as beta decay. Owing to the lack of time-reversal symmetry in the expanding Poincaré patch, we find signatures of intrinsic T-violation in polarized beta decay. The observable we study begins at $O(H)$. The possibility of T-violation was examined theoretically in the 1950s by Jackson, Treiman, and Wyld in flat space and has been probed more recently in the emiT experiment, with the purpose of examining fundamental T-violation coming from additional interactions in the Lagrangian. Our analysis places a lower bound on the intrinsic T-violation in the expanding Poincaré patch. At $O(H)$, we find both energy conserving and energy non-conserving contributions. Surprisingly, the energy-violating piece, in principle, can give large T-violation at fine-tuned values of the kinematical variables. 
\end{abstract}

\hypersetup{linkcolor = black}

\newpage
\tableofcontents
\hypersetup{linkcolor = blue}

\section{Introduction}

According to our present understanding of the standard model of cosmology, the primordial universe can be considered as approximately de Sitter with a huge uncertainty in the Hubble constant \cite{smatrixmarathon}:
\begin{equation}
    10^{-24}GeV<H<10^{13}GeV\,.
\end{equation}
Furthermore, experiments point to an accelerated phase for our present day universe with \cite{pdg}
\begin{equation}
    H_0\sim 10^{-42}GeV\,.
\end{equation}

Over the last several years, it has become a worthwhile endeavor to study quantum field theory in de Sitter space and develop techniques to calculate various observable quantities efficiently \cite{arkanimalda, arkani2, arkani1, bau1, bau2}. The focus has mainly been on cosmological correlators, which carry information about the end of inflation via so-called in-in formalism (see \cite{smatrixmarathon} for a summary and a list of references). Sometimes it is calculationally more efficient to compute the cosmological wavefunction~\cite{Goodhew:2020hob,Jazayeri:2021fvk,Melville:2021lst,Hillman:2021bnk,Lee:2023jby,Anninos_2015, suvrat}. More recently, various avatars of the de Sitter S-matrix \cite{marolf1, marolf2, pimentel1, pajer} have been introduced. Besides offering the exciting possibility of leveraging ideas from the flat-space S-matrix theory, it also contains information about the in-in cosmological correlators. At the same time, there are several conceptual difficulties that rear their heads---for example, the validity of the adiabatic approximation to define asymptotic states needs to be checked (see \cite{pimentel1} for a preliminary discussion). The current proposals of de Sitter S-matrices can be summarized into 3 categories, shown in the figure below:
\begin{figure}[htbp]
  \centering
  \begin{subfigure}[b]{0.35\textwidth}
    \includegraphics[width=\textwidth]{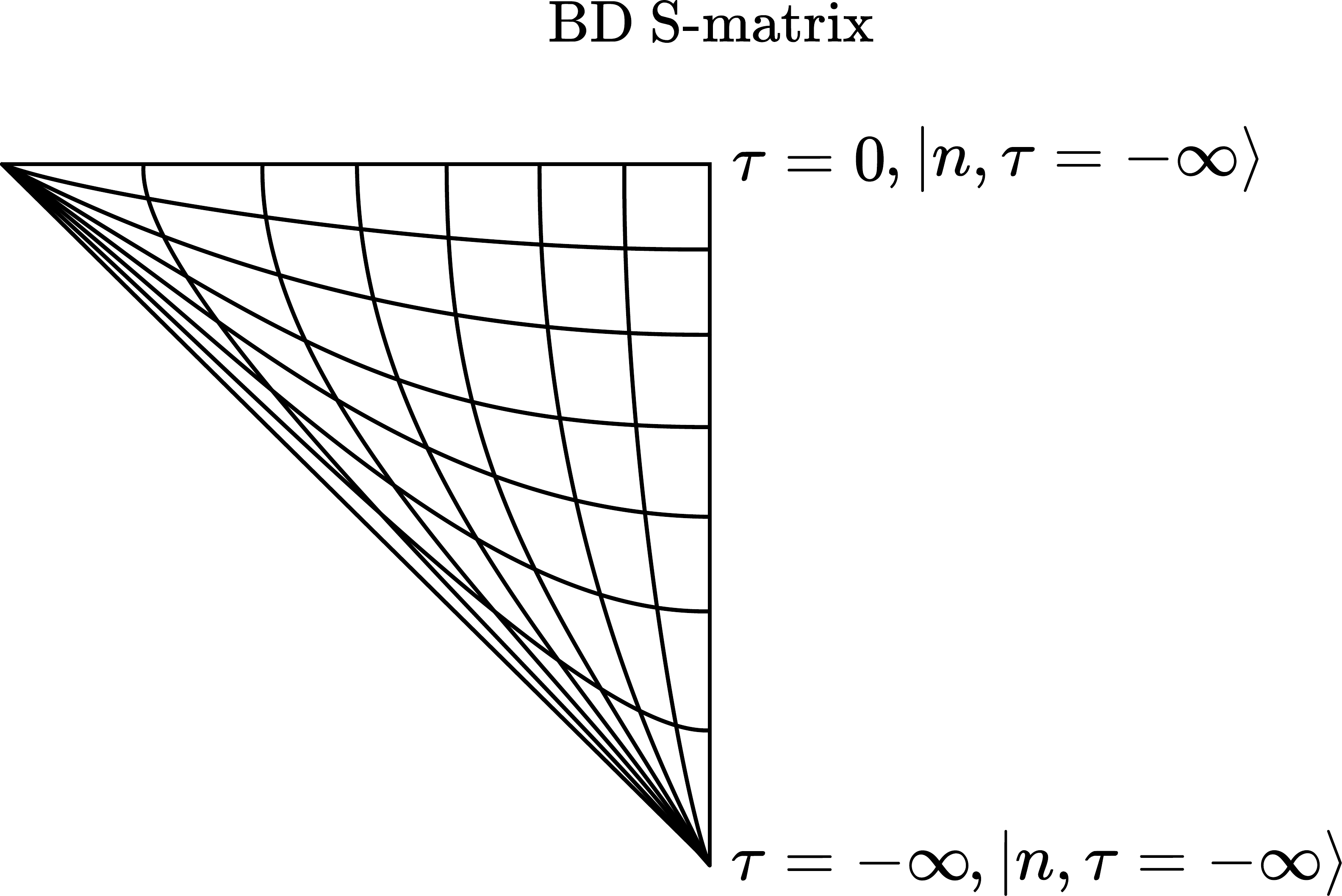}
    \caption{BD}
  \end{subfigure}
  \begin{subfigure}[b]{0.35\textwidth}
    \includegraphics[width=\textwidth]{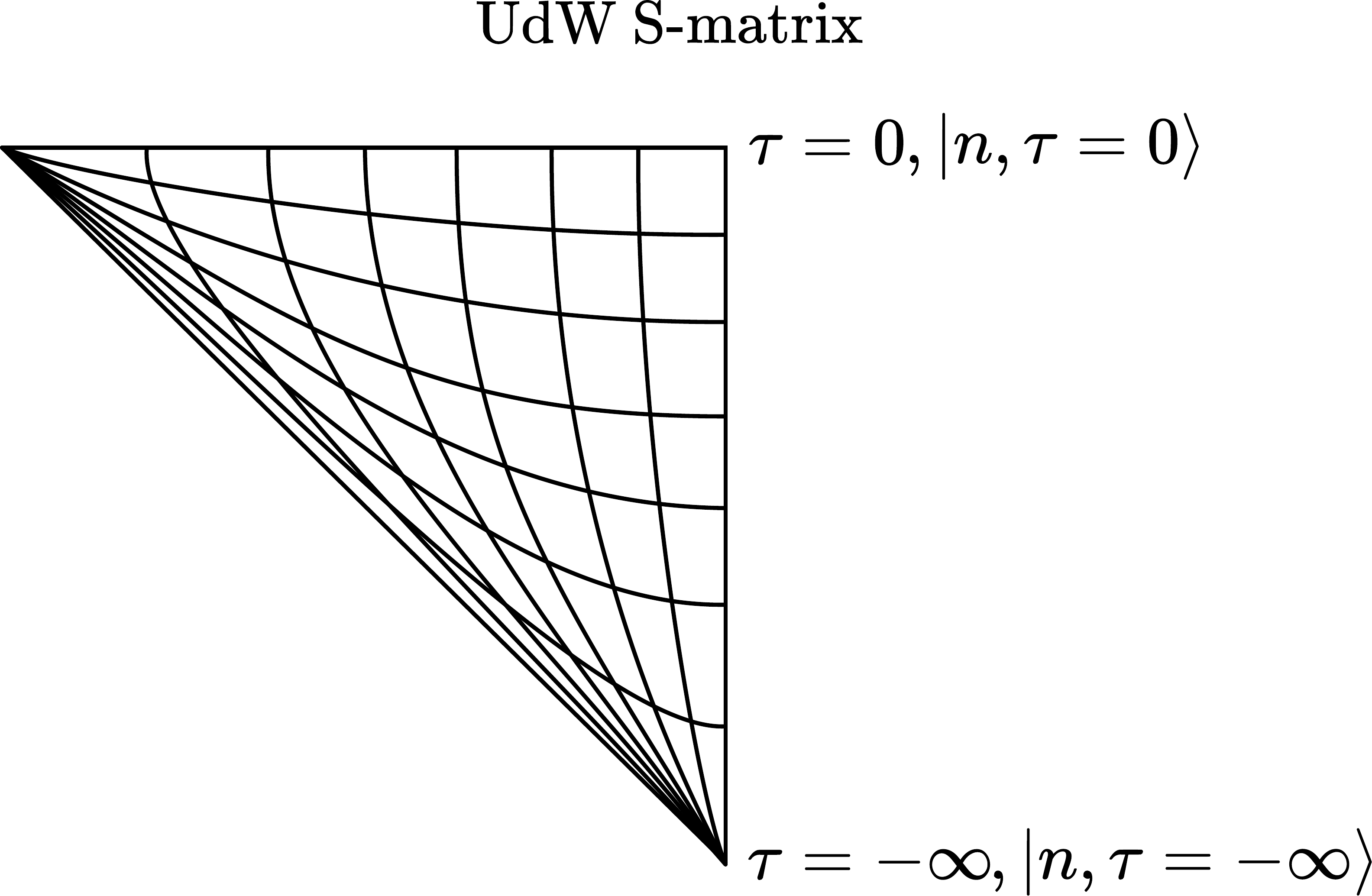}
    \caption{UdW}
 \end{subfigure} 
  \begin{subfigure}[b]{0.25\textwidth}
    \includegraphics[width=\textwidth]{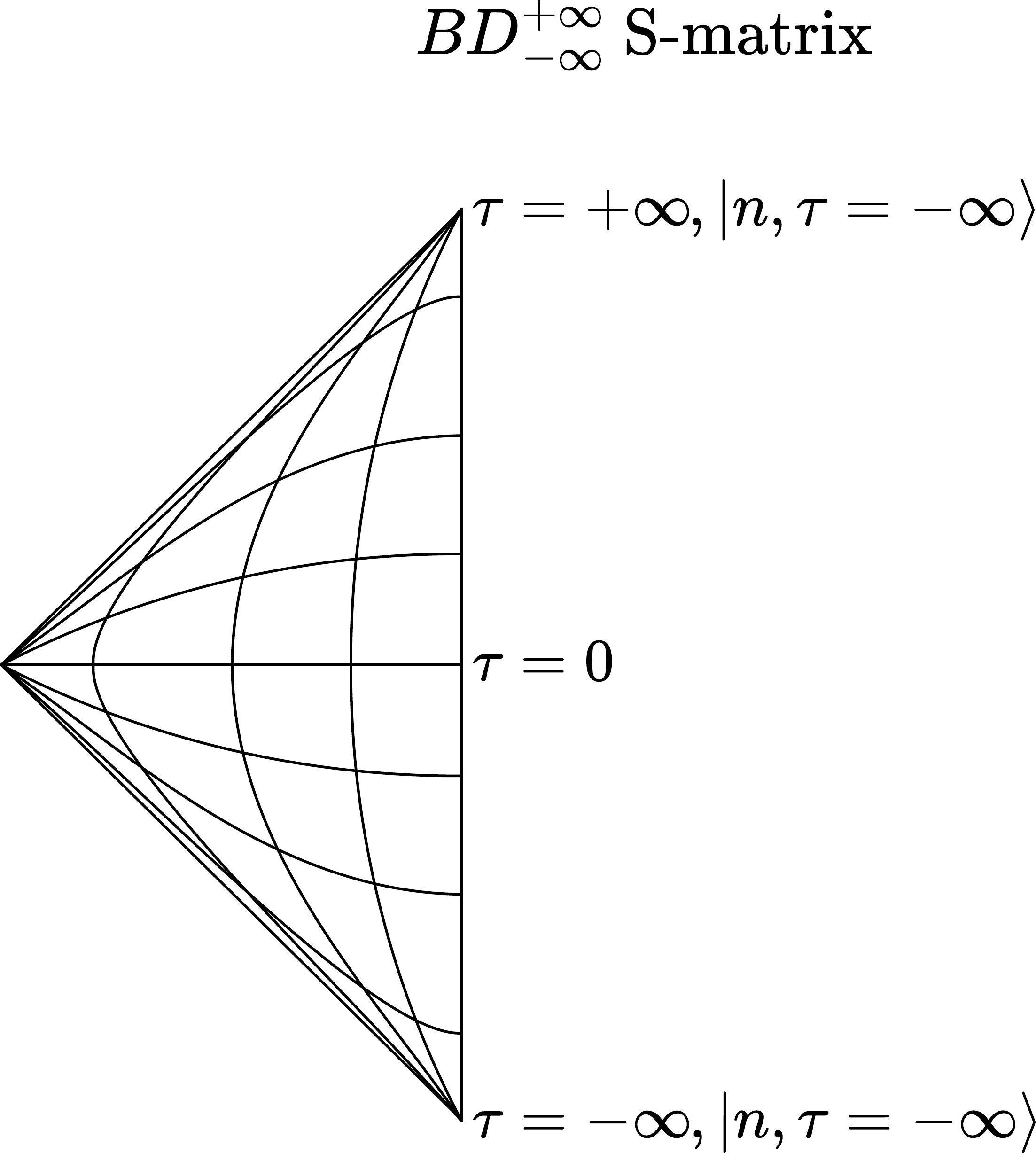}
    \caption{BD$_{-\infty}^{+\infty}$}
    \label{fig: pajer in-out}
  \end{subfigure}
  \caption{(a), (b) were introduced in \cite{pimentel1} while (c) was introduced in \cite{pajer}. $\tau$ here is the conformal time.}
  \label{fig:three_figures}
\end{figure}
These are the Bunch-Davies (BD) S-matrix, Unruh-de Witt (UdW) S-matrix \cite{pimentel1}, and the flat space analog Bunch-Davies (we will use the shorthand notation BD$_{-\infty}^{+\infty}$) considered in \cite{pajer} where the last one can be thought of as the `S-matrix way' of calculating in-in correlators. The first two are defined in the expanding Poincaré patch (EPP) while the last one is defined in the EPP glued with the contracting Poincaré patch (CPP)---the resulting figure resembles the flat-space Penrose diagram. The BD and BD$_{-\infty}^{+\infty}$ S-matrices have certain common features with the flat space S-matrices that are being explored \cite{appl3,appl2,appl1}. In particular, the BD$_{-\infty}^{+\infty}$ in-out correlators were argued to be the same as the in-in correlators in \cite{pajer} and provide calculational advantages over the cumbersome techniques used for the in-in correlators. The last statement assumes the absence of dissipation---for recent attempts to treat cosmological correlators using open quantum systems, see \cite{loga}.

\begin{figure}[H]
	\centering
	\includegraphics[scale=0.3]{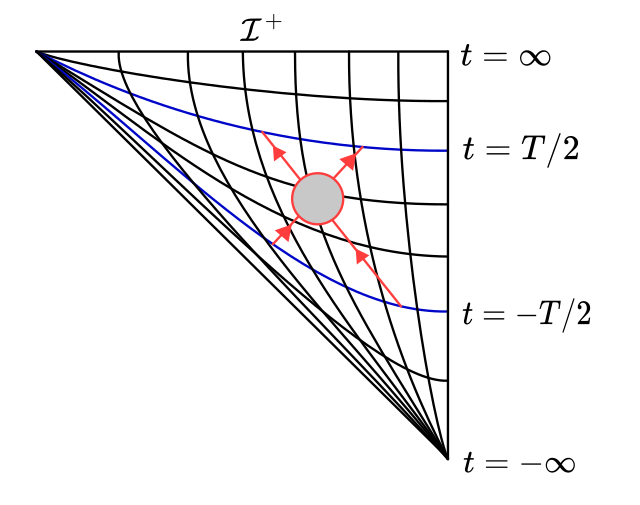}
	\caption{ The UdW$_T$ S-matrix. Scattering in our conventions happens between $-T/2$ and $T/2$. $t$ here is the co-moving time.}\label{epp}
\end{figure}

In this paper, we will also consider the finite time version of the UdW S-matrix, which we will call UdW$_T$. This is depicted in fig. (\ref{epp}). 

One motivation for doing this is to enable consideration of particle physics processes at present times (or in the early universe under certain assumptions), which take place over very short time-scales, compared to the scale set by $1/H$. Although most of our formalism will be general, our focus will be to extract the $O(H)$ corrections to flat space answers\footnote{In \cite{altflat}, a novel flat space limit has been proposed for cosmological correlators. Our starting point is different from theirs, as our interest is in objects that look like particle physics S-matrices. It will be interesting to examine the similarities and differences in the future.}. It will turn out that for processes taking place over time-scales much smaller than $1/H$, the $O(H)$ answers arising from the UdW$_T$ are the same as those from BD. Our considerations should also be useful to consider similar processes in the early universe, in the situation when $H$ was small compared to all other mass-scales in the problem. The existing discussions of the de Sitter S-matrix as introduced in \cite{pimentel1, pajer} focus on scalars. We will extend these results to include fermions \cite{fermi1,Schaub:2023scu}. In particular, we will recast the in-in correlators in the language of the fermionic S-matrix elements following \cite{pimentel1}. Further, we will also show, following \cite{pajer}, using the Fermi theory as an example, that in-in is equal to in-out as in the fig.(\ref{fig: pajer in-out}).  Such calculations may have bearing on our understanding of fermionic processes in the early universe, when $H/m\ll 1$. More interestingly, however, we can ask how they affect terrestrial experiments.

Typically, in order to explain terrestrial scattering processes, we neither take into account Earth's (or Sun's or Moon's) gravity, nor the expansion of the universe. Both of these, of course, are not just for simplicity---they are expected to be tiny effects, which will be overwhelmed by experimental noise. But how small is small? And, are there situations where the $O(H)$ becomes the dominant effect? We will now try to give some rough answers to these questions. 

Let us assume that the metric near the Earth's surface is fairly modeled by the Schwarzschild-de Sitter metric\footnote{In our conventions, we have $R_{ab}-1/2 R g_{ab}-3 H^2 g_{ab}=0$. Since the detectors are stationary w.r.t the Earth, we will ignore the effect of the Earth's rotation. Another important assumption is that we are ignoring the motion of our galaxy.}. This can be conveniently written as:
\begin{equation}
    ds^2=f(r,t)^2 dt^2- g(r,t)^2 e^{2H t} (dx^2+dy^2+dz^2)\,,
\end{equation}
where
\begin{equation}
    f(r,t)=\left(\frac{1-\frac{r_s}{r} e^{-H t}}{1+\frac{r_s}{r}e^{-H t}}\right)^2\,,\quad g(r,t)=\left(1+\frac{r_s}{r}e^{-H t}\right)^2\,,
\end{equation}
where $r_s$ is the Schwarzschild radius of the Earth (which can be taken to be $\sim 1 cm$) and $H$ is the present day Hubble constant ($\sim 10^{-42} GeV$). Since the Hubble constant is so tiny, it is usually expected that no terrestrial experiment will be able to detect this, since, in any event, even the Earth's tiny gravitational contribution will overwhelm its effect. While this reasoning usually makes the inclusion of such effects in particle physics quantities of academic interest only, it is still interesting to ask if, by some chance, there could be observable quantities which start at $H$ and have no dominating flat space or $r_s/r$ contribution. First, let us expand the metric by assuming $r_s/r$ to be small. This is fine since $r_s/r\sim 10^{-9}$ on approximating the earth's radius to be $10^4 km$ (over time-scales of say $\tau$-seconds scattering experiments, $H t\sim 10^{-18}\tau$ which is the next approximation we will do)\footnote{A similar rough estimate will put the sun's gravity to contribute at $10^{-11}$ and moon's at $10^{-12}$.}. This presents the metric as:
\begin{equation}
    ds^2=\left(1-4\frac{r_s}{r}\right)dt^2-\left(1+4\frac{r_s}{r}\right)e^{2H t}(dx^2+dy^2+dz^2)\,.
\end{equation}
Now, most terrestrial experiments can be approximated as being carried out with $r$ fixed to be the radius of the Earth; hence, one may think that gradients with respect to $r$ can be ignored. This is not quite true; the gradients will go as $r_s\Delta r/r^2$. For $O(H)$ to have the opportunity to dominate, we will need $r_s \Delta r/r^2 \ll H t$. This gives $\Delta r\ll 10^{-2}\tau$. This means that the detectors should be placed on the same horizontal plane with this precision, just to have a shot at removing the contribution from Earth's gravity! Thus, a priori, it seems pointless to consider $O(H)$ effects in terrestrial experiments. An interesting possibility is that if the T-reversal symmetry is violated, in which case in a suitable observable, the $O(H)$ becomes the dominant term. We will be interested in such observables. Thus, there are at least three reasons\footnote{Well, these depend on individual tastes!} why documenting $O(H)$ effects will be fruitful, which we reiterate here:
\begin{enumerate}
\item In the context of bootstrapping cosmological correlators \cite{snowmass,penedones}, one may want to solve the bootstrap equations in an $H$ expansion and examine the flat space limit. 

\item In the early universe, there is a huge uncertainty in $H$, and for processes where $H/m\ll 1$, such calculations could be of phenomenological relevance. Furthermore, our calculations are also valid when the scale factor $a(t)\sim 1+ H t+ O(H^2 t^2)$ and the $O(H)$ answers are expected to hold more generally whenever $H t\ll 1$. 

\item Since $t\rightarrow -t$ on its own is no longer a symmetry, as is evident from the metric---the expanding Poincaré patch (EPP) is mapped to the contracting Poincaré patch (CPP)---one may wonder if there are observables in particle physics which vanish due to this symmetry but would lead to a $O(H)$ nonvanishing result in a de Sitter space. Indeed, we will find such observables. Then it is interesting to ask what kind of precision would be needed to harbor any hope of having a local experiment to measure $H$.
\end{enumerate}

In fact, the last point has already an experimental motivation. In the 1950s, Jackson (cf Jackson's Classical Electrodynamics!), Treiman, and Wyld examined terms in the Lagrangian that would violate the T-reversal symmetry \cite{jackson}. They considered  beta decay and allowed for terms such as $\bar\psi_p \sigma^{\mu\nu}\psi_n \bar\psi_e \sigma_{\mu\nu}\psi_{\nu_e}$ which violated T if the coefficient was allowed to be complex. The reason was to look for new sources of CP violation. Similar terms in the associated Yang-Lee Lagrangian, allowing for complex coefficients, lead to what we refer to as fundamental T-violation in flat space. In polarized beta decay, such effects contributed terms called the parity-even triple correlation function $D \,\vec{J}\cdot \vec{p_e}\times \vec{p_{\nu}}$, where $\vec{J}$ is the spin of the neutron and $\vec{p_e}$ and $\vec{p_\nu}$ are the 3-momenta of the electron and neutrino, respectively. The neutron is assumed to be in the rest frame.  The emiT experiment \cite{emit} used a clever experimental setup for polarized beta decay to directly measure such T-violating terms. The present limit on this violation is $D<10^{-4}$, which is not significant to contribute enough CP violation to solve the baryogenesis problem. In an expanding universe, even without adding such explicit T-violating terms, we will find analogous effects in the beta decay. One naively expects that the dimensionless quantity that governs this violation is $H_0/m_n\sim 10^{-42}$, a hopelessly small number to see in any experiment. We find that an explicit calculation using Fermi theory (without adding any T-violating terms) shows that in an expanding universe, there are intrinsic T-violations, which are indeed suppressed by the ratio $H_0/m_n$. However, there are two distinguishing features of our results. First, one may have worried that if there are fundamental T-violations through the presence of terms like the one shown above, there would be no point in trying to calculate the $O(H)$ corrections. Our findings are subtly different from \cite{jackson}. Importantly, \cite{jackson} and \cite{emit} look for effects in the rest frame of the neutron. Our T-violating effects are non-zero only if the neutron is in motion and lead to terms like $\hat D \,\vec{J}\cdot\vec{p_n}\times \vec{p_\nu}$. The reason for this difference is that \cite{jackson} assumes flat space symmetries, while in our case, we do not have boosts, and hence reference frames are different from flat space. Another added ingredient in our calculations is that, due to energy non-conservation, for specific (extremely) fine-tuned kinematical configurations, the effect can be enhanced. The amount of fine-tuning\footnote{Explicit numbers are given later. What we find is that for $H_0\sim 10^{-42}GeV$, for fixed positions of the detectors, the 3-momentum (in $GeV$) resolution of the neutron needs $\sim 42$ digits to reach a $O(1)$ T-violation.} is commensurate with the smallness of $H_0$. Now, such fine-tuned kinematics is not possible to resolve in an experiment, and this explains why they have never been detected. Nevertheless, it is useful to document these for some brilliant future experimentalist(s) to find a workaround!

With these motivating thoughts in mind, we will embark on our journey. Fermionic quantum fields in de Sitter space were investigated in the 1960s, for example, by the work of Gursey, Lee \cite{dSGursey:1963ir} and Nachtmann \cite{dSNachtmann1967}. Nevertheless, to the best of our knowledge, there has been no documentation of the kind of corrections we will calculate in this paper. This is presumably because the correction that we consider vanishes when doing spin sums and only shows up in polarized decays. There are also technical difficulties even for this simple set-up, which arise due to the absence of time translation symmetry. We will explain how these are overcome. For now, we make the following comment. The measurable decay probability in polarized beta decay schematically takes the form ${\rm coupling}^2\times {\rm phase-space-integral}\times {\rm combination~of~} |M|^2$. In what we consider, the last factor already starts at $O(H)$. Hence, we will not have to worry about $O(H)$ corrections to the other factors. This will reduce the computational complexity involved.

Our paper is organized as follows. In Section 2, we set up the Fermionic dS S-matrix in various avatars and connect with cosmological correlators.  In Section 3, we warm up by considering scalar EFTs where we find that there are no $O(H)$ corrections in $|\mathcal{M}|^2$. In Section 4, we turn to fermionic examples. In Section 5, we discuss the T-violation in the polarized beta decay calculation in light of the emiT experiment. We conclude in Section 6. There are appendices with helpful background and additional material.

\section{Fermions in de Sitter: Cosmological correlators and S-matrix}

We want to consider a scattering process of the kind depicted in fig.(\ref{epp}). For this purpose, we find the recent discussion in \cite{pimentel1, pimentel2} especially suitable. We will begin with a brief review of this formalism before adapting it to our case.
\subsection{A brief review}
For defining the \(S\)-matrix in de Sitter, we need to define asymptotic states in the far past/future. While the \(S\)-matrix itself is an operator that exists independently of basis, the choice of basis allows one to write down matrix elements explicitly that describe how the \(S\)-matrix acts on that basis. In flat space, the choice of so-called ``in'' and ``out'' states is straightforward, where they correspond to states that evolve to states of the free theory in the far past and future, respectively~\cite{Schwartz:2014sze,Peskin:1995ev}. In de Sitter, the ambiguity is that even in free theory, the same basis does not diagonalize the Hamiltonian at all times, unlike free theory in flat space. For instance, the vacuum state at some time will be full of particles at a later time. This is the so-called `cosmological particle production' in free quantum field theory in de Sitter. However, by a suitable choice of ``in'' and ``out'' states, this additional complexity can be hidden, and we can define an \(S\)-matrix, which is diagonal in the free theory. In \cite{pimentel1, pimentel2}, these concepts were established using the scalar field as an example. For our examples, which focus on realistic particle physics processes, we will revisit this and then extend the formalism for Dirac fields in de Sitter.
First, we describe the asymptotic states in the interacting theory.

In \cite{pimentel1, pimentel2} two notions of de Sitter S-matrices have been introduced. First, let us use the conformal coordinates $\tau = -\frac{1}{H}e^{-Ht}$:
\begin{equation}
ds^2=\frac{1}{\tau^2 H^2}(d\tau^2-\sum_{i=1}^d dx_i dx_i)
\end{equation}
where the range of $\tau$ is from $-\infty$ to 0. The $t=-T/2, T/2$ in fig.(\ref{epp}) get mapped to $\tau_{i,f}=-(1/H) e^{\pm H T/2}$ respectively. Let us denote by $|n,\tau_*\rangle$ the $n$-particle state (the associated quantum numbers are suppressed) at $\tau=\tau_*$. The boundary conditions for the mode functions are such that the Hamiltonian is diagonal. 

\begin{enumerate}
\item Bunch-Davies S-matrix: This is defined by identifying the in and out states as follows:
\begin{equation}\label{eq: BD asymptotic states}
    |n\rangle_{\text{in}} \equiv |n,-\infty\rangle\,, \quad |n\rangle_{\text{out}}\equiv |n,-\infty\rangle\,, 
\end{equation}
In other words, we use the Bunch-Davies vacuum at $\tau=-\infty$ in both cases. The Bunch-Davies S-matrix (BD S-matrix for short) has the advantage that it coincides with the usual flat-space S-matrix in the $H\rightarrow 0$ limit. 
\item Unruh de Witt S-matrix: This is defined by identifying the in and out states as follows:
\begin{equation}
    |n\rangle_{\text{in}} \equiv |n,-\infty\rangle\,, \quad |n\rangle_{\text{out}}\equiv |n,0\rangle\,. 
\end{equation}
In other words, the out state is defined using a different vacuum at $\tau=0$. We should note that for both Bunch-Davies and Unruh-de Witt S-matrices, the scattering process happens from strictly past infinity (\(\tau=-\infty\)) to strictly future infinity (\(\tau=0\)).

\item Finite-\(T\) S-matrix/ UdW$_T$ S-matrix: Now we introduce a notion of an S-matrix where the scattering happens between times \(t=-T/2\) to \(t=+T/2\). And for `in' and `out' states we choose appropriate states that diagonalize the Hamiltonian before and after the scattering,
\begin{equation}
    |n\rangle_{\text{in}} \equiv \left|n,-\frac{1}{H}e^{H T/2}\right\rangle\,, \quad |n\rangle_{\text{out}}\equiv \left|n,-\frac{1}{H}e^{-H T/2}\right\rangle\,. 
\end{equation}

\end{enumerate}
From the point of view of fig.(\ref{epp}), it would appear that the detectors at the initial and final times should use the vacua at these respective times to define particles. It should be noted that the vacuum keeps changing with time due to the explicit time dependence of the Hamiltonian. The particle states belonging to the Fock space built on the vacuum at a particular time, we call `physical particle states'. The mathematical condition for choosing such vacuum at arbitrary times is presented in Eq.(\ref{eq: boundary condition scalar mode-function}). Hence, the UdW$_T$ S-matrix may appear to be the natural choice. One point to recall is that $t=\pm T/2$ in fig.(\ref{epp}) get mapped to $\tau_{i,f}=-(1/H)e^{\pm H T/2}$. Since we will be interested in the small-$H$ expansion, this corresponds to $\tau_{i,f}$ being close to $-\infty$~\footnote{The fact that \(\tau_{i,f}\) is close to infinity does not limit us from observing the corrections that we talk about in our paper. Our results apply when the comoving time \(t\) satisfies \(H t \ll 1\) which implies \(H\tau \approx -1 + H t \). Therefore, our finite time scale \(-T/2 < t <T/2\) is really embedded around \(\tau = -1/H\) which is very large (compared to \(T\)) as \(H\) is small even though \( \tau_f-\tau_i = \Delta t = T \ll 1/H\).}.  We will show shortly that at $O(H)$ and at weak coupling\footnote{For instance, in a scalar theory we mean $\phi^m$ is suppressed w.r.t. $\phi^n$ for $m>n$.}, there is no distinction between the UdW$_T$ S-matrix and the usual BD S-matrix with the same asymptotic states defined as in Eq.(\ref{eq: BD asymptotic states}) on the time-slices \(t=\pm T/2\). 



\subsection{\texorpdfstring{Bogoliubov transformations and $O(H)$ expansion}{Bogoliubov transformations and O(H) expansion}}

In the sub-horizon expansion/flat-space expansion of the mode-functions, we have to use the approximation \(e^{Ht}\approx 1 + H t + H^2t^2/2\) wherever necessary. So we \textit{cannot} consider the scattering processes from \(t=-\infty\) to \(t=+\infty\). We must confine ourselves to scattering experiments performed on time scales \(t\) satisfying \(Ht\ll 1\). Therefore, we consider scattering of incoming particles created at \(t=-T/2\) to outgoing particles to be detected at \(t=+T/2\) (with of course \(HT\ll 1\)). In the \(\tau\) coordinate this implies scattering from \(\tau_i = (-1/H -T/2)\) to \(\tau_f = (-1/H + T/2)\). So the scattering is happening `deep' inside the Poincar\'e patch.

The UdW$_T$ \(S\)-matrix can be formally computed as follows. First, express the physical `in' state in terms of (superposition) of different Bunch-Davies `in' states, and similarly for the `out' states. Finally, the UdW$_T$ \(S\)-matrix elements can be expressed as a linear combination of the Bunch-Davis \(S\)-matrix elements. Going from the Bunch-Davis \(S\)-matrix elements to the UdW$_T$ \(S\)-matrix elements can be thought of as a change of basis. Here we will show by general arguments that the UdW$_T$ \(S\)-matrix elements are the same as the Bunch-Davis \(S\)-matrix elements of the same process at leading \(H\) and at leading order in perturbation theory. Therefore, they are \textit{physically equivalent} at leading order.
Before giving the argument, we describe how this Bogoliubov transformation works for the scalars and the spinors. 

\paragraph{Bogoliubov transformation for scalar fields}
We will begin by reviewing well-known material concerning Bogoliubov transformations. Then we will compute the Bogoliubov coefficients relevant for the UdW$_T$ S-matrix. 
In the spatial momentum space, the quantized free scalar field has form,
\begin{equation}
    \phi(\mathbf k,\tau) = f_{\mathbf k}(\tau)\, a_{\mathbf k} + \bar f_{\mathbf k}(\tau)\, a_{-\mathbf k}^\dagger\,.
\end{equation}
In a different basis, the same field can be expanded in terms of different mode-functions,
\begin{equation}
    \phi(\mathbf k,\tau) = f'_{\mathbf k}(\tau)\, a'_{\mathbf k} + \bar f'_{\mathbf k}(\tau)\, a_{-\mathbf k}'^\dagger\,.
\end{equation}
Suppose that the mode functions are related by
\begin{equation}
\begin{aligned}
    f'_{\mathbf k}(\tau) = \alpha\,f_{\mathbf k}(\tau) + \beta\,\bar f_{\mathbf k}(\tau)\,,\qquad
    \bar f'_{\mathbf k}(\tau) = \alpha^*\,\bar f_{\mathbf k}(\tau) + \beta^*\, f_{\mathbf k}(\tau)\,.
\end{aligned}
\end{equation}
Using these, we will have the relation among creation-annihilation operators,
\begin{equation}
    \begin{aligned}
        a_{\mathbf k} = \alpha\,a'_{\mathbf k} + \beta^*\,a_{-\mathbf k}'^\dagger\,,\qquad
        a_{-\mathbf k}^\dagger = \alpha^* a_{-\mathbf k}'^\dagger +\beta\, a'_{\mathbf k}\,,
    \end{aligned}
\end{equation}
(\(\alpha,\beta\) only depends on \(|\mathbf k|\) for scalars due to isotropy).

Demanding that the commutation relation (\([a_{\mathbf k}\, ,a_{\mathbf k'}^\dagger]=\delta_{\mathbf k,\mathbf k'}\)) holds among the creation-annihilation operators, we get the following condition:
\begin{equation}
    |\alpha|^2 - |\beta|^2 = 1\,.
\end{equation}
We can now have the expression of \(a'_{\mathbf k}\)'s in terms of \(a_{\mathbf k}\)'s
\begin{equation}
    \begin{aligned}
        a'_{\mathbf k} = \alpha^*\,a_{\mathbf k} - \beta^*\,a_{-\mathbf k}^\dagger\,,\qquad
        a_{-\mathbf k}'^\dagger = \alpha\, a_{-\mathbf k}^\dagger -\beta\, a_{\mathbf k}\,.
    \end{aligned}
\end{equation}
Now we have to find the new vacuum \(\ket{0}'\) in terms of the previous particle states. This vacuum satisfies
\begin{equation}
\begin{aligned}
    a_{\mathbf k}' \ket{0}' = 0\,,\qquad
    or,\,\left(\alpha^*\,a_{\mathbf k} - \beta^*\,a_{-\mathbf k}^\dagger\right) \ket{0}' = 0\,.
\end{aligned}
\end{equation}
This last equation is sufficient to solve for the (normalized) new vacuum in terms of the previous particle states:
\begin{equation}
    \ket{0}' =\frac{1}{|\alpha|} e^{\frac{\beta^*}{\alpha^*}\,a_{\mathbf k}^\dagger a_{-\mathbf k}^\dagger} \ket{0}=\frac{1}{|\alpha|}\sum_{n=0}^\infty \left(\frac{\beta^*}{\alpha^*}\right)^n \ket{n}_{\mathbf k} \ket{n}_{-\mathbf k}\,. 
\end{equation}
Therefore, a single particle state in the new basis is given by
\begin{equation}
    \begin{aligned}
        \ket{1}_{\mathbf k}'
        = a_{\mathbf k}'^\dagger \ket{0}'
        = \frac{1}{|\alpha|}\left( \alpha\, a_{\mathbf k}^\dagger -\beta\, a_{-\mathbf k}\right) \sum_{n=0}^\infty \left(\frac{\beta^*}{\alpha^*}\right)^n \ket{n}_{\mathbf k} \ket{n}_{-\mathbf k} \,.
    \end{aligned}
\end{equation}
Now we shall use the formalism of Bogoliubov transformation to express the UdW\(_T\) \(S\)-matrix in terms of the BD \(S\)-matrix. To choose a basis that diagonalizes the Hamiltonian at \(\tau = \tau_*\) (we shall eventually consider \(\tau_* = \tau_{i,f}\)), \(f_{\mathbf k}'\) must satisfy the boundary condition Eq.(\ref{eq: boundary condition scalar mode-function}). This condition gives us the following
\begin{equation}
    \frac{\beta}{\alpha} = - \frac{[\partial_\tau+ i \omega_k (\tau)] f_{\mathbf k}(\tau)}{[\partial_\tau+ i \omega_k (\tau)]\bar f_{\mathbf k}(\tau)}{\bigg |}_{\tau =\tau_*}\,.
\end{equation}
Using the properties of the Hankel functions, we finally get
\begin{equation}\label{eq: beta by alpha scalar}
    r=\frac{\beta}{\alpha} =  - e^{\frac{i\pi}{2}} \frac{[H_{i\mu-1}^{(1)}(-k\tau_*)+\frac{i}{k}\left(\frac{\mu}{\tau_*}-\omega_k(\tau_*)\right)H_{i\mu}^{(1)}(-k\tau_*)]}{[H_{i\mu-1}^{(1)*}(-k\tau_*)+\frac{i}{k}\left(-\frac{\mu^*}{\tau_*}-\omega_k(\tau_*)\right)H_{i\mu}^{(1)*}(-k\tau_*)]}\,.
\end{equation}
Here \(\alpha,\beta\) are implicitly functions of \(\tau_*\) and \(k\) (also mass \(m\)). By the normalization condition, we have
\begin{equation}
    \alpha = \frac{1}{\sqrt{1-|r|^2}}\, ,\quad \beta = r\,\alpha\,.
\end{equation}
For \(\tau_*\rightarrow -\infty\), \(r\rightarrow 0\) (Bunch-Davies vacuum), and for \(\tau_*\rightarrow 0\), \(|r|\rightarrow e^{-\pi\mu}\) (Unruh-deWitt vacuum). Similarly, for our set-up at initial time \(\tau_*=\tau_i\) we will have some Bogoliubov coefficients which we denote by \(\alpha^{\text{in}},\beta^{\text{in}}\). And for the final time \(\tau_*=\tau_i\), we have \(\alpha^{\text{out}}\, ,\beta^{\text{out}}\). Now in Eq.(\ref{eq: beta by alpha scalar}), we make a small \(H\) approximation, and get\footnote{We have stripped off the phase \(e^{2i\phi/H}\) where \(\phi=\omega_k -m \sinh^{-1}(\frac{m}{k})\) from \(r\) since while taking flat space limit we strip off \(e^{i\phi/H}\) from the mode functions.}
\begin{equation}
    r = i e^{-2i \omega_k t_*}\frac{k^2 H}{4 \omega_k^3} + O(H^2)\,,\quad\text{ for our case }\,\,\ t_* = \pm \frac{T}{2}\,.
\end{equation}
This implies that we can take \(\alpha=1+ O(H^2)\) and \(\beta\) is \(O(H)\). Therefore, the `true in-vacuum' is given by,
\begin{equation}
    \ket{0,\tau_i}_{\text{in}} = \frac{1}{\Pi_{\mathbf k}|\alpha_k^{\text{in}}|}e^{\sum_{\mathbf k} \frac{\beta^{\text{in} *}_k}{2\,\alpha^{\text{in} *}_k}a_{\mathbf k}^\dagger a_{-\mathbf k}^\dagger} \ket{0,-\infty}_{\text{in}} \approx \ket{0,-\infty}_{\text{in}} + \sum_{\mathbf k} \frac{\beta^{\text{in} *}_k}{2} \ket{\{\mathbf k,-\mathbf k\},-\infty}_{\text{in}}.
\end{equation}
Similarly, the `true out-vacuum' is given by
\begin{equation}
    \ket{0,\tau_f}_{\text{out}} = \frac{1}{\Pi_{\mathbf k}|\alpha_k^{\text{out}}|} e^{\sum_{\mathbf k} \frac{\beta^{\text{out} *}_k}{2\,\alpha^{\text{out} *}_k}a_{\mathbf k}^\dagger a_{-\mathbf k}^\dagger} \ket{0,-\infty}_{\text{out}} \approx \ket{0,-\infty}_{\text{out}} + \sum_{\mathbf k} \frac{\beta^{\text{out} *}_k}{2} \ket{\{\mathbf k,-\mathbf k\},-\infty}_{\text{out}}.
\end{equation}

Now, as an example, the UdW\(_T\) scattering amplitude of scalars can be expressed in terms of BD \(S\)-matrix elements as follows
\begin{equation}
    \begin{aligned}
        S^{\text{UdW}_T}_{\mathbf k_1 \mathbf k_2 \rightarrow \mathbf k_3 \mathbf k_4 }
        &= {}_{\text{out}}\bra{\{\mathbf k_3,\mathbf k_4\},\tau_f} \{\mathbf k_1,\mathbf k_2\},\tau_i\rangle_{\text{in}}\\
        &= {}_{\text{out}}\bra{\{\mathbf k_3,\mathbf k_4\},-\infty} \{\mathbf k_1,\mathbf k_2\},-\infty\rangle_{\text{in}} + \sum_{\mathbf k} \frac{\beta^{\text{in} *}_k}{2}{}_{\text{out}}\bra{\{\mathbf k_3,\mathbf k_4\},-\infty} \{\mathbf k_1,\mathbf k_2,\mathbf k,-\mathbf k\},-\infty\rangle_{\text{in}}\\
        & \quad + \sum_{\mathbf k} \frac{\beta^{\text{out}}_k}{2}{}_{\text{out}}\bra{\{\mathbf k_3,\mathbf k_4,\mathbf k,-\mathbf k\},-\infty} \{\mathbf k_1,\mathbf k_2\},-\infty\rangle_{\text{in}} + O(H^2)\,.
    \end{aligned}
\end{equation}
The first term is the BD \(S\)-matrix of the same process. The second and third terms are the BD \(S\)-matrix elements with extra particles in the `in' state and `out' state, respectively. To see some explicit calculations, we assume the following kind of EFT Lagrangian, \(\mathcal L_{eff} = \sum_{n\geq 4} \frac{\lambda_n}{n!}\phi^n\). Then the first term is (taking proper normalization of the creation-annihilation operators),
\begin{equation}
\begin{aligned}
    {}_{\text{out}}\braket{\{\mathbf k_3,\mathbf k_4 \},-\infty | \{\mathbf k_1,\mathbf k_2\},-\infty}_{\text{in}}\quad \quad \quad &\\
    =\frac{\lambda_4 (2\pi)^d \delta(\mathbf k_1+\mathbf k_2-\mathbf k_3-\mathbf k_4)}{V^2}&\int dt\, (e^{- Ht})^d f_{\mathbf k_1} (t) f_{\mathbf k_2} (t) \bar f_{\mathbf k_3} (t) \bar f_{\mathbf k_4} (t)\,.
\end{aligned}
\end{equation}
A typical contribution in the second term is,
\begin{equation}
    \begin{aligned}
        \frac{\beta^{\text{in} *}_k}{2}{}_{\text{out}}\bra{\{\mathbf k_3,\mathbf k_4\},-\infty} \{\mathbf k_1,\mathbf k_2,\mathbf k,-\mathbf k\},-\infty\rangle_{\text{in}}\quad \quad \quad &\\
        = \frac{\beta^{\text{in} *}_k}{2} \frac{\lambda_6 (2\pi)^d \delta(\mathbf k_1+\mathbf k_2-\mathbf k_3-\mathbf k_4)}{V^3} &\int dt\,  f^{(0)}_{\mathbf k_1} (t) f^{(0)}_{\mathbf k_2} (t) f^{(0)}_{\mathbf k} (t) f^{(0)}_{-\mathbf k} (t) \bar f^{(0)}_{\mathbf k_3} (t) \bar f^{(0)}_{\mathbf k_4} (t)\,.
    \end{aligned}
\end{equation}
There is an important point to be noted. The energy-conserving delta function coming from the zeroth-order contribution of the first term is \(\delta(\omega_1+\omega_2-\omega_3-\omega_4)\), but that from the typical second term is \(\delta(\omega_1+\omega_2+2\omega_k-\omega_3-\omega_4)\) and from the typical third term is \(\delta(\omega_1+\omega_2-2\omega_k-\omega_3-\omega_4)\). Noting the spatial momentum-conserving delta function, all the above terms have different support in the space of kinematic configurations. Since \(\omega_k \geq m \gg H\), all the cross terms vanish in the scattering probability, and there is no contribution of the second and third term at \(O(H)\).

\paragraph{Bogoliubov transformation for spinor fields}

If we write the spinor field in the momentum space in two different ways, as,
\begin{equation}
    \begin{aligned}
        \psi(\mathbf k,\tau) = \sum_r \left(u^r_{\mathbf k}(\tau)\,b^r_{\mathbf k} + v^r_{-\mathbf k}(\tau)\,d^{r\dagger}_{-\mathbf k}\right)
        = \sum_r \left(u'^r_{\mathbf k}(\tau)\,b'^r_{\mathbf k} + v'^r_{-\mathbf k}(\tau)\,d'^{r\dagger}_{-\mathbf k}\right)\,,
    \end{aligned}
\end{equation}
then the Bogoliubov transformation of the mode functions is given by\footnote{We choose a basis in the spinor space such that for different values of \(r\), the basis does not mix under the Bogoliubov transformation, that is, the \(r\) label is `conserved' under the Bogoliubov transformation. For our case, it will be the helicity basis, to be shown later.},
\begin{equation}
    \begin{aligned}
        u'^r_{\mathbf k}(\tau) = \alpha_{\mathbf k}^{(1)} u^r_{\mathbf k}(\tau) + \beta_{\mathbf k}^{(1)} v^r_{-\mathbf k}(\tau)\,,\qquad
    v'^r_{-\mathbf k}(\tau) = \alpha_{-\mathbf k}^{(2)} v^r_{-\mathbf k}(\tau) + \beta_{-\mathbf k}^{(2)} u^r_{\mathbf k}(\tau)\,.
    \end{aligned}
\end{equation}
Correspondingly, we get the Bogoliubov transformation of the creation-annihilation operators as
  $      b^r_{\mathbf k} = \alpha_{\mathbf k}^{(1)} b'^r_{\mathbf k} + \beta_{-\mathbf k}^{(2)}d'^{r\dagger}_{-\mathbf k}\,,
        d^{r\dagger}_{-\mathbf k} = \alpha_{-\mathbf k}^{(2)} d'^{r\dagger}_{-\mathbf k} + \beta_{\mathbf k}^{(1)} b'^r_{\mathbf k} $
while the canonical anti-commutation relation gives,
 $   |\alpha_{\mathbf k}^{(1)}|^2 + |\beta_{-\mathbf k}^{(2)}|^2 =1,\,\,\,|\alpha_{-\mathbf k}^{(2)}|^2 + |\beta_{\mathbf k}^{(1)}|^2 = 1\,.$
Using \(\{b^r_{\mathbf k},d^{r}_{-\mathbf k}\}=0\),
and from the requirement \(v'^{r\dagger}_{-\mathbf k}(\tau)u'^{r}_{\mathbf k}(\tau) = 0\),
we finally find (since we can choose the \(\alpha\)'s to be real),
\begin{equation}
    \beta_{\mathbf k}^{(1)} = \beta_{-\mathbf k}^{(2)}\,,\quad\alpha_{-\mathbf k}^{(2)} =  \alpha_{\mathbf k}^{(1)}\,,\quad \beta_{\mathbf k}^{(1)*}=-\beta_{\mathbf k}^{(1)}\,.
\end{equation}
Therefore, we really have one set of Bogoliubov coefficients,
\begin{equation}
        b^r_{\mathbf k} = \alpha_{\mathbf k}^{(1)} b'^r_{\mathbf k} + \beta_{\mathbf k}^{(1)}d'^{r\dagger}_{-\mathbf k} \,, \quad 
        d^{r\dagger}_{-\mathbf k} =  \alpha_{\mathbf k}^{(1)}  d'^{r\dagger}_{-\mathbf k} + \beta_{\mathbf k}^{(1)} b'^r_{\mathbf k}\,. 
\end{equation}
These equations can be inverted to give,
\begin{equation}
        b'^r_{\mathbf k} = \alpha_{\mathbf k}^{(1)} b^r_{\mathbf k} -\beta_{\mathbf k}^{(1)} d^{r\dagger}_{-\mathbf k}\, , \quad 
        d'^{r\dagger}_{-\mathbf k} = \alpha_{\mathbf k}^{(1)} d^{r\dagger}_{-\mathbf k} - \beta_{\mathbf k}^{(1)} b^r_{\mathbf k}\,.
\end{equation}
The next important step is to find the new vacuum state in terms of the previous particle states. The new vacuum \(\ket{0}'\) must satisfy
\begin{equation}
        b'^r_{\mathbf k} \ket{0}' = \left(\alpha_{\mathbf k}^{(1)} b^r_{\mathbf k} -\beta_{\mathbf k}^{(1)} d^{r\dagger}_{-\mathbf k}\right) \ket{0}' = 0\, , \quad
        d'^{r}_{-\mathbf k} \ket{0}' = \left( \alpha_{\mathbf k}^{(1)} d^{r}_{-\mathbf k} + \beta_{\mathbf k}^{(1)} b^{r\dagger}_{\mathbf k}\right)\ket{0}' = 0\,.
\end{equation}
The solution is
\begin{equation}
    \ket{0}' = \alpha_{\mathbf k}^{(1)}\ket{0} + {\beta_{\mathbf k}^{(1)}}\ket{1}_{\mathbf k}\ket{\bar 1}_{-\mathbf k}\,.
\end{equation}
But for single particle states, from explicit calculation we can derive that (suppressing the \(r\) label),
\begin{equation}
    \ket{1}'_{\mathbf k} = b'^\dagger_{\mathbf k} \ket{0}'
    = \ket{1}_{\mathbf k}\,.
\end{equation}
There is no difference between the single-particle states even after the Bogoliubov transformation. However, the new vacuum has an overlap with the previous two-particle state with the amplitude proportional to \(\beta\).

 We are interested in the Bogoliubov transformation between the Bunch-Davis particle basis and the basis that diagonalizes the Hamiltonian at some time \(\tau=\tau_*\). The new spinors \(u'_{\mathbf k}\) and \(v'_{\mathbf k}\) involve the primed mode functions mentioned in Eq.(\ref{eq: new spinor mode functions}) and Eq.(\ref{eq: relation between spinor bogoliubovs}), which we expand in terms of the BD spinors,
\begin{equation}
    u'^r_{\mathbf k}(\tau) = \sum_s \alpha^r_s u^s_{\mathbf k}(\tau) + \sum_s \beta^r_s v^s_{-\mathbf k}(\tau)\,.
\end{equation}
The explicit forms of the spinors in both RHS and LHS (involving coefficients \(\alpha,\beta\)) are known. Therefore, it suffices to use the orthonormality condition of the spinors to determine the coefficients \(\alpha^r_s\) and \(\beta^r_s\). We find that
\begin{equation}
    \alpha_s^r = u^{s\dagger}_{\mathbf k}(\tau)u'^{r}_{\mathbf k}(\tau)  =\alpha\, \delta_s^r\,,
\end{equation}
\begin{equation}\label{eq: beta r,s}
    \beta_s^r = v^{s\dagger}_{-\mathbf k}(\tau)u'^{r}_{\mathbf k}(\tau)  = \beta\,\left(\chi_{-\mathbf k}^{s\dagger}\frac{\mathbf \gamma \cdot\mathbf k}{k}\xi^r_{\mathbf k}\right)\,.
\end{equation}
This can be made proportional to \(\delta_s^r\) too if we choose helicity spinor basis as in Eq.(\ref{eq: helicity spinors}). It can be seen from the helicity operator (defined in Eq.(\ref{eq: helicity operator})) that
\begin{equation}
    \frac{\mathbf \gamma \cdot\mathbf k}{k} = 2 \begin{pmatrix}
        0 & \hat h_{\mathbf k}\\
        - \hat h_{\mathbf k} &0
    \end{pmatrix}
\end{equation}
which simplifies Eq.(\ref{eq: beta r,s}) to
\begin{equation}
    \beta_s^r = - r\,\beta \,\delta_s^r
\end{equation}
where \(r,s \in \{+,-\}\) are now helicity labels. The coefficients \(\alpha,\beta\) at some \(\tau=\tau_*\) can be determined from Eq.(\ref{eq: beta by alpha spinors}) and Eq.(\ref{eq: spinor alpha beta normalization}). For \(\tau_*\rightarrow 0\) we have \(|\beta/\alpha| \rightarrow e^{-\pi m/H}\), relevant for the UdW \(S\)-matrix. Finally, we can evaluate the explicit forms of \(\alpha\) and \(\beta\) to \(O(H)\) using the asymptotic expansion of Hankel functions. The result is\footnote{Stripping off phase \(e^{2i\phi/H}\) where \(\phi=\omega_k - m\sinh^{-1}(\frac{m}{k})\)},
\begin{equation}
    \frac{\beta}{\alpha}=r_+ = \pm \frac{i e^{-2i\omega_k t_*}\, m}{4 \omega_k^2}   H + O(H^2) = - r_-\,.
\end{equation}
As we have the normalization condition \(|\alpha|^2+|\beta|^2 = 1\), we have \(\alpha = 1+ O(H^2),\,\beta = r_+ \). Now, just like in the scalar case, there will be a \(O(H)\) contribution to the UdW$_T$ \(S\)-matrix due to Bogoliubov transformation. But those will be amplitudes involving a larger number of particles and a non-overlapping delta function with respect to the zeroth order term, hence will not contribute to the amplitude-squared.

\paragraph{The final argument}

In the exact \(H\rightarrow0\) limit, the Bunch-Davies particle states coincide exactly with the physical particle states. Therefore, the Bogoliubov coefficients are \(\alpha=1\) and \(\beta=0\) in this limit. If we now consider small \(H\) corrections, then we must have
\begin{equation}
    \alpha = 1+c_\alpha H\,,\quad \beta = c_\beta H\,.
\end{equation}
But the coefficients must satisfy (-ve sign for scalars and +ve sign for fermions),
\begin{equation}
    |\alpha|^2 \pm |\beta|^2 =1\,.
\end{equation}
Since \(|\beta|^2\) contributes at order \(H^2\), we have to \(O(H)\) (we can always choose \(\alpha\) to be real),
\begin{equation}
    \alpha =1\,.
\end{equation}

As \(\beta\) is \(O(H)\), we can write the physical vacuum (for modes \(\pm\mathbf k\)) for scalar as
\begin{equation}
\begin{aligned}
    \ket{0}' &\approx \frac{1}{|\alpha|} \left(\ket{0} + \frac{\beta^*}{\alpha^*}\ket{1}_{\mathbf k}\ket{1}_{-\mathbf k}\right)\\
    &= \ket{0} + \beta^* (2\text{-particle state})\,.
\end{aligned}
\end{equation}
For fermions, it is,
\begin{equation}
\begin{aligned}
    \ket{0}' &= \alpha^* \ket{0} - \beta\ket{1}_{\mathbf k}\ket{\bar 1}_{-\mathbf k}\\
    &= \ket{0} -\beta (2\text{-particle state, fermion-anti-fermion pair})\,.
\end{aligned}
\end{equation}
A physical one-particle state of a scalar field has a form
\begin{equation}
    \begin{aligned}
        \ket{1}'_{\mathbf k}
        &\approx \frac{1}{|\alpha|} \left( \alpha\, a_{\mathbf k}^\dagger -\beta\, a_{-\mathbf k}\right) \left(\ket{0} + \frac{\beta^*}{\alpha^*}\ket{1}_{\mathbf k}\ket{1}_{-\mathbf k}\right)\\
        &\approx  \ket{1}_{\mathbf k} + \beta^* (3\text{-particle state})\,.
    \end{aligned}
\end{equation}
The physical one-particle state of a fermionic field is insensitive to the change of basis, \(\ket{1}'_{\mathbf k}=\ket{1}_{\mathbf k}\).

Since \(\alpha\) is just 1 in the leading order, a \(n\rightarrow n'\) scattering amplitude in the physical basis will be same as the \(n\rightarrow n'\) scattering amplitude  \textit{plus} \(n+2\rightarrow n'\) and \(n\rightarrow n'+2\) scattering amplitudes in the Bunch-Davies \(S\)-matrix. However, these extra contributions involving more particles generally contribute to higher orders in perturbation theory or do not contribute at all, depending on the details of the interaction Lagrangian. Therefore, we can discard these contributions, and it turns out that the physical \(S\)-matrix element that describes scattering from some initial state in the physical basis to some final state in the physical basis is equivalent to the corresponding Bunch-Davies \(S\)-matrix element at \(O(H)\).


Due to the lack of time translation invariance in this space-time, the mode functions in the time domain do not generally have a simple exponential form in general; instead they involve complicated Hankel functions. The propagators also involve these Hankel functions. In this scenario, the time integrals for each internal vertex remain (the space integrations give spatial momentum-conserving delta functions just like flat space), which cannot be done explicitly with the currently developed technologies, except for very special values of mass and space-time dimensions. Therefore, for most general cases, it is not possible to obtain an explicit analytical answer for the \(S\)-matrix elements. We will focus on the sub-horizon expansion, where we take the Hubble parameter \(H\) to be much smaller than all the relevant energy scales of the problem, and do a systematic small \(H\) expansion. A sanity check is that the $H=0$ answer is the expected flat space answer. Our task is to obtain the first non-trivial correction at \(O(H)\) for various scattering amplitudes and decay processes. 

\subsection{\texorpdfstring{Fermionic In-In Correlator from the \(S\)-matrix}{Fermionic In-In Correlator from the S-matrix}}

The Bunch-Davies \(S\)-matrix is an important quantity as it be can be related to the `in-in' cosmological correlators. For the scalar case, this has already been shown in \cite{pimentel1}. Here we will attempt the spinor case and illustrate with an explicit example. We recall that the `in-in' correlator is defined in the following way,
\begin{equation}
  \lim_{\tau\rightarrow 0} {}_{\text{in}}\bra{0} \psi(\mathbf k_1,\tau)\bar\psi(\bar{\mathbf k}_1,\tau)\psi(\mathbf k_2,\tau)\bar\psi(\bar{\mathbf k}_2,\tau)\cdots \ket{0}_{\text{in}}\,.
\end{equation}
The trick to establish a relation with the \(S\)-matrix is to write the operator product in the above expression in the basis of the particle states in the out basis as follows:
\begin{equation}
    \psi(\mathbf k_1,\tau)\bar\psi(\bar{\mathbf k}_1,\tau)\psi(\mathbf k_2,\tau)\bar\psi(\bar{\mathbf k}_2,\tau)\cdots = \sum_{j,\bar j;j'\bar j'} \ket{j,\bar j}_{\text{out}}{}_{\text{out}}\bra{j',\bar j'}\, C_{j,\bar j;j'\bar j'}\,.
\end{equation}
Let's explain the notation here carefully. The label \(j\) label denotes \(j\)-particle state with particular momenta \(\mathbf p_a\) and spin-label \(r_a\), compactly written as the set \(\{\mathbf p_a^{r_a}\}_{a=1}^j\). On the other hand, \(\sum_j\) should be interpreted as \(\Pi_{a=1}^j \sum_{r_a} \int \frac{d^3\mathbf p_a}{(2\pi)^3}\). The label \(\bar j\) is similar, just for anti-particles. The \(C\)-coefficients can be determined from
\begin{equation}
    C_{j,\bar j;j'\bar j'} = \frac{1}{j!\bar j! j' ! \bar j'!} {}_{\text{out}}\bra{j,\bar j} \psi(\mathbf k_1,\tau)\bar\psi(\bar{\mathbf k}_1,\tau)\psi(\mathbf k_2,\tau)\bar\psi(\bar{\mathbf k}_2,\tau)\cdots \ket{j',\bar j'}_{\text{out}}\,.
\end{equation}

Now the `in-in' correlator can be expressed as
\begin{equation}
\begin{aligned}
   {}_{\text{in}}\bra{0} \psi(\mathbf k_1,\tau)\bar\psi(\bar{\mathbf k}_1,\tau)\psi(\mathbf k_2,\tau)\bar\psi(\bar{\mathbf k}_2,\tau)\cdots \ket{0}_{\text{in}} &= \sum_{j,\bar j;j'\bar j'} {}_{\text{in}}\langle 0\ket{j,\bar j}_{\text{out}}{}_{\text{out}}\bra{j',\bar j'} 0\rangle_{\text{in}}\, C_{j,\bar j;j'\bar j'}\\
   &= \sum_{j,\bar j;j'\bar j'} S_{0\rightarrow j,\bar j}^* S_{0\rightarrow j',\bar j'}\, C_{j,\bar j;j'\bar j'}\,.
\end{aligned}
\end{equation}

Let us see in detail the example of the fermion 2-point function. We need to evaluate the following (we are suppressing the coordinate \(\tau\) which we need set to \(0\) at the end),
\begin{equation}
    {}_{\text{out}}\bra{j,\bar j} \psi(\mathbf k_1)\bar\psi(\bar{\mathbf k}_1)\ket{j',\bar j'}_{\text{out}}\,.
\end{equation}
To evaluate this, the key step is to expand (we are assuming that the quantum fields are free at future infinity),
\begin{equation}
     \psi(\mathbf k_1)\bar\psi(\bar{\mathbf k}_1) = \sum_{r,s}\left(u^r_{\mathbf k_1}\bar u^s_{\bar{\mathbf k}_1} b^r_{\mathbf k_1} b^{s\dagger}_{\bar{\mathbf k}_1}+v^r_{-\mathbf k_1}\bar v^s_{-\bar{\mathbf k}_1} d^{r\dagger}_{-\mathbf k_1} d^{s}_{-\bar{\mathbf k}_1}+v^r_{-\mathbf k_1}\bar u^s_{\bar{\mathbf k}_1} d^{r\dagger}_{-\mathbf k_1} b^{s\dagger}_{\bar{\mathbf k}_1}+u^r_{\mathbf k_1}\bar v^s_{-\bar{\mathbf k}_1} b^r_{\mathbf k_1} d^{s}_{-\bar{\mathbf k}_1}\right)\,.
\end{equation}
There are a few things we need to keep in mind. From fermion number conservation, there can be only an equal number of particles and anti-particles for a non-zero \(S\)-matrix. Also, there is no non-vanishing \(S_{0\rightarrow 1,\bar 1}\) kind of \(S\)-matrix. So at the lowest order, we need to care about only \(C\)-coefficients of type \(C_{0,\bar 0;0,\bar 0}\) and \(C_{2,\bar 2;2,\bar 2}\). The easiest one is,
\begin{equation}
    C_{0,\bar 0;0,\bar 0} = {}_{\text{out}}\bra{0,\bar 0} \psi(\mathbf k_1)\bar\psi(\bar{\mathbf k}_1) \ket{0,\bar 0}_{\text{out}} = (2\pi)^3 \delta(\mathbf k_1 - \bar{\mathbf k}_1) \sum_r u^r_{\mathbf k_1}\bar u^r_{{\mathbf k}_1}\,.
\end{equation}
This is the contribution, which is the free-theory value. The first non-trivial correction that can come from an interacting theory is
\begin{equation}
    \begin{aligned}
        &{}_{\text{in}}\bra{0} \psi(\mathbf k_1,\tau)\bar\psi(\bar{\mathbf k}_1,\tau)\ket{0}_{\text{in}}\\
        &\supset \frac{1}{4} (2\pi)^3 \delta(\mathbf k_1 - \bar{\mathbf k}_1) \sum_r u^r_{\mathbf k_1}\bar u^r_{{\mathbf k}_1} \sum_{r_1,r_2,\bar r_1,\bar r_2} \int \frac{d^3\mathbf p_1}{(2\pi)^3}\frac{d^3\mathbf p_2}{(2\pi)^3}\frac{d^3\bar{\mathbf p}_1}{(2\pi)^3} \frac{d^3\bar{\mathbf p}_2}{(2\pi)^3} |S_{0\rightarrow \mathbf p_1^{r_1},\mathbf p_2^{r_2},\bar{\mathbf p}_1^{\bar r_1},\bar{\mathbf p}_2^{\bar r_2}}|^2\\
        &-\frac{1}{4} \sum_{r,s} u^r_{\mathbf k_1}\bar u^s_{\bar{\mathbf k}_1}\sum_{r_2,\bar r_1,\bar r_2} \int \frac{d^3\mathbf p_2}{(2\pi)^3}\frac{d^3\bar{\mathbf p}_1}{(2\pi)^3} \frac{d^3\bar{\mathbf p}_2}{(2\pi)^3} S_{0\rightarrow \bar{\mathbf k}_1^s,\mathbf p_2^{r_2},\bar{\mathbf p}_1^{\bar r_1},\bar{\mathbf p}_2^{\bar r_2}}^* S_{0\rightarrow {\mathbf k}_1^r,\mathbf p_2^{r_2},\bar{\mathbf p}_1^{\bar r_1},\bar{\mathbf p}_2^{\bar r_2}}\\
        &+ \frac{1}{4} \sum_{r,s} v^r_{-\mathbf k_1}\bar v^s_{-\bar{\mathbf k}_1}\sum_{r_1, r_2,\bar r_2} \int \frac{d^3\bar{\mathbf p}_2}{(2\pi)^3}\frac{d^3{\mathbf p}_1}{(2\pi)^3} \frac{d^3{\mathbf p}_2}{(2\pi)^3} S_{0\rightarrow {\mathbf p}_1^{r_1},\mathbf p_2^{r_2},-{\mathbf k}_1^{r},\bar{\mathbf p}_2^{\bar r_2}}^* S_{0\rightarrow {\mathbf p}_1^{r_1},\mathbf p_2^{r_2},-\bar{\mathbf k}_1^{s},\bar{\mathbf p}_2^{\bar r_2}}
    \end{aligned}  
\end{equation}

\subsection{In-In $=$ In-Out for fermions}

It has been argued recently in \cite{pajer, smatrixmarathon} that in-in correlators can be computed using the in-out formalism where a CPP is glued to the EPP as shown in fig. (\ref{fig: pajer in-out}). Equivalence is shown when the system is closed (in our example, no poles in the lower half of the complex-$\tau$ plane) and does not have any IR divergences at $\tau \to 0$. 

We use the same primed convention as given in \cite{pajer} to exclude the spatial momentum conserving Dirac delta function, i.e., 
\begin{equation}
    \braket{\prod_a^n \phi(\tau_a, \vec{k}_a) } \equiv (2\pi)^d \delta\left(\sum^n_a \vec{k}_a\right)\braket{\prod_a^n \phi(\tau_a, \vec{k}_a)}{}'.
\end{equation}
Now, there are two ways to calculate the time-ordered correlator: the in-in correlator and the in-out correlator. Consider the product $\mathcal{O}(\{\tau, \vec{x}\})$ of a set of local operators inserted at positions $\{\tau_a, \vec{x}_a\}$ for $a = 1,\dots,n$. Then the in-in correlator is defined as  

\begin{equation}
    B_{ \text{in-in}} \equiv \braket{ 0| \bar{T} \left[ e^{ + i\int_{ -\infty (1 + i\epsilon)}^{\tau_0} H_{ \text{int} } d\tau } \right] T\left[\mathcal{O}(\{\tau, \vec{x}\}) e^{- i\int_{-\infty (1 - i\epsilon)}^{\tau_0} H_{ \text{int} } d\tau}\right]|0} \! {}',
\end{equation}
where $\ket{0}$ is the Bunch-Davies vacuum, $\bar{T}$ indicates \emph{anti-time} ordering, and $\tau_0 > \tau_a$. The in-out correlator is defined as 
\begin{equation}
    B_{\text{in-out}} \equiv \frac{ \braket{0|T\left[ \mathcal{O}(\{\tau, \vec{x}\}) e^{- i\int_{-\infty (1 - i\epsilon)}^{+\infty (1 - i\epsilon)} H_{ \text{int} } d\tau} \right]|0} \! {}' }{ \braket{0|T\left[ e^{- i\int_{-\infty (1 - i\epsilon)}^{+\infty (1 - i\epsilon)} H_{ \text{int} } d\tau} \right]|0} \! {}' }.
\end{equation}

We want to show the equivalence of in-in and in-out correlators for spinor fields for a contact interaction term. The fields are inserted at times $\tau_a \leq 0$ with $\mathcal{O} = \bar{\psi}_1(\tau_1)\psi_2(\tau_2)\bar{\psi}_3(\tau_3)\psi_4(\tau_4)$. 
The in-in correlator up to first order in perturbation theory (contact diagram) is given by 
\begin{equation}
    B_{ \text{in-in}} \equiv -i \int_{-\infty (1 - i\epsilon)}^{0} d\tau \braket{ 0|T\left[\mathcal{O} H_{ \text{int} } \right]|0} \! {}' + i\int_{ -\infty (1 + i\epsilon)}^{0}  d\tau\braket{ 0| \bar{T} \left[ H_{ \text{int} } \right] T\left[\mathcal{O}\right]|0} \! {}',
\end{equation}
For the interaction \(H_{\text{int}}\propto\left(\bar\psi\psi\right)\left(\bar\psi\psi\right)\) this is\footnote{We have redefined the fields as specified in (\ref{dirac field quantization}).} 
\begin{align}
    B_{\text{in-in}} &\propto\; \int_{-\infty (1 - i\epsilon)}^{0} d\tau\, (-\tau H)^2\, 
    \left[G^F_{\vec{k}_1}(\tau,\tau_1)G^F_{\vec{k}_2}(\tau_2,\tau)\}G^F_{\vec{k}_3}(\tau,\tau_3)G^F_{\vec{k}_{{4}}}(\tau_4,\tau)\right] \notag \\
     - \int_{-\infty (1 + i\epsilon)}^{0} &d\tau\, (-\tau H)^2\, 
    \left[G_{\vec{k}_1}(\tau, \tau_1; +1)G_{\vec{k}_2}(\tau, \tau_2; -1) G_{\vec{k}_3}(\tau, \tau_3; +1) G_{\vec{k}_{\textcolor{blue}{4}}}(\tau, \tau_4; -1) \right]\,,
\end{align}
where 
\begin{align}
    \{G^F_{\vec{k}}(\tau, \tau')\}^{\alpha\beta} \equiv 
    \bra{0} \mathrm{T}\psi^\alpha(\tau,\vec{k}) \, \bar{\psi}^\beta(\tau', \vec{k}) \ket{0}\,,
\end{align} 
and  
\begin{equation}
    \begin{aligned}
        \{G_{\vec{k}}(\tau, \tau'; +1)\}^{\alpha\beta} &\equiv \bra{0} \psi^\alpha(\tau,\vec{k}) \, \bar{\psi}^{\beta}(\tau', \vec{k}) \ket{0} = u^\alpha(\tau, \vec{k})\bar{u}^\beta(\tau', \vec{k})\,, \\
        \{G_{\vec{k}}(\tau, \tau'; -1)\}^{\alpha\beta} &\equiv \bra{0} \bar{\psi}^\alpha(\tau,\vec{k}) \, \psi^\beta(\tau', \vec{k}) \ket{0} = v^\alpha(\tau',\vec{-k})\bar{v}^\beta(\tau, \vec{-k})\,.
    \end{aligned}
\end{equation}
The in-out correlator up to first order in perturbation theory (contact diagram) is given by 
\begin{equation}
    B_{\text{in-out}} \propto \int_{-\infty (1 - i\epsilon)}^{+\infty (1 - i\epsilon)} d\tau (-\tau H)^2 \left[G^F_{\vec{k}_1}(\tau,\tau_1)G^F_{\vec{k}_2}(\tau_2,\tau)\}G^F_{\vec{k}_3}(\tau,\tau_3)G^F_{\vec{k}_{{4}}}({{4}},\tau)\right]\,.
\end{equation}
This leads to 
\begin{equation}
    \begin{aligned}
    B_{\text{in-out}} - B_{\text{in-in}} & \propto \bigg(\int_{0}^{+\infty (1 - i\epsilon)} d\tau (-\tau H)^2 [\prod_{a=1}^4 G_{\vec{k}_a}(\tau, \tau_a; (-1)^{a+1})] 
    \\ & \qquad \qquad \qquad + \int_{ -\infty (1 + i\epsilon)}^{0} d \tau (-\tau H)^2 [\prod_{a=1}^4 G_{\vec{k}_a}(\tau, \tau_a; (-1)^{a+1})] \bigg) 
    \\ & \propto \int_{ -\infty (1 + i\epsilon)}^{+\infty (1 - i\epsilon)} d \tau (-\tau H)^2 [\prod_{a=1}^4 G_{\vec{k}_a}(\tau, \tau_a; (-1)^{a+1})].
    \end{aligned} 
\end{equation}
This implies that 
\begin{equation}
    B_{\text{in-out}} - B_{\text{in-in}} \propto \int_{ -\infty (1 + i\epsilon)}^{+\infty (1 - i\epsilon)} d \tau (-\tau H)^2 u_{1\alpha_1}(\tau, \vec{k_1})\bar{v}_{2\alpha_2}(\tau, -\vec{k_2})u_{3\alpha_3}(\tau, \vec{k_3})\bar{v}_{4\alpha_4}(\tau, -\vec{k_4})
\end{equation}
where $\alpha_i$'s labels the spinor components. As the mode functions (\ref{modefunc definition}) are finite in the limit $\tau \to 0 $, the integrand is convergent in this limit, which means that there is no IR divergence present. To evaluate the above integral, we close the contour in the lower half of the complex plane.

\begin{figure}[h!]
    \centering
    \includegraphics[width=0.5\linewidth]{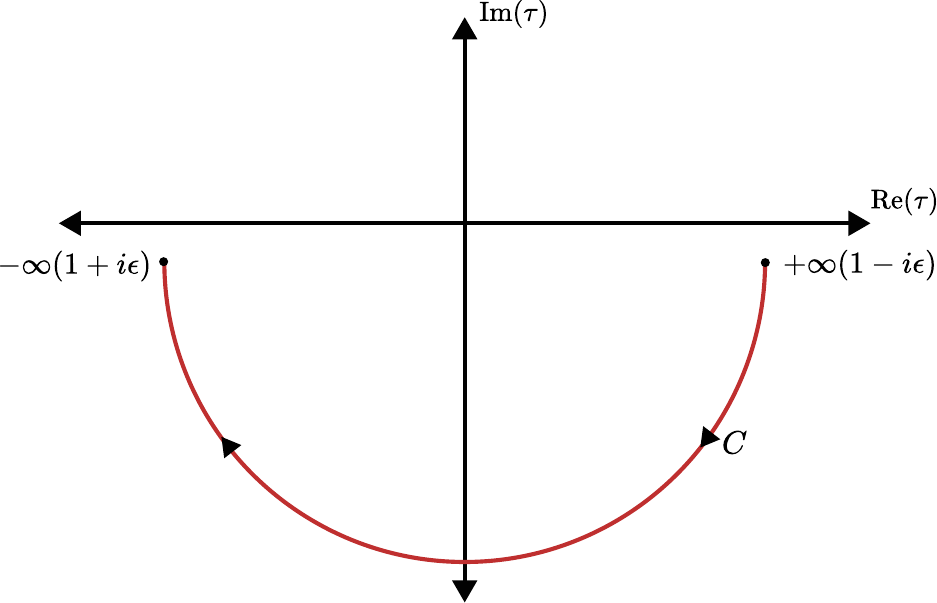}
    \caption{Contour for integral}
    \label{fig:enter-label}
\end{figure}

In the limit $|\tau| \to \infty$ $ u(\tau , \vec{k}) \propto e^{-i k \tau} $ and $v(\tau, \vec{k}) \propto e^{i k \tau }$ therefore, the arc's contribution is proportional to 
\begin{equation}
    \int_C d\tau \, \tau^2 e^{-i k_T\tau}\,, \quad \text{where}\quad k_T = k_1 + k_2 + k_3 + k_4 .
\end{equation}
We want to put bounds on the above integral. Going to polar coordinates, we have
\begin{equation}
    I  \equiv \lim_{R \to \infty} R^3 i \int^{-\epsilon}_{-\pi + \epsilon} d \theta  \, e^{i 3 \theta}e^{-i R  k_T \cos \theta + R k_T \sin \theta}.
\end{equation}
Using the inequality 
\begin{equation}
    \left|\int_a^b f(x) dx\right|\leq \int_a^b|f(x)|dx 
\end{equation}
and replacing $\theta$ with $-\theta$ we get
\begin{equation}
    |I| \leq \lim_{R \to \infty} R^3\int_\epsilon^{\pi - \epsilon} d \theta \,e^{- R k_T \sin \theta } = \lim_{R \to \infty} 2 R^3\int_\epsilon^{\pi/2} d \theta \,e^{- R k_T \sin \theta } .
\end{equation}
As for $\theta \in [0,\pi/2],\, \sin \theta \geq 2 \theta / \pi $ therefore
\begin{equation}\label{jordan}
    |I| \leq \lim_{R \to \infty} 2 R^3\int_\epsilon^{\pi/2} d \theta \,e^{-2 R k_T \theta/ \pi } = \lim_{R \to \infty} \pi \frac{R^2}{k_T}(e^{-2 R k_T \epsilon /\pi} - e^{- R k_T }) = 0\,.
\end{equation}
So, the arc contribution is zero. We can choose our mode functions such that they are analytical in the lower half of the complex plane, so the integral over the closed contour goes to zero, which implies that the difference of in-in and in-out correlator is zero!  Note the importance of the regulator $\epsilon$ in arriving at this result. In eq.(\ref{jordan}), had $\epsilon$ been set to zero, before taking the $R\rightarrow \infty$ limit, the rhs would go to infinity. This is exactly what is expected from Jordan's lemma. The presence of $\epsilon$ and the order in which the limits are taken (first $R\rightarrow \infty$ followed by $\epsilon\rightarrow 0$) are vital. 

The interaction term mentioned here ensures that there are no IR divergences present as $\tau \to 0 $. We can generalize this for a generic $n$ point fermionic contact interaction which involves $n_{\del_i}$ spatial and $n_{\del_\tau}$ time derivatives, following \cite{pajer}. For massless fermions, the condition for no IR divergences at $\tau = 0$ is $nd/2 + n_{\del_i} + 2n_{\del_\tau} > d$ ($d$ is the number of spatial dimensions). This can be easily derived by noting the scaling dimension of the fermionic field as $\tau \to 0$, and counting factors of $\tau$ coming from the inverse metric, which does the contraction of the derivatives. Since a fermionic interaction term always has $n > 2$, this condition is always satisfied in any number of dimensions.

\subsection{Sub-horizon expansion of the mode functions}\label{small h}

We have seen that under certain assumptions, the UdW$_T$ \(S\)-matrix elements are equivalent to the corresponding Bunch-Davies \(S\)-matrix elements at the leading order in Hubble parameter \(H\). Thus, we can use the Bunch-Davies states to simplify our calculations. As mentioned earlier, the amplitude involves the integration of products of Hankel functions, which turn out to be very complicated. Hence, to make our lives easier, we will work with a perturbative approximation of the Hankel functions at \(O(H)\).

Bunch-Davies mode functions for the scalars as well as spinors involve Hankel functions with complex order: for scalar mode functions, it is \(ip = i\sqrt{m^2/H^2 - d^2/4}\), while for spinor mode functions it is \(ip = im/H \pm 1/2 \), where \(m\) denotes the mass of the fields. In the small \(H\) limit, the complex orders acquire large imaginary parts. Another observation is that in the small $H$ limit the argument of Hankel functions \(x=-|\mathbf k|\tau=(|\mathbf k|/H) e^{-Ht}\) also becomes a large real number. But the quantity \(p/x\) assumes a finite value in this limit. Therefore, we need an asymptotic expansion of Hankel functions \(H_{ip}(x)\) with large \(p\) and large \(x\) keeping \(p/x\) finite. Indeed, such an asymptotic expansion exits~\cite{bateman_bateman_manuscript_project_1953}. The required expansion for the Hankel function of the first kind is
\begin{equation}
\begin{split}
     H_{ip}^{(1)}(x) \sim  \sqrt{\frac{2}{\pi}}\frac{e^{i\sqrt{p^2 + x^2}} \left(-\frac{p}{x} + \sqrt{1 + \frac{p^2}{x^2}}\right)^{ip}}{(p^2 + x^2)^{\frac{1}{4}}}e^{\left(\frac{p\pi}{2} - i\frac{\pi}{4}\right)}\\
     \times \left[\sum_{n=0}^{N} (-i)^n 2^n b_n \Gamma\left(n + \frac{1}{2}\right) (p^2 + x^2)^{-\frac{n}{2}} + O(x^{-N-1})\right]\,.
\end{split}
\end{equation}

\noindent where $b_n$'s are given as follows 
\begin{equation}
    \begin{split}
        b_0 = 1,\quad b_1 = \frac{1}{8} - \frac{5}{24}\left(1 + \frac{x^2}{p^2}\right)^{-1},\quad b_2 = \frac{3}{128} - \frac{77}{576}\left(1 + \frac{x^2}{p^2}\right)^{-1} +  \frac{385}{3456}\left(1 + \frac{x^2}{p^2}\right)^{-2}, \cdots\,.
    \end{split}
\end{equation}
Since \((p^2 + x^2)^{-\frac{n}{2}}\sim H^n\) in the small \(H\) limit, for the purpose of first-order corrections, we only need to keep terms \(n=0,1\) in the summation appearing in the asymptotic expansion. It is now straightforward to calculate the mode functions, including the \(O(H)\) correction. We state the results below.

\paragraph{Scalar mode-function:} The small \(H\) expansion of the scalar mode-function (we now completely shift to using comoving time coordinate \(t\)) is given by\footnote{We have stripped off the phase \(e^{i\phi/H}\) where \(\phi=\omega_k -m \sinh^{-1}(\frac{m}{k})\) while taking the \(H\rightarrow 0\) limit},
\begin{equation}
    f_{\mathbf k}(t) = \frac{e^{-i\omega_k t}}{\sqrt{2\omega_k}}\left(1+i c_0^B H + c_1^B Ht + i c_2^B H t^2 \right) = \frac{e^{-i\omega_k t}}{\sqrt{2\omega_k}}(1+ f_{\mathbf k}^{(1)}(t))\,
\end{equation}
where
\begin{equation}
    c_0^B = \frac{1}{8}\frac{d^2}{m}\sinh^{-1}(m/k) + \frac{5m^2 - 3\omega_k^2}{24\omega_k^3}\,,\quad c_1^B = \frac{k^2}{2\omega_k^2}\,,\quad c_2^B = \frac{k^2}{2\omega_k}\,.
\end{equation}
The first order correction of the scalar mode-function has an interesting property,
\begin{equation}
    \bar f_{\mathbf k}^{(1)} (t) = -f_{\mathbf k}^{(1)}(-t)\,.
\end{equation}

\paragraph{Scalar propagator:} The scalar propagator will be useful for computing tree-level exchange diagrams in de Sitter spacetime. In momentum space, the propagator is defined as follows
\begin{equation}
    G_{\mathbf k} (t,t') = \int d^3\mathbf x\, e^{-i \mathbf k\cdot \mathbf x} \langle 0|\phi (\mathbf x, t) \phi (\mathbf 0, t')|0\rangle \Theta (t-t') + \int d^3\mathbf x\, e^{-i \mathbf k\cdot \mathbf x} \langle 0| \phi (\mathbf 0, t') \phi (\mathbf x, t)|0\rangle \Theta (t'-t)\,.
\end{equation}

We now use the mode decomposition of the scalar fields, which will give
\begin{equation}
\begin{aligned}
    G_{\mathbf k} (t,t') &= f_{\mathbf k} (t) \bar f_{\mathbf k} (t') \Theta (t-t') + f_{-\mathbf k} (t') \bar f_{-\mathbf k} (t) \Theta (t'-t)\\
    &= f_{\mathbf k} (t) \bar f_{\mathbf k} (t') \Theta (t-t') + f_{\mathbf k} (t') \bar f_{\mathbf k} (t) \Theta (t'-t)\,.
\end{aligned}
\end{equation}
Going to the last line we have used that, \(f_{-\mathbf k} (t) = f_{\mathbf k} (t)\) for scalar mode-functions. We have,
\begin{equation}
    f_{\mathbf k} (t) \bar f_{\mathbf k} (t') = \frac{e^{-i\omega_k (t-t')}}{2\omega_k} \left( 1 + c_1^B H (t+t')  + i c_2^B H (t^2 - t'^2)  \right)\,.
\end{equation}
Then we use the following representation of the \(\Theta\)-function,
\begin{equation}
    \Theta (t-t')\,e^{-i\omega_k (t-t')} = \int \frac{d k^0}{2\pi} \frac{i}{k^0 - \omega_k + i \epsilon}\,\, e^{-i k^0 (t-t')},\,\, \epsilon \rightarrow 0^+\,.
\end{equation}The integration range of the variable \(k^0\) is \(-\infty\) to \(+\infty\). The above relation can be checked by doing a simple contour integration. Thus we have,
\begin{equation}
    \begin{aligned}
    G_{\mathbf k} (t,t')
    &= f_{\mathbf k} (t) \bar f_{\mathbf k} (t') \Theta (t-t') + f_{\mathbf k} (t') \bar f_{\mathbf k} (t) \Theta (t'-t)\\
    &= \int \frac{d k^0}{2\pi } e^{- i k^0 (t-t')} \frac{i}{(k^0)^2 - \omega_k^2 + i \epsilon} \left(1 + c_1^B H (t+t') + \frac{i k^0 }{\omega_k} c_2^B H (t^2 - t'^2) \right)\\
    &= G_{\mathbf k}^{(0)} (t,t') + G_{\mathbf k}^{(1)} (t,t')\,.
\end{aligned}
\end{equation}
This representation explicitly shows the correction to the scalar propagator in de Sitter spacetime at first order in \(H\).

\paragraph{Spinor mode-functions:} The de Sitter spinor mode functions can be written as \cite{Schaub:2023scu}
\begin{equation}
        u^r_{\mathbf k}(\tau) = \sqrt{k} \left(g_{i\nu_+}(\tau)-\frac{\mathbf \gamma\cdot \mathbf k}{k}g_{i\nu_-}(\tau)\right)\xi^r,\quad v^r_{\mathbf k}(\tau) = \sqrt{k} \left((g_{i\nu_+}(\tau))^*+\frac{\mathbf \gamma\cdot \mathbf k}{k}(g_{i\nu_-}(\tau))^* \right)\chi^r\,.
\end{equation}
where \(\gamma_0 \xi^r = \xi^r\), \(\gamma_0 \chi^r = -\chi^r\), and we have the following small \(H\) expansion of the Hankel function involved\footnote{We have stripped off the phase \(e^{i\phi/H}\) where \(\phi=\omega_k -m \sinh^{-1}(\frac{m}{k})\) while taking \(H\rightarrow 0\) limit.},
\begin{equation}\label{mode function g approximation}
        g_{i\nu_+} = \frac{e^{-i\omega_k t}}{2\omega_k} \sqrt{\frac{\omega_k+m}{k}} \left(1+\mathcal O (H, t) \right)\,,\quad
        g_{i\nu_-} = \frac{e^{-i\omega_k t}}{2\omega_k} \sqrt{\frac{k}{\omega_k+m}} \left(1+\mathcal O' (H, t) \right)\,.
\end{equation}
The explicit forms of the corrections are
\begin{equation}
        \mathcal O(H, t) = i c_0^F H + c_1^F H t +  i c_2^F H t^2 \,, \quad
        \mathcal O'(H, t) = i c_0'^F H + c_1'^F H t +  i c_2'^F H t^2
\end{equation}
with,
\begin{equation}
    \begin{aligned}
        c_0^F = \frac{m (5 m - 6 \omega_k)}{24\omega_k^3}\,,\quad c_1^F = \frac{m (\omega_k - m)}{\omega_k^2}\,,\quad c_2^F = \frac{k^2}{2\omega_k}\,;
    \end{aligned}
\end{equation}
\begin{equation}
    \begin{aligned}
        c_0'^F = \frac{m (5 m + 6 \omega_k)}{24\omega_k^3}\,,\quad c_1'^F = -\frac{m (\omega_k + m)}{\omega_k^2}\,,\quad c_2'^F = \frac{k^2}{2\omega_k}\,.
    \end{aligned}
\end{equation}
So, de Sitter spinor mode functions in the small \(H\) expansion can be written as 
\begin{equation}
    \begin{aligned}
        u^r_{\mathbf k} (t) &= \frac{e^{-i\omega_k t}}{\sqrt{2\omega_k}} (\omega_k + m)^{-1/2} \left((\omega_k + m)(1+\mathcal O (H,t))-\mathbf \gamma\cdot \mathbf k (1+\mathcal O'(H,t))\right)\xi^r\,,\\
        v^r_{\mathbf k} (t) &= \frac{e^{+i\omega_k t}}{\sqrt{2\omega_k}} (\omega_k + m)^{-1/2} \left((\omega_k + m)(1+(\mathcal{O} (H,t))^*)+\mathbf \gamma\cdot \mathbf k (1+(\mathcal{O}'(H,t))^*)\right)\chi^r\,.
    \end{aligned}
\end{equation}
In the limit $H \to 0$, these are the exact flat space spinors.
While computing \(|\mathcal M|^2\) of amplitudes involving external spinors, we are required to evaluate projectors of form \(u\bar u\) or \(v\bar v\). Therefore, it is useful to compute corrections to these projectors at \(O(H)\).

 We denote the \(2\times2\) matrix for the \(r\)-spin projector to be \(P_r\), where \(r\) can either be up or down (along some chosen unit vector). So we have
\begin{equation}
    \xi^r \xi^{r\dagger} = 
    \begin{pmatrix}
        P_r & 0 \\
        0  & 0
    \end{pmatrix}\,,\quad 
    \chi^r \chi^{r\dagger} = 
    \begin{pmatrix}
        0 & 0 \\
        0  & P_{-r}
    \end{pmatrix}\,.
\end{equation}
(If \(r\) is up/down, then \(-r\) is down/up). Using these, the spinor projectors can be written as
\begin{equation}\label{uu vv first order expansion}
\begin{aligned}
    u^r_{\mathbf k}(t) \bar u^r_{\mathbf k}(t') = \frac{e^{-i\omega_k (t-t')}}{2\omega_k} (\mathcal P^r_u + \alpha^r_u (t,t'))\,,\quad
    v^r_{\mathbf k}(t) \bar v^r_{\mathbf k}(t') = \frac{e^{+i\omega_k (t-t')}}{2\omega_k} (\mathcal P^r_v + \alpha^r_v (t,t'))\,,
\end{aligned}
\end{equation}
where \(\mathcal P\) denotes the flat-space part and \(\alpha\) denotes the first order time-dependent corrections.

We have \(\sum_r \mathcal P^r_u = \slashed{k} + m \), and \(\sum_r \mathcal P^r_v = \slashed{k} - m \), with \(\slashed{k} = \gamma^0\omega_k - \mathbf \gamma\cdot \mathbf k\). Their explicit forms in the Dirac representation are given by,
\begin{equation}
\begin{aligned}
    \mathcal P^r_u &=\frac{1}{(\omega_k + m)}  
    \begin{pmatrix}
        (\omega_k+m)^2P_r & -(\omega_k+m) P_r (\mathbf \sigma \cdot \mathbf k) \\
        (\omega_k+m)(\mathbf \sigma \cdot \mathbf k) P_r & - (\mathbf \sigma \cdot \mathbf k) P_r (\mathbf \sigma \cdot \mathbf k)
    \end{pmatrix},\\ \mathcal P^r_v &=\frac{-1}{(\omega_k + m)}\begin{pmatrix}
        - (\mathbf \sigma \cdot \mathbf k) P_{-r} (\mathbf \sigma \cdot \mathbf k) & (\omega_k+m)(\mathbf \sigma \cdot \mathbf k) P_{-r} \\
        -(\omega_k+m)P_{-r} (\mathbf \sigma \cdot \mathbf k) & (\omega_k+m)^2P_{-r}
    \end{pmatrix}.
\end{aligned}
\end{equation}
The explicit forms of first order corrections are
\begin{equation}
\begin{aligned}
    \alpha_u^r &= \frac{1}{(\omega_k + m)}\begin{pmatrix}
        (\omega_k+m)^2P_r d_1 & - (\omega_k+m)P_r (\mathbf \sigma \cdot \mathbf k) d_2 \\
        (\omega_k+m)(\mathbf \sigma \cdot \mathbf k) P_r d_3 & - (\mathbf \sigma \cdot \mathbf k) P_r (\mathbf \sigma \cdot \mathbf k) d_4
    \end{pmatrix}, \\
    \alpha^r_v &= \frac{-1}{(\omega_k+m)}\begin{pmatrix}
         - (\mathbf \sigma \cdot \mathbf k) P_{-r} (\mathbf \sigma \cdot \mathbf k)\bar d_4 & (\omega_k+m)(\mathbf \sigma \cdot \mathbf k) P_{-r} \bar d_3 \\
        -(\omega_k+m)P_{-r} (\mathbf \sigma \cdot \mathbf k)\bar d_2 & (\omega_k+m)^2P_{-r} \bar d_1
    \end{pmatrix}\,,
\end{aligned}
\end{equation}
where
\begin{equation}
\begin{aligned}
    &d_1 = \mathcal O (H,t) + \bar{\mathcal O} (H,t')\,,\quad d_2 = \mathcal O (H,t) + \bar{\mathcal O'} (H,t') \,, \\
    &d_3=\mathcal O' (H,t) + \bar{\mathcal O} (H,t') \,, \quad d_4 = \mathcal O' (H,t) + \bar{\mathcal O}' (H,t') \,.
\end{aligned}
\end{equation}
It will turn out later in Sec. \ref{fermionic examples} that we will eventually have to perform the substitution $t' \to -t$. Under this substitution 
\begin{equation}
    \begin{aligned}
        & d_1 = 0\,, \quad d_2 = - \frac{H m (i - 2 \omega_k t)}{2 \omega^2_k } \,,
        \\ & d_3 = \frac{H m (i - 2 \omega_k t)}{2 \omega^2_k } \,, \quad d_4 = 0 \,.
    \end{aligned}
\end{equation}
The relevant projectors are
\begin{equation}
\begin{aligned}
    P_{r} = 
    \begin{pmatrix}
        \cos^2 (\frac{\theta}{2}) & \sin(\frac{\theta}{2}) \cos(\frac{\theta}{2}) e^{-i\phi}\\
        \sin(\frac{\theta}{2}) \cos(\frac{\theta}{2}) e^{i\phi} & \sin^2(\frac{\theta}{2})
    \end{pmatrix},\,\,
    P_{-r} = 
    \begin{pmatrix}
        \sin^2 (\frac{\theta}{2}) & -\sin(\frac{\theta}{2}) \cos(\frac{\theta}{2}) e^{-i\phi}\\
        -\sin(\frac{\theta}{2}) \cos(\frac{\theta}{2}) e^{i\phi} & \cos^2(\frac{\theta}{2})
    \end{pmatrix},
\end{aligned}
\end{equation}
where, \(\theta\) and \(\phi\) denote the direction in which the spin is measured. A sanity check for the projection operators is as follows:
\begin{equation*}
    P_{\uparrow} + P_{\downarrow} = 1.
\end{equation*}
We give the fermionic propagator at $O(H)$ in the appendix.

\subsection{\texorpdfstring{$\mc{CPT}$ transformation in de Sitter}{CPT transformation in de Sitter}}

In this section, we investigate the aspects of charge conjugation (\(\mc C\)), parity (\(\mc P\)), and time-reversal (\(\mc T\)) transformation. The first two are unitary but the time reversal is anti-unitary. In flat-space free theory, these are separately the symmetries of the theory. In the presence of interactions, individual transformations may not be a symmetry of the theory, but \(\mc{CPT}\) is always a symmetry of a local Lorentz-invariant QFT in flat space~\cite{Schwartz:2014sze,Peskin:1995ev}. In case of Poincar\'e patch de Sitter spacetime, \(\mc C \text{ and }\mc P\) transformations are exactly similar as in flat-space but time-reversal \(\mc T\) transformation has to be understood carefully, because of the directionality of time in de-Sitter space-time. Aspects of \(\mc{CPT}\) theorem in cosmology is discussed in detail with implications on cosmological wavefunctions in Ref.~\cite{goodhew2024cosmologicalcpttheorem} and constraints on parity violation in Ref.~\cite{thavanesan2025nogotheoremcosmologicalparity}. In this paper we would mainly focus on the implications of time-reversal on scattering amplitudes and probabilities at \(O(H)\). In fact, there are two notions of time-reversal. One in {\(\tau\)-coordinate \(\tau\rightarrow-\tau\) that is equivalent to \(t\rightarrow-t\) along with \(H \rightarrow -H\), which we denote by \(\mc T_\tau\)}. Another is the time-reversal in the {\(t\)-coordinate, denoted by \(\mc T_t\)}, which is \textit{not} a symmetry of the free theory in de Sitter. As we shall see, \(\mc T_t\) maps the free theory in the EPP to the same theory in the CPP (flipping the sign of \(H\)).

\subsubsection{\texorpdfstring{Charge conjugation \(\mc C\)}{Charge conjugation C}}

For real scalars, the charge conjugation transformation is trivial,
\begin{equation}
    \mc C\,a_{\mb k}\,\mc C^{-1} = a_{\mb k},\quad \mc C\,a_{\mb k}^\dagger\,\mc C^{-1} = a_{\mb k}^\dagger.
\end{equation}
Therefore, the scalar field also transforms trivially,
\begin{equation}
    \mc C\,\phi(\mb x,\tau)\,\mc C^{-1} = \phi(\mb x,\tau).
\end{equation}
For spinors, it exchanges particles and anti-particles, i.e.,
\begin{equation}
    \mc C\,b_{\mb k}^r\,\mc C^{-1} = d_{\mb k}^r, \quad \mc C\,d_{\mb k}^r\,\mc C^{-1} = b_{\mb k}^r
\end{equation}
(\(\mc C\) satisfies \(\mc C=\mc C^{-1}=\mc C^\dagger\)).\\

\noindent Now using the properties in Eq.(\ref{eq: charge conugation on gamma}) and Eq.(\ref{charge conjugation on uv}), we can find,
\begin{equation}
    \mc C\,\psi(\mb x,\tau)\,\mc C^{-1} = C\, (\bar\psi(\mb x,\tau))^T,\quad \mc C\,\bar \psi(\mb x,\tau)\,\mc C^{-1} =  -(\psi(\mb x,\tau))^T\,C^{-1}.
\end{equation}
where the matrix \(C\) is defined in Eq.(\ref{eq: charge conjugation}) and has the property \(C^2 = -1\). By direct calculation, one can find that the free Dirac Lagrangian in de Sitter
\begin{equation}
    \mc L_{0} = \bar \psi \left(i\slashed{\partial} - \frac{m}{(-\tau H)}\right)\psi.
\end{equation}
is invariant under charge conjugation, that is \(\mc C\, \mc L_{0}\, \mc C^{-1} = \mc L_{0}\).

\subsubsection{Parity \(\mc P\)}
\(\mc P\) acts as follows on the scalar field creation-annihilation operators,
\begin{equation}
    \mc P\, a_{\mathbf k}\,\mc P^{-1} = a_{-\mb k}\,,\quad \mc P\, a_{\mathbf k}^\dagger\,\mc P^{-1} = a_{-\mb k}^\dagger.
\end{equation}
Therefore, on the scalar field, it acts as
\begin{equation}
    \mc P\,\phi(\mb x,\tau)\,\mc P^{-1} = \phi(-\mb x,\tau).
\end{equation}
So, the free scalar field Lagrangian in de Sitter is invariant under \(\mc P\).

\noindent On the Dirac field creation-annihilation operators, the action is
\begin{equation}
    \mc P\, b_{\mathbf k}^r\,\mc P^{-1} = \eta_b\, b_{-\mb k}^r\,,\quad \mc P\, d_{\mathbf k}^{r \dagger}\,\mc P^{-1} = -\eta_b \,d_{-\mb k}^{r \dagger}.
\end{equation}
(It is a requirement that the intrinsic parities are opposite for particles and anti-particles, also we have the restriction \(\eta_b^2 = \pm 1\)). Using \(\gamma^0 u_{\mb k}^r (\tau) = + u_{-\mb k}^r (\tau) \text{ and } \gamma^0 v_{\mb k}^r (\tau) = - v_{-\mb k}^r (\tau) \), we can find the transformation of the Dirac field under parity,
\begin{equation}
     \mc P\,\psi(\mb x,\tau)\,\mc P^{-1} = \eta_b \gamma^0\psi(-\mb x,\tau) = \eta_b P\,\psi(-\mb x,\tau).
\end{equation}
The unitary parity matrix \(P\)  has properties,
\begin{equation}
    P \gamma^0 P^{-1} = \gamma^0\,,\quad P\gamma^i P^{-1} = -\gamma^i.
\end{equation}
From these properties, it can be shown that the free Dirac action is invariant under parity in de Sitter.

\subsubsection{Time reversal \(\mc T\)}\label{time reversal}

Here we will investigate \(\mc T_\tau\), which differs from the notion of time reversal in Minkowski spacetime as mentioned earlier. We now see exactly how it is different. The index \(\tau\) has been dropped in this section for convenience. Under time-reversal transformation, both 3-momentum and spin flips in sign. For scalar creation-annihilation operator we have
\begin{equation}
    \mc T\,a_{\mb k}\,\mc T^{-1} = a_{-\mb k}\,,\quad \mc T\,a_{\mb k}^\dagger \,\mc T^{-1} = a_{-\mb k}^\dagger.
\end{equation}
On the scalar field it acts as (we have shown the explicit dependence of Hubble parameter \(H\), which will be important),
\begin{equation}
    \begin{aligned}
        \mc T \phi^H(\mb x,\tau)\mc T^{-1}
        &=\int \frac{d^3\mb k}{(2\pi)^3} \left(\bar f_{\mb k}^H(\tau) e^{-i\mb k\cdot \mb x} a_{-\mb k} + h.c.\right)\\
        &= \int \frac{d^3\mb k}{(2\pi)^3} \left(\bar f_{-\mb k}^H(\tau) e^{+i\mb k\cdot \mb x} a_{\mb k} + h.c.\right)\\
        &= \int \frac{d^3\mb k}{(2\pi)^3} \left( f_{-\mb k}^{-H}(-\tau) e^{+i\mb k\cdot \mb x} a_{\mb k} + h.c.\right)\\
        &= \phi^{-H}(\mb x,-\tau).
    \end{aligned}
\end{equation}
(Going to the third line we have used the property of the mode-functions, and \(-H\) is to be understood as \(e^{-i\pi} H\) for the positive energy solution and \(e^{+i\pi} H\) for the negative energy solution, if we are using the analytic expressions Eq.(\ref{modefunc definition}). Also, \(-\tau\) should be understood as \(-\tau\,(1-i 0^+)\) for a positive energy solution and \(-\tau\,(1+i 0^+)\) for negative energy solution. This subtlety of analytic continuation is also addressed in Ref.~\cite{goodhew2024cosmologicalcpttheorem}.) 

If we write the free scalar field action in Eq.(\ref{eq: action of rescaled free scalar field}) as \(S_H[\phi] = \int d^d\mb x\int_{-\infty}^0 d\tau\, \mc L_{0}(\phi^H(\mb x,\tau);H,\tau)\), then under \(\mc T\),
\begin{equation}
    \mc T\, S_H [\phi]\,\mc T^{-1} = \int d^d\mb x\int_{0}^\infty d\tau\, \mc L_{0}(\phi^{-H}(\mb x,\tau);-H,\tau) = S_{-H}[\phi].
\end{equation}
It shows that the free-scalar-field action in the EPP is mapped to the free-scalar-field action in the CPP.

On the fermion creation-annihilation operator, the action of \(\mc T\) is as follows,
\begin{equation}
    \mc T\,b_{\mb k}^r\,\mc T^{-1} = \eta^r_b b^{-r}_{-\mb k},\quad \mc T\,d_{\mb k}^{r\dagger}\,\mc T^{-1} = \eta^{r*}_d d^{-r\dagger}_{-\mb k}.
\end{equation}
where the condition is that the phases for up and down spins must be opposite but are otherwise arbitrary. Here, we make a convenient choice \(\eta_b^{\uparrow}=1,\eta_b^{\downarrow}=-1,\eta_d^{\uparrow}=1,\eta_d^{\downarrow}=-1\), to be justified in the following.

On the quantized Dirac field, time reversal acts as
\begin{equation}
    \begin{aligned}
        \mc T\,\psi^H(\mb x,\tau)\,\mc T^{-1}
        &= \int \frac{d^3\mb k}{(2\pi)^3} \sum_r\left([ u_{\mb k}^{H\,r}(\tau)]^* e^{-i\mb k\cdot \mb x}\eta_b^r b^{-r}_{-\mb k} + [v_{\mb k}^{H\,r}(\tau)]^* e^{+i\mb k\cdot \mb x}\eta_d^{r*} d^{-r\dagger}_{-\mb k}\right)\\
        &= \int \frac{d^3\mb k}{(2\pi)^3} \sum_r\left([ u_{-\mb k}^{H\,-r}(\tau)]^* e^{i\mb k\cdot \mb x}\eta_b^{-r} b^{r}_{\mb k} + [v_{-\mb k}^{H\,-r}(\tau)]^* e^{-i\mb k\cdot \mb x}\eta_d^{-r*} d^{r\dagger}_{\mb k}\right)\\
        &= T \,\psi^{-H}(\mb x,-\tau),
    \end{aligned}
\end{equation}
where the time reversal matrix is \(T = \gamma^1\gamma^3\), having properties,
\begin{equation}
    T\,\gamma^{0*}\,T^{-1} = \gamma^0,\quad T\,\gamma^{i*}\,T^{-1} = -\gamma^i.
\end{equation}
We have used \(g^{-H}_{i\nu_\pm}(-\tau) = \bar g_{i\nu_\pm}^{H}(\tau),\,f^{-H}_{i\nu_\pm}(-\tau) = \bar f_{i\nu_\pm}^{H}(\tau)\), to prove the following.
\begin{equation}
    T\,u^{-H\,r}_{\mb k} (-\tau) = \eta_b^{-r} [u^{H\,-r}_{-\mb k} (-\tau)]^*,\quad T\,v^{-H\,r}_{\mb k} (-\tau) = \eta_d^{-r *} [v^{H\,-r}_{-\mb k} (-\tau)]^*.
\end{equation}
Under time reversal transformation the free Dirac action in the EPP maps to the free Dirac action in the CPP,
\begin{equation}
\begin{aligned}
    \mc T &\left(\int d^d\mb x\int_{-\infty}^0d\tau\, \bar\psi^H (\mb x,\tau) \left(i\slashed{\partial}-\frac{m}{(-\tau H)}\right)\psi^H (\mb x,\tau)\right) \mc T^{-1}\\
    &=\int d^d\mb x\int_{0}^\infty d\tau\, \bar\psi^{-H} (\mb x,\tau) \left(i\slashed{\partial}-\frac{m}{(-\tau (-H))}\right)\psi^{-H} (\mb x,\tau).
\end{aligned}
\end{equation}

\section{Scalar examples}

\subsection{Scalar tree-level 2-to-2 Scattering in de Sitter}

In scattering amplitudes involving only scalar fields, first, we discuss the case of effective field theories where only contact diagrams contribute to scattering. After that, we discuss the cases of exchange diagrams that contribute to the scattering of scalar fields. We find that in all cases the \(O(H)\) correction to the matrix element is purely imaginary, and therefore there is no correction \(O(H)\) to the probability of scattering.
 We show by detailed calculations and general arguments that this is the case.

\subsubsection{Scattering in effective field theory}

Our first goal is to analyze the contribution of effective field theory vertices in 2-2 scalar scattering in de Sitter spacetime at leading order in \(H\). We consider the scattering process \(\phi_1\phi_2 \rightarrow \phi_1\phi_2 \). 
First, we shall start with non-derivative contact interactions and then move to higher derivative contact interactions. In what follows, the time-integral always runs from $-T/2$ to $+T/2$. We will refrain from explicitly showing the limits.

 The simplest possibility is that of a non-derivative contact interaction contributing to the above-mentioned scattering process. In this case, we can assume interactions of the form,
\begin{equation}
    \mathcal L_{EFT} \supset \frac{\lambda_1}{(2!)^2} \phi_1^2\phi_2^2\,.
\end{equation}
The matrix element for \(\phi_1\phi_2 \rightarrow \phi_1\phi_2 \) is given by,
\begin{equation}
    \mathcal M_1 = \lambda_1 \int dt\, (e^{- Ht})^d f_{1,\mathbf p_1} (t)f_{2,\mathbf p_2} (t) \bar f_{1,\mathbf p_3} (t) \bar f_{2,\mathbf p_4} (t)\,.
\end{equation}
The zeroth-order part is
\begin{equation}
    \mathcal M_1^{(0)} = \frac{\lambda_1}{\Pi_i \sqrt{2\omega_i}} 2\pi \delta(\Delta \omega) \, n.
\end{equation}
We show the first-order correction coming from the first mode-function,
\begin{equation}
    \mathcal M_1^{(1)} \supset \frac{\lambda_1}{\Pi_i \sqrt{2\omega_i}} \int dt\, (i c_0^B H + c_1^B H t + i c_2^B Ht^2)\,e^{-i\Delta\omega\,t}
\end{equation}
where we have used 
\begin{equation}
    f_{\mathbf k} (t) = \frac{e^{-i\omega_k t}}{\sqrt{2\omega_k}} (1+i c_0^B H+c_1^B H t + i c_2^B Ht^2)
\end{equation}
(where \(c_0^B,c_1^B\) and \(c_2^B\) are real functions of \(\omega_k\) and mass \(m\)).

Similar first-order corrections will come from the other mode-functions as well. Then by explicit calculation, we can show that,
\begin{equation}
    (\mathcal M_1^{(1)})^* = - \mathcal M_1^{(1)}\,.
\end{equation}
We have noted that the first-order correction from \((e^{-Ht})^\#\) is also imaginary. This shows that the first-order correction is imaginary where the zeroth-order term is real. Therefore, there will not be any first-order correction in the probability \(|\mathcal M|^2\). As it turns out, there is no such correction in the cases of higher derivative corrections, also, as illustrated in the Appendix \ref{scalar} through an explicit calculation involving two-derivative contact interactions and a general argument applicable for all higher derivative contact interactions.

\subsubsection{Scattering via tree-level exchange}
Now, we analyze the case for tree-level scalar exchange for 2-2 scattering of scalar particles. At first, we will focus on non-derivative interaction, and then on derivative interaction.

First, we show explicitly how the first-order contribution coming from the simplest possible exchange interaction vanishes. Consider the following interaction Lagrangian involving two scalar fields \(\phi_1\) and \(\phi_2\) (with masses \(m_1\) and \(m_2\) respectively),
\begin{equation}
    \mathcal L_{int} = \frac{\lambda}{2!} \phi_1^2 \phi_2\,.
\end{equation}
We are interested in the following scattering, \(\phi_1\phi_1 \rightarrow \phi_1 \phi_1\). In this scattering \(s,t,u\)-channels will contribute via \(\phi_2\) exchange. We denote the incoming momenta by \(\mathbf p_1,\mathbf p_2\) and outgoing momenta by \(\mathbf p_3,\mathbf p_4\). The matrix element for \(s\)-channel exchange is given by
\begin{equation}
    \begin{aligned}
        i\mathcal M_{(s)} = (i\lambda)^2 \int dt_1\,(e^{- Ht_1})^{d/2} \int dt_2 (e^{- Ht_2})^{d/2}  f_{\mathbf p_1} (t_1)f_{\mathbf p_2} (t_1) G_{\mathbf p} (t_1,t_2) \bar f_{\mathbf p_3} (t_2)\bar f_{\mathbf p_4} (t_2)
    \end{aligned}
\end{equation}
where \(\mathbf p = \mathbf p_1 + \mathbf p_2 = \mathbf p_3 + \mathbf p_4\) (by 3-momentum conservation).

There will be different corrections coming from the mode functions as well as the internal propagator. The zeroth order term can be calculated to be
\begin{equation}
    \begin{aligned}
        \mathcal M_{(s)}^{(0)} &= \frac{+ i\lambda^2}{\Pi_i \sqrt{2\omega_i}} \int dt_1 \int dt_2 \int \frac{dp^0}{2\pi} e^{-i  (\omega_1 + \omega_2 + p^0) t_1} e^{+i  (\omega_3 + \omega_4 + p^0) t_2}  \frac{i}{(p^0)^2 - \omega_p^2 + i \epsilon}\\
        &= \frac{2\pi \lambda^2 \delta(\Delta \omega)}{\Pi_i \sqrt{2\omega_i}} \frac{1}{\omega_p^2 - (\omega_1+\omega_2)^2-i\epsilon}
    \end{aligned}
\end{equation}
(here, \(\omega_p^2 = \mathbf p^2 + m_2^2\)). It should be noted that as we work with a finite time interval, it is not an exact Dirac delta function; instead, we have to use its representation, which is given by $\delta(x) \equiv \int^{T/2}_{-T/2}dt\frac{e^{ixt}}{2 \pi} = \frac{\sin(xT/2)}{T\pi}$. This will be implicitly assumed from now on for all the other calculations in this paper.

Let us now analyze the first-order correction coming from the mode function of one incoming \(\phi_1\) particle.
\begin{equation}
    \begin{aligned}
        \mathcal M_{(s)}^{(1)} &\supset \frac{+ i\lambda^2}{\Pi_i \sqrt{2\omega_i}} \int dt_1 \int dt_2 \int \frac{dp^0}{2\pi} e^{-i  (\omega_1 + \omega_2 + p^0) t_1} e^{+i  (\omega_3 + \omega_4 + p^0) t_2}  f_{\mathbf p_1}^{(1)} (t_1) \frac{i}{(p^0)^2 - \omega_p^2 + i \epsilon}\\
        &= \frac{\lambda^2}{\Pi_i \sqrt{2\omega_i}} \frac{1}{\omega_p^2 - (\omega_3 + \omega_4)^2 - i\epsilon} \int dt_1 e^{-i\Delta\omega\,t_1} f_{\mathbf p_1}^{(1)} (t_1) \,.
    \end{aligned}
\end{equation}
Away from the resonance (\(\omega_p \neq \omega_3 + \omega_4\)) we can set \(\epsilon\) to 0, and we find that this first-order correction is purely imaginary,
\begin{equation}
    \left( \int dt_1 e^{-i\Delta\omega\,t_1} f_{\mathbf p_1}^{(1)} (t_1) \right)^*
    = -  \int dt_1 e^{-i\Delta\omega\,t_1} f_{\mathbf p_1}^{(1)} (t_1)\,.
\end{equation}
We have used \(\bar f_{\mathbf p_1}^{(1)} (-t_1)=-f_{\mathbf p_1}^{(1)} (t_1)\), which is a property of the first-order corrections. This calculation shows that all first-order corrections coming from the external scalar-mode functions are all imaginary. Now we proceed to understand the correction coming from the internal scalar propagator. We recall that the first-order correction to the scalar propagator is
\begin{equation}
    \begin{aligned}
        G_{\mathbf p}^{(1)}(t_1,t_2) = \int \frac{d p^0}{2\pi } e^{- i p^0 (t_1-t_2)} \frac{i}{(p^0)^2 - \omega_p^2 + i \epsilon} \left(c_1^B H (t_1+t_2) + \frac{i p^0 }{\omega_p} c_2^B H (t_1^2 - t_2^2) \right)
    \end{aligned}
\end{equation}
(note that \(c_1^B\) and \(c_2^B\) are functions of \(\omega_p\) and \(m_2\)). It contributes to the \(S\)-matrix element by
\begin{equation}
    \begin{aligned}
        \mathcal M_{(s)}^{(1)} &\supset \frac{+ i\lambda^2}{\Pi_i \sqrt{2\omega_i}} \int dt_1 \int dt_2 \int \frac{dp^0}{2\pi} e^{-i  (\omega_1 + \omega_2 + p^0) t_1} e^{+i  (\omega_3 + \omega_4 + p^0) t_2}  \frac{i}{(p^0)^2 - \omega_p^2 + i \epsilon} \\
        &\times \left(c_1^B H (t_1+t_2) + \frac{i p^0 }{\omega_p} c_2^B H (t_1^2 - t_2^2) \right).
    \end{aligned}
\end{equation}
 On performing one of the time integrals for each term,
\begin{equation}
    \begin{aligned}
        \mathcal M_{(s)}^{(1)} &\supset \frac{\lambda^2 H}{\Pi_i \sqrt{2\omega_i}} \left(\int dt_1\,e^{-i\Delta\omega t_1}\frac{c_1^B t_1 - i (\omega_3+\omega_4)c_2^B t_1^2/\omega_p }{\omega_p^2 - (\omega_3+\omega_4)^2} \right. \\
        &\hspace{4 pt}\left. + \int dt_2\,e^{-i\Delta\omega t_2}\frac{c_1^B t_2 + i (\omega_1+\omega_2)c_2^B t_2^2/\omega_p }{\omega_p^2 - (\omega_1+\omega_2)^2}\right).
    \end{aligned}
\end{equation}
This contribution is also purely imaginary, as can be seen that \(\int dt\, e^{- i u t} t \) is purely imaginary and \(\int dt\, e^{- i u t} t^2 \) is purely real.\\

\begin{tcolorbox}
 Therefore, all the corrections to the S-matrix element at first order in \(H\) are purely imaginary, where the zeroth order terms are purely real. Thus, there is no \(O(H)\) correction to \(|\mathcal M_{(s)}|^2\). This will also be true for the \(t,u\)-channel. Since all the zeroth-order terms are purely real, even \(|\mathcal M_{(s)}+\mathcal M_{(t)}+\mathcal M_{(u)}|^2\) will not have any \(O(H)\) correction. 
 \end{tcolorbox}
 The same conclusion holds for derivative exchange interactions, also, as detailed in the Appendix \ref{scalar}.

\subsection{Symmetry arguments for explaining results in scalar scattering}

In the last subsection, we have found by explicit calculations that there are no \(O(H)\) corrections to \(|\mathcal M|^2\) involving scalar particles. This can be explained by symmetry arguments. It has been shown in \ref{time reversal} that the scalar field transforms nicely under \(\tau\)-reversal,
\begin{equation}
    \mathcal T_\tau \phi_H(\mathbf x, t) \mathcal T_\tau^{-1} = \phi_{-H}(\mathbf x,-t)\,;
\end{equation}
\begin{equation}
    \int_{-T/2}^{T/2} dt\,e^{\alpha H t}\mathcal L(\phi_H(\mathbf x, t)) \rightarrow \int_{-T/2}^{T/2} dt\,e^{\alpha H t}\mathcal L(\phi_{-H}(\mathbf x, -t)) = \int_{-T/2}^{T/2} dt\,e^{-\alpha H t}\mathcal L(\phi_{-H}(\mathbf x, t))\,.
\end{equation}
Therefore, the theory in the expanding Poincaré patch (EPP) is mapped to the contracting Poincaré patch (CPP). Therefore, this \(\tau\)-reversal `invariance' tells us that a process in the EPP will have the same probability as the time-reversed process in the CPP. Now suppose that there is an \(O(H)\) correction to \(|\mathcal M|^2\), 
\begin{equation}
    |\mathcal M|^2 = |\mathcal M|^2_{(0)} + H \,|\mathcal M|^2_{(1)}\,. 
\end{equation}
Since the full correction term must be invariant under \( t \rightarrow -t,\, H \rightarrow -H\), \(|\mathcal M|^2_{(1)}\) must be odd under \(t\)-reversal \(\mathcal T_t\). Since the scalar fields do not have any spin structure, only possible \(\mathcal T_t\) violating terms have form,
\begin{equation}
    \mb p_i\cdot (\mb p_j \times \mb p_k)
\end{equation}
(where \(\mb p_i\) etc. are external momenta involved in the process).

However, such \(\mathcal T_t\) violating terms must also be parity \(\mathcal P\) violating. But the scalar interactions we have taken are all \(\mathcal P\) conserving. Therefore, demanding \(\mathcal P\) conservation, the only possibility left is that
\begin{equation}
    |\mathcal M|^2_{(1)} = 0 \,.
\end{equation}
This is precisely what we have found in our explicit calculations.

\section{Fermionic examples}\label{fermionic examples}

\subsection{3pt process: charged pion decaying to two fermions}

In this section, we shall analyze the decay rate of a charged pion decaying into the lepton-lepton-neutrino channel via the weak force. The main goal will be to understand how this decay rate is affected by the expanding universe at the leading order in the Hubble parameter \(H\). Specifically, we shall analyze the decay of \(\pi^+\)\footnote{ Experimental value of mass is \(m_{\pi^+}=139.57039\pm 0.00017\) MeV, and lifetime is \(\tau_{\pi^+} = (2.6033\pm 0.005)\times 10^{-8} s\) (data taken from Particle Data Group/PDG collaboration~\cite{PhysRevD.110.030001}). We are interested in the following decay of \(\pi^+\)}, which is a pseudo-scalar composed of up-quark (\(u\)) and anti-down quark (\(\bar d\)). ,
\begin{equation}
    \pi^+ \rightarrow \ell^+ \nu_\ell\,.
\end{equation}
The dominant decay channel is \(\pi^+\rightarrow \mu^+\nu_\mu\) with \(99.98770(4)\%\) branching ratio. This kind of leptonic decay at tree level is mediated by the \(W^+\) boson in the \(s\)-channel. However, at low energies, this process effectively becomes a contact interaction with coupling proportional to Fermi coupling \(G_F\)(=\(1.1663787(6)\times 10^{-5}\) GeV\({}^{-2}\)), the CKM matrix element \(V_{ud}\)(\(=0.97435 \pm 0.00016\)) and the pion decay constant \(F_\pi\)(\(=130.2\pm 0.8\) MeV) defined by,
\(\bra{0}v_{d}\gamma^\mu (1-\gamma_5)u_{u}\ket{\pi^+} = i F_\pi q^\mu\).

Therefore, effectively, we have a 3-pt vertex where the \(\pi^+\) couples through derivative interaction.  We are going to assume that $\pi^+$ is an anti-particle.

\begin{figure}[H]
    \centering
    \includegraphics[width=0.3\linewidth]{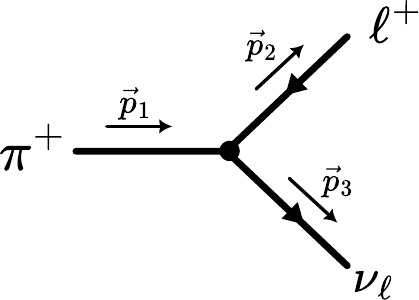}
    \caption{Leptonic decay of \(\pi^+\) via weak interaction.}
\end{figure}
\vspace{\baselineskip}We shall now analyze charged pion decay in de Sitter spacetime. We start with the effective interacting action in the $3+1$ Minkowski spacetime,

\begin{equation}
    S_{\text{int}} = \int dt d^3 \vec{x} \, i \lambda \bar{\psi}_{\nu_\ell} \gamma^\mu(1 - \gamma_5)\psi_\ell \del_\mu \pi^+,
\end{equation}
where \(\lambda = G_F V_{ud}/\sqrt{2}\). We extend this action to general curved spacetime coordinates. The action is given by
\begin{equation}
    S_{\text{int}} = \int dt d^3\mathbf x \sqrt{-g} \,\mathcal L_{\text{int}} = \int dt d^3 \vec{x} \sqrt{-g} \, i \lambda \bar{\psi}_{\nu_\ell} \gamma^\mu(1 - \gamma_5)\psi_\ell \del_\mu \pi^+ .
\end{equation}
At this point, $\gamma^\nu$ are the curved space gamma matrices, not the usual Dirac gamma matrices. The curved space versions are related to the flat space ones by $\gamma^\mu = \tns{e}{^\mu_a}\gamma^a$, where $\gamma^a$ are the usual flat space Dirac gamma matrices and $\tns{e}{^\mu_a}$ are the frame fields. More details are given in (\ref{dirac field quantization}). In de Sitter we have the relation $\tensor{e}{^\mu_a} = -\tau H \tns{\delta}{^\mu_a}$. All Dirac matrices with a Latin index are the usual Dirac matrices. Therefore, in de Sitter the net interacting action in $3 + 1 $ spacetime dimensions is given by 
\begin{equation}
    S_{\text{int}} = \int d\tau d^3 \vec{x} \sqrt{-g} \, i \lambda \bar{\psi}_{\nu_\ell} (-\tau H )\tns{\delta}{^\mu_a}\gamma^a(1 - \gamma_5)\psi_\ell \del_\mu \pi^+.
\end{equation}
We now do the rescaling of fields, $ \{\bar\psi_{\nu_\ell} , \psi_\ell, \pi^+\} \rightarrow (-\tau H)^{3/2} \{\bar\psi_{\nu_\ell} , \psi_\ell, \pi^+\}.$ The Lagrangian is given by,
\begin{equation}
    \mathcal{L}_{\text{int}} = (-\tau H )^{3/2} i \lambda \bar{\psi}_{\nu_\ell}\left(\gamma^i(1 - \gamma_5)\psi_\ell \del_i \pi^+ +\gamma^0(1 - \gamma_5)\psi_\ell (\del_\tau \pi^+ + 3/(2\tau) \pi^+\right).
\end{equation}
Using this, we can write the S-matrix element for the pion decay process, using the LSZ scheme, which is (spin labels are suppressed for clarity).
\begin{align}
\braket{\nu_\ell \ell^+ |\hat{S}| \pi^+} = -\int & d\tau d^3
    \vec{x}\,(-\tau H )^{3/2} \lambda \bar u_{\nu_\ell,\mathbf p_3}(\tau)e^{i(\vec{p_1}-\vec{p_2}-\vec{p_3}) \cdot\vec{x}} \notag \\
    \big(&(ip_1^i \gamma_i)(1 - \gamma_5)v_{\ell,\mathbf p_2}(\tau) \bar{f}_{\vec{p_1}}(\tau) 
    +\gamma^0(1 - \gamma_5)v_{\ell,\mathbf p_2}(\tau) (\del_\tau \bar{f}_{\vec{p_1}}(\tau) + 3/(2\tau) \bar{f}_{\vec{p_1}}(\tau)\big).
\end{align}
We can integrate out $\vec{x}$ to get a 3-momentum conserving delta function. Factoring out the delta function and changing the integration variable to $t$, we can write the $\mathcal{M}$-matrix element as  
\begin{align}
    \mathcal{M} = -\int dt e^{-3Ht/2} \lambda \bar u_{\nu_\ell,\mathbf p_3}(t)\gamma^a(1 - \gamma_5)v_{\ell,\mathbf p_2}(t) \bar{F}_a(t).
\end{align}
where $\bar{F}_a(t) = \left(e^{Ht}(\del_t \bar{f}_{\vec{p_1}}(t) - 3H \bar{f}_{\vec{p_1}}(t)/2), -i\vec{p_1}\bar{f}_{\vec{p_1}}(t)\right))$. This leads to 
\begin{equation}
    |\mathcal{M}|^2 = \int dt dt' e^{-3H(t + t')/2} \lambda^2 Tr\left[\gamma^a(1 - \gamma_5)u_{\nu_\ell,\mathbf p_3}(t')\bar u_{\nu_\ell,\mathbf p_3}(t)\gamma^b(1 - \gamma_5)v_{\ell,\mathbf p_2}(t) \bar{v}_{\ell,\mathbf p_2}(t') \right]F_a(t')\bar{F}_b(t).
\end{equation}
As we work in a limited time interval, the limits of the time integral are $-T/2$ to $T/2$.

\subsubsection{Unpolarized Decay}
We sum over the spin polarizations of the final products. Therefore, 
\begin{equation}\label{unpolarized pion M^2 full expression}
     |\mathcal{M}|^2 = \int dt dt' e^{-3H(t + t')/2} \lambda^2 Tr\left[\sum_{s_\ell,s_{\nu_\ell}}\gamma^a(1 - \gamma_5)u_{\nu_\ell,\mathbf p_3}(t')\bar u_{\nu_\ell, \mathbf p_3}(t)\gamma^b(1 - \gamma_5)v_{\ell,\mathbf p_2}(t) \bar{v}_{\ell, \mathbf p_2}(t') \right]F_a(t')\bar{F}_b(t).
\end{equation}
We now use the first-order expansions for $u\bar{u}$ and $v\bar{v}$ given in (\ref{uu vv first order expansion}).
As for scalars, we have 
\begin{equation}
    f_{\vec{p}}(\tau) \sim \frac{e^{i\omega_p t}}{\sqrt{2\omega_p}}(1 + f^{(1)}_{\vec{p}}(t)).
\end{equation}
Therefore, up to first-order 
\begin{equation}
    F_a(t) = \frac{e^{i\omega_p t}}{\sqrt{2\omega_p}}\left[(i \omega_p , -i \vec{p}) + \varepsilon_a(t)\right]
\end{equation}
where
\begin{equation}
    \varepsilon_a(t) = (-3H/2 + i \omega_p f^{(1)}_{\vec{p}}(t) + i\omega_p H t + \del_t(f^{(1)}_{\vec{p}}(t)), -i\vec{p}f^{(1)}_{\vec{p}}(t)).
\end{equation}
By explicitly plugging in the leading order expressions for mode functions, one can observe that the time dependence in $|\mathcal{M}|^2$ is of the form $f(t) + g(t')$ \footnote{Here $f$ and $g$ should not be confused with mode functions.}. Therefore, the integral can be simplified as
\begin{align}\label{M matrix trick of removing one time integral}
    \iint dt dt' e^{- i\Delta \omega (t - t')}(f(t) + g(t')) =&\,2\pi\delta(\Delta \omega)\left(\int dt e^{-i\Delta\omega t}f(t) + \int dt' e^{i\Delta\omega t'}g(t')\right)  \notag\\ 
    =&\,2\pi\delta(\Delta \omega)\int dt e^{-i\Delta\omega t}\left(f(t) + g(-t)\right).
\end{align}
Therefore, in $|\mathcal{M}|^2$ we can integrate $t'$ to get a finite version of the $\delta$ function, with the understanding to replace $t'$ with $-t$. 
Explicit calculations show that the correction terms are proportional to $\vec{p_1}\cdot (\vec{p_2}\times \vec{p_3})$. As a three-body collision/decay is planar in nature, this quantity goes to zero.

\subsubsection{Polarized decay}

Now we consider the case where we can differentiate the two polarizations of the lepton ($\ell^+$). Therefore,  
\begin{equation}
     |\mathcal{M}_{s_2}|^2 = \int dt dt' e^{-3H(t + t')/2} \lambda^2 Tr\left[\sum_{s_{\nu_\ell}}\gamma^a(1 - \gamma_5)u_{\nu_\ell, \mathbf p_3}(t')\bar u_{\nu_\ell, \mathbf p_3}(t)\gamma^b(1 - \gamma_5)v^{s_2}_{\ell, \mathbf p_2}(t) \bar{v}^{s_2}_{\ell, \mathbf p_2}(t') \right]F_a(t')\bar{F}_b(t).
\end{equation}
Again, using the trick of $t' \to -t $ and doing the calculations, we found that there are two correction terms in this case. The contribution from the lepton is given by 
\begin{align}
    -2 H m_\ell (\vec{\hat{j}} \cdot (\vec{p_1}\times \vec{p_2}))\frac{(2 (p_1 \cdot p_3)  + m_\pi^2 )(1 - 2 i \omega_2 t)}{\omega_2^2},
\end{align}
and the contribution from the pion is
\begin{equation}
    4H m_\ell(\vec{\hat{j}}.(\vec{p_1} \times \vec{p_2}))\left(3 - \frac{|\vec{p_1}|^2}{2 \omega_1^2} + 2 i \frac{|\vec{p_1}|^2}{2 \omega_1} t + 2i \omega_1 t\right).
\end{equation}
Therefore, after performing the integration, the correction to $|\mathcal{M}|^2$ takes the form 
\begin{equation}
    |\mathcal{M}_{s_2}|^2_{(1)} = \frac{(2\pi)^2}{\Pi_i (2\omega_i)} 2\lambda^2 H m_\ell(\vec{\hat{j}}.(\vec{p_1} \times \vec{p_2})) (C_0 \delta_T^2 (\Delta \omega)- C_1 \delta_T (\Delta \omega)\delta_T' (\Delta \omega) )
\end{equation}
where
\begin{equation}
    C_0=6 - \frac{|\vec{p_1}|^2}{ \omega_1^2}-\frac{2 (p_1 \cdot p_3)  + m_\pi^2 }{\omega_2^2}\,,\quad C_1=2  \frac{|\vec{p_1}|^2}{ \omega_1}  + 2 \omega_1 + 2 \frac{(2 (p_1 \cdot p_3)  + m_\pi^2) }{\omega_2}.
\end{equation}
Our conclusions can be summarized as follows.

\begin{tcolorbox}
Therefore, there is no first-order correction to the decay probability of the pion when we sum over all polarizations of the decay products. However, there is T-violation in polarized decay. 
\end{tcolorbox}

\subsection{4pt process: with four external fermions}
In this section, we illustrate our method of calculating the \(O(H)\) correction to de Sitter amplitudes in scattering involving four external fermions. After analyzing corrections to the scattering of 2-2 in a toy model, we move on to determining the \(O(H)\) corrections to muon decay and beta decay in realistic scenarios.
\subsubsection{Tree-level \(2\rightarrow 2\) scattering}

First, we will analyze scattering involving two species of fermions of \(\psi_1\) and \(\psi_2\) (with masses \(m_1\) and \(m_2\), respectively). Suppose that \(\psi_1\) annihilates particle \(a\) (and creates \(\bar a\)) while \(\psi_2\) annihilates particle \(b\) (and creates \(\bar b\)). We will be interested in the scattering \(a\bar a \rightarrow b\bar b\). The 3-momenta are \(\mathbf p_1,\mathbf p_2, \mathbf p_3\), and \(\mathbf p_4\) respectively.
There can be two kinds of interaction that mediate such a process: contact and exchange. The interaction terms for the Lagrangian is
\begin{equation}
    \mathcal L_{int} = \lambda_1\left( \bar \psi_1 \gamma^\mu (1-g_A^{(1)}\gamma_5) \psi_2\right) \left( \bar \psi_2 \gamma_\mu (1-g_A^{(1)}\gamma_5) \psi_1  \right) + \lambda_2 \left(\bar \psi_2 (1-g_A^{(2)}\gamma_5)\psi_1\,\phi + h.c.\right).
\end{equation}
The interaction involves a scalar field \(\phi\), mediating the exchange. This will contribute to the \(t\)-channel for the above-mentioned scattering process. We have introduced two parameters \(g_A^{(1)}\) and \(g_A^{(2)}\) to control the strength of the axial interaction.

\begin{figure}[!h]
    \centering
    \includegraphics[width=0.5\linewidth]{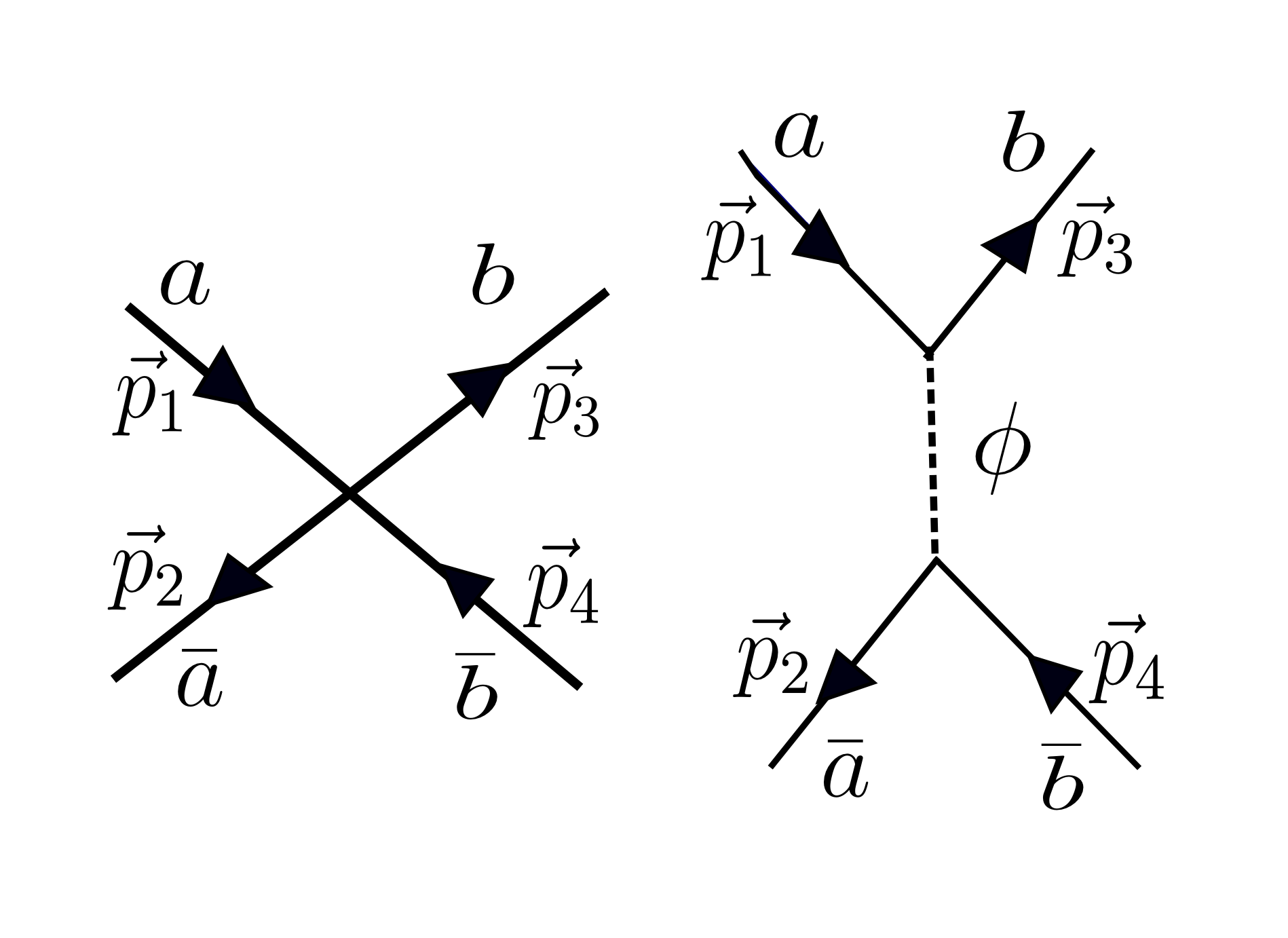}
    \caption{{2-2 scattering in de Sitter spacetime Feynman diagrams}}
    \label{fig:enter-label}
\end{figure}

 We shall always average over the spins of the incoming particles, so we show only the spin labels \(s_3\) and \(s_4\) of the outgoing particles. The matrix element of this scattering process has two contributions from the two different forms of interactions. This process is given by
\begin{equation}
    \mathcal M_{s_3,s_4} = \mathcal M_{s_3,s_4}^{(con)} + \mathcal M_{s_3,s_4}^{(ex)}.
\end{equation}
The contact diagram contribution is\footnote{We strip off 3-momentum conserving delta function \((2\pi)^2\delta(\Delta\mathbf p)\) from the matrix element.}
\begin{equation}
    \begin{aligned}
        \mathcal M_{s_3,s_4}^{(con)} = \lambda_1 \int dt\, \left(e^{-H t}\right)^{d}\,\left( \bar u_{2,\mathbf p_3}^{s_3} (t) \gamma_{\mu} (1-g_A^{(1)}\gamma_5) u_{1,\mathbf p_1} (t)\right)\left( \bar v_{1,\mathbf p_2}(t) \gamma^{\mu} (1-g_A^{(1)}\gamma_5) v_{2,\mathbf p_4}^{s_4} (t) \right).
    \end{aligned}
\end{equation}
The exchange diagram contribution is
\begin{equation}
    \begin{aligned}
        \mathcal M_{s_3,s_4}^{(ex)} = - i \lambda_2^2 \int dt_1\,\left(e^{-H t_1}\right)^{d/2} \int &dt_2\,\left(e^{-H t_2}\right)^{d/2} \left( \bar u_{2,\mathbf p_3}^{s_3} (t_1) (1-g_A^{(2)}\gamma_5) u_{1,\mathbf p_1} (t_1)\right)\\
        &\times G_{\mathbf p} (t_1,t_2) \left( \bar v_{1,\mathbf p_2}(t_2) (1-g_A^{(2)}\gamma_5) v_{2,\mathbf p_4}^{s_4} (t_2) \right)
    \end{aligned}
\end{equation}
where \(\mathbf p = \mathbf p_3-\mathbf p_1\). The expression for the full \(|\mc M|^2\) is
\begin{equation}
    |\mathcal M_{s_3,s_4}|^2= |\mathcal M_{s_3,s_4}^{(con)}|^2 + |\mathcal M_{s_3,s_4}^{(ex)}|^2 + 2 Re\left( \mathcal M_{s_3,s_4}^{(con)} \mathcal M_{s_3,s_4}^{(ex) *} \right).
\end{equation}

\paragraph{Unpolarized \(|\mathcal M|^2\) calculation}
\mbox{}\\[1ex]
First we analyze the case where we sum over the spins \(s_3\) and \(s_4\), i.e., we are not measuring the spin of the outgoing states. We will only look at the first-order corrections to \(|\mc M|^2\).

\paragraph{First order corrections with \(\lambda_1 \neq 0,\,\,\lambda_2 =0\)}

We first consider corrections to the contribution of the contact interaction. Possible first-order corrections come from the mode functions of the external fermions involved in the scattering process. The correction coming from the incoming \(a\) particle is,
\begin{equation}
    \begin{aligned}
        &\sum_{s_3,s_4}|\mathcal M_{s_3,s_4}^{(con)}|^2_{(1)} \\
        &\supset \frac{\lambda_1^2\,H}{\Pi_{i}(2\omega_i)} \int dt\int dt' \frac{32 g_A^{(1)}((g_A^{(1)})^2-1)m_1m_2}{\omega_1^2} (1+2it\omega_1) (\mathbf p_1\cdot (\mathbf p_4\times\mathbf p_2))\,e^{-i\Delta\omega (t-t')}\,.
    \end{aligned}
\end{equation}
The correction coming from the incoming \(\bar a\) particle is,
\begin{equation}
    \frac{\lambda_1^2\,H}{\Pi_{i}(2\omega_i)} \int dt\int dt' \frac{32 g_A^{(1)}((g_A^{(1)})^2-1)m_1m_2}{\omega_2^2} (1+2it\omega_2) (\mathbf p_2\cdot (\mathbf p_1\times\mathbf p_3))\,e^{-i\Delta\omega (t-t')}\,.
\end{equation}
The correction coming from the outgoing \(b\) particle is,
\begin{equation}
    \frac{\lambda_1^2\,H}{\Pi_{i}(2\omega_i)} \int dt\int dt' \frac{32 g_A^{(1)}((g_A^{(1)})^2-1)m_1m_2}{\omega_3^2} (1-2it\omega_3) (\mathbf p_3\cdot (\mathbf p_2\times\mathbf p_4))\,e^{-i\Delta\omega (t-t')}\,.
\end{equation}
The correction coming from the outgoing \(\bar b\) particle is,
\begin{equation}
    \frac{\lambda_1^2\,H}{\Pi_{i}(2\omega_i)} \int dt\int dt' \frac{32 g_A^{(1)}((g_A^{(1)})^2-1)m_1m_2}{\omega_4^2} (1-2it\omega_4) (\mathbf p_4\cdot (\mathbf p_3\times\mathbf p_1))\,e^{-i\Delta\omega (t-t')}\,.
\end{equation}
All these corrections are \(SO(3)\) invariant (as they should be) but violate parity. However, there are a few important points to note:
\begin{itemize}
    \item We need both \(m_1\neq 0\) and \(m_2 \neq 0\) to obtain the corrections. That is, chiral symmetry is already broken in free theory.
    \item The corrections are zero whenever \(g_A^{(1)}=\pm1\) or \(g_A^{(1)}=0\), i.e., for interactions violating chirality maximally or not violating chirality at all.
\end{itemize}
Even in the most general case, where there are corrections, they can have vanishing effects for particular momentum configurations, for instance \(\mathbf p_2 = -\mathbf p_1\), \(\mathbf p_3 = -\mathbf p_4\). Therefore, for such choice of kinematics, the unpolarized amplitude will not have any \(O(H)\) corrections.

\paragraph{First order corrections with \(\lambda_1 =0 ,\,\,\lambda_2 \neq 0\)} In this case (only scalar current exchange diagram) by explicit calculations, it has been found that the corrections coming from spinor mode functions as well as the corrections from the scalar propagator are vanishing. The fact that the correction to the scalar propagator does not contribute can be seen by observing that this first-order correction is proportional to
\begin{equation}
    \begin{aligned}
        &\lambda_2^4 H \int dt_1\int dt_1'\int dt_2\int dt_2' \int \frac{dp^0}{2\pi}\int \frac{dp'^0}{2\pi}\,e^{-i(\omega_1-\omega_3+p^0)t_1} e^{i(\omega_4-\omega_2+p^0)t_2}\times\\
        &\times e^{i(\omega_1-\omega_3+p'^0)t'_1}e^{-i(\omega_4-\omega_2+p'^0)t'_2}\frac{c_1^B(t_1+t_2+t_1'+t_2')+i(c_2^B/\omega_p)(p^0(t_1^2-t_2^2)-p'^0(t_1'^2-t_2'^2))}{((p^0)^2 -\omega_p^2)((p'^0)^2 -\omega_p^2)}\,.
    \end{aligned}
\end{equation}
The integration result can be seen to vanish identically with some clever change of variables. This cancellation will still persist even if we take the fermions to be polarized as we are only dealing with linear corrections. Therefore, from now on we will not need to worry about first-order corrections that are coming from the modification of scalar propagators due to nonzero \(H\) even for the polarized \(|\mathcal M|^2\) calculation that we will take care of in the next subsection.

\paragraph{Polarized \(|\mathcal M|^2\) calculation}
\mbox{}\\[1ex]
Now we will do our calculations by taking the outgoing \(\psi_2\) to be polarized with spin label \(s_3\) (but we will sum over spins \(s_1,s_2,s_4\) as before). We will discuss the cases of contact interaction and exchange interaction separately. We will take the direction of spin polarization along some unit vector \(\mathbf n\) and take it perpendicular to the 3-momenta of the particles. Due to the presence of \(SO(3)\) invariance we can take \(\mathbf n\) to be along the \(z\)-direction for the calculation.

\paragraph{Kinematics and spin-asymmetry in the differential cross-section} 
We shall choose kinematic configuration \(\mathbf p_2 = - \mathbf p_1\), which, by momentum conservation, implies that \(\mathbf p_4 = -\mathbf p_3\). Due to this choice of kinematics we have \(\omega_1=\omega_2\) and \(\omega_3=\omega_4\).
Spin-asymmetry in the scattering cross section is given by
\begin{equation}
    \sigma_\uparrow - \sigma_\downarrow =  \frac{\omega_1}{|\mathbf p_1|}\int \frac{d^3\mathbf p_3}{(2\pi)^3} \frac{|\mathcal M|^2_{\uparrow}-|\mathcal M|^2_{\downarrow} }{T}.
\end{equation}
Then we will have spin asymmetry in the differential scattering cross-section
\begin{equation}
    \begin{aligned}
        \frac{d\sigma_\uparrow}{dp_{3z}d\phi}-\frac{d\sigma_\downarrow}{dp_{3z}d\phi} = \frac{\omega_1}{|\mathbf p_1|} \int \frac{|\mathbf p_3|d|\mathbf p_3|}{(2\pi)^3} \frac{|\mathcal M|^2_{\uparrow}-|\mathcal M|^2_{\downarrow} }{T}
    \end{aligned}
\end{equation}
where \(\phi\) is the angle between \(\mathbf p_1\) and \(\mathbf p_3\).  This integration will involve the energy-conserving delta function as well as the derivative of the energy-conserving delta function. The method for dealing with this is detailed in the Appendix. In the notation of the Appendix, the variable \(x\) is \(|\mathbf p_3|\) here. The energy function is \(u(x) =2\omega_1 - 2\sqrt{x^2+m_2^2}\). Solving for \(x\) in terms of \(u\), we get
\begin{equation}
    x = \sqrt{\frac{(2\omega_1-u)^2}{4} - m_2^2}\,.
\end{equation}
We also need
\begin{equation}
    |u'(x)| = \frac{2 x}{\sqrt{x^2+m_2^2}}\,.
\end{equation}
From the kinematic configurations we have,
\begin{equation}
        \mathbf p_1 = (|\mathbf p_1|,0,0)=-\mathbf p_2\,,\quad
        \mathbf p_3 = (x\cos\phi,x\sin\phi,0)=-\mathbf p_4\,.
\end{equation}

\paragraph{Scattering through contact interaction}

First, we will find the quantity \(|\mathcal M|_\uparrow^2-|\mathcal M|^2_\downarrow\) for the case of contact interactions. There are several contributions to this quantity at \(O(H)\). The first correction comes from the polarized mode function of the outgoing \(b\) particle. This is
\begin{equation}
    \begin{aligned}
        \frac{1}{4} &\frac{\lambda_1^2 H}{\Pi_i (2\omega_i)} (2\pi)^2 \delta(\Delta\omega)(\delta(\Delta\omega)+2\omega_3 \delta'(\Delta\omega)) \frac{16 m_2 (p_{1y}p_{3x}-p_{3y}p_{1x})}{\omega_3^2} \\
        &\times \left[m_1m_2\left(1-(g_A^{(1)})^4\right)+\left((g_A^{(1)})^4+6(g_A^{(1)})^2+1\right)(\omega_1 \omega_3 +\mathbf p_1\cdot\mathbf p_3)\right],
    \end{aligned}
\end{equation}
(we have already made use of the special choice of kinematic configuration and also the \(1/4\) factor is due to the spin averaging of the incoming particles).\\

\noindent Then we have the correction from the mode function of the incoming \(a\) particle. This correction is
\begin{equation}
\begin{aligned}
    \frac{1}{4} &\frac{\lambda_1^2 H}{\Pi_i (2\omega_i)} (2\pi)^2 \delta(\Delta\omega) (\delta(\Delta\omega)-2\omega_1 \delta'(\Delta\omega)) \frac{32 m_1 (p_{1y}p_{3x}-p_{3y}p_{1x}) }{\omega_1 (\omega_3+m_2)}\\
    &\times \left((g_A^{(1)})^4-1\right) \left[m_2 (\omega_3+m_2)+|\mathbf p_3|^2\right].
\end{aligned}
\end{equation}
The correction coming from the mode function of the outgoing \(\bar b\) particle is
\begin{equation}
    \begin{aligned}
        -\frac{1}{4} &\frac{\lambda_1^2 H}{\Pi_i (2\omega_i)} (2\pi)^2 \delta(\Delta\omega) (\delta(\Delta\omega)+2\omega_3 \delta'(\Delta\omega)) \frac{16 m_1^2 m_2 (p_{1y}p_{3x}-p_{3y}p_{1x})}{\omega_3^2}\,.
    \end{aligned}
\end{equation}
The correction coming from the mode function of the incoming \(\bar a\) particle is
\begin{equation}
    -\frac{1}{4} \frac{\lambda_1^2 H}{\Pi_i (2\omega_i)} (2\pi)^2 \delta(\Delta\omega) (\delta(\Delta\omega)-2\omega_1 \delta'(\Delta\omega)) \frac{16 m_1^2 m_2 (p_{1y}p_{3x}-p_{3y}p_{1x})}{\omega_1^2}\,.
\end{equation}
All these corrections add up to give the total correction to \(|\mathcal M|_\uparrow^2-|\mathcal M|^2_\downarrow\). The corresponding differential scattering cross-section has been plotted. \begin{figure}
    \centering
    \includegraphics[width=0.48\linewidth]{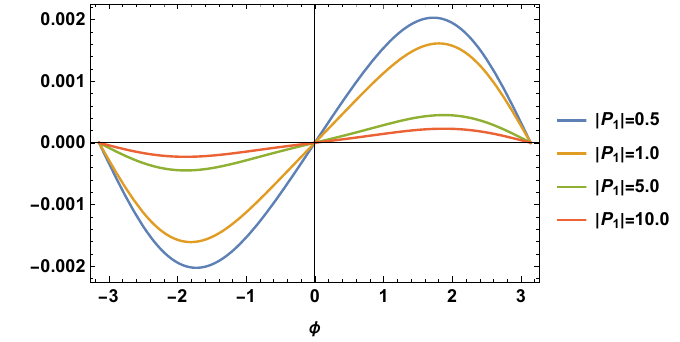}
    \includegraphics[width=0.48\linewidth]{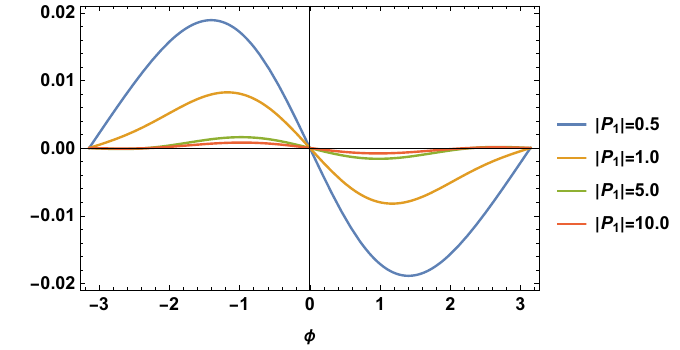}
    \caption{Spin-asymmetry in the differential scattering cross-section in contact interaction for different values of incoming momentum \(|\mathbf p_1|\). \textbf{ Left}: \(m_1=1\) and \(m_2=0.1\), \textbf{Right}: \(m_1=1\) and \(m_2=1\). \(g_A^{(1)}=0.5\) for these plots. The vertical axis the spin-asymmetry in the differential cross-section with the dimensionfull pre-factor \(\lambda_1^2 H\) stripped off.}
\end{figure}In the \(\mathbf p_1 \rightarrow 0\) limit we have the corresponding asymmetry as
\begin{equation}
    \lambda_1^2 H\frac{
\left[\footnotesize
\left(1 - \left(g_A^{(1)}\right)^4\right) m_1^3 
- 8 \left(g_A^{(1)}\right)^2 m_1^2 m_2 
- 3 \left(1 - \left(g_A^{(1)}\right)^4\right) m_1 m_2^2 
- \left(1 - 6 \left(g_A^{(1)}\right)^2 + \left(g_A^{(1)}\right)^4\right) m_2^3
\right]
\sin(\phi)
}{
32 m_1^2 \sqrt{(m_1 - m_2)(m_1 + m_2)} \pi^2
},
\end{equation}

with the following special cases,
\begin{equation}
    \begin{cases}
        \lambda_1^2 H\frac{
\left(m_1^3 - 3 m_1 m_2^2 - m_2^3\right) \sin(\phi)
}{
32 m_1^2 \sqrt{(m_1 - m_2)(m_1 + m_2)} \pi^2
}, \text{ for } g_A^{(1)} = 0,\\
-\lambda_1^2 H\frac{
m_2 \left(2 m_1^2 - m_2^2\right) \sin(\phi)
}{
8 m_1^2 \sqrt{(m_1 - m_2)(m_1 + m_2)} \pi^2
}, \text{ for } g_A^{(1)} = 1.
    \end{cases}
\end{equation}

\begin{tcolorbox}
    It is observed that there is an enhancement in the low initial momentum limit.
\end{tcolorbox}

\paragraph{Scattering through exchange interaction}

In this case, spin-asymmetry in \(|\mathcal M|^2\) has a contribution from the mode-function of the outgoing polarized \(b\) particle. This contribution is,
\begin{equation}
    \begin{aligned}
        \frac{1}{4} &\frac{\lambda_2^4 H}{\Pi_i (2\omega_i)} \frac{2\pi \delta(\Delta\omega)}{((\omega_3-\omega_1)^2-\omega_p^2)((\omega_2-\omega_4)^2-\omega_p^2)}\times 8 ((g_A^{(2)})^2+1)\frac{m_2 (p_{1y}p_{3x}-p_{3y}p_{1x})}{\omega_3^2}\\
        &\times 2\pi (\delta(\Delta\omega)+2\omega_3\delta'(\Delta\omega))\times \left[((g_A^{(2)})^2-1)m_1 m_2 - ((g_A^{(2)})^2+1)(p_2\cdot p_4)\right].
    \end{aligned}
\end{equation}
Another contribution comes from the mode function of the incoming \(a\) particle. This contribution is
\begin{equation}
    \begin{aligned}
        \frac{1}{4} &\frac{\lambda_2^4 H}{\Pi_i (2\omega_i)} \frac{2\pi \delta(\Delta\omega)}{((\omega_3-\omega_1)^2-\omega_p^2)((\omega_2-\omega_4)^2-\omega_p^2)}\times 8 ((g_A^{(2)})^2-1)\frac{m_1 (p_{1y}p_{3x}-p_{3y}p_{1x})}{\omega_1^2}\\
        &\times 2\pi (\delta(\Delta\omega)-2\omega_1\delta'(\Delta\omega))\times \left[((g_A^{(2)})^2-1)m_1 m_2 - ((g_A^{(2)})^2+1)(p_2\cdot p_4)\right].
    \end{aligned}
\end{equation}
We note that in this case \(\omega_p^2 = |\mathbf p_1 - \mathbf p_3|^2 + m_\phi^2 =|\mathbf p_1|^2 + x^2 - 2 |\mathbf p_1|x\cos\phi + m_\phi^2\). Now we can calculate the spin asymmetry in the differential scattering cross section as a function of angle \(\phi\) for these exchange diagram contributions. The results are shown in the plots, where we show the dependence on different parameters involved. \begin{figure}[H]
    \centering
    \includegraphics[width=0.48\linewidth]{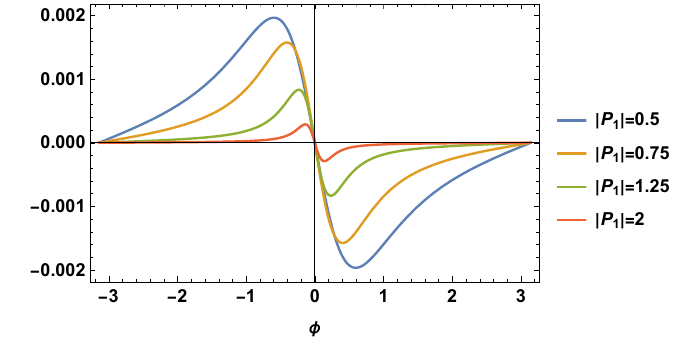}
    \includegraphics[width=0.48\linewidth]{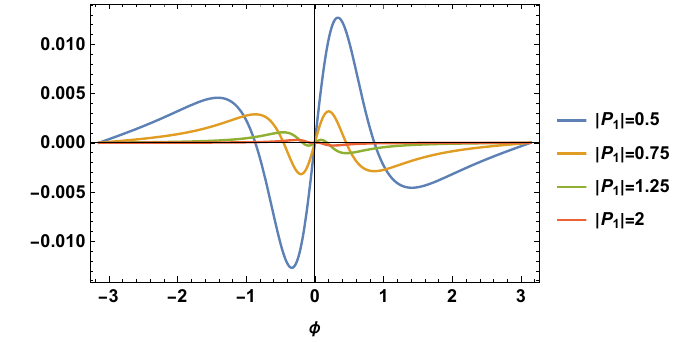}
    \caption{Spin asymmetry in the differential scattering cross-section in exchange interaction for different incoming momentum \(|\mathbf p_1|\). \textbf{Left}: \(m_1=1,m_2=0.1\), \textbf{Right}: \(m_1=1,m_2=1\). \(g_A^{(2)}=0.5\) and \(m_\phi=0.5\) for these plots.}
\end{figure}In the initial low-momentum limit,
\begin{equation}
    \begin{cases}
        -\lambda_2^4 H\frac{
(m_1 + m_2)^{3/2} \left[ (5 m_1 - 2 m_2)(m_1^2 - m_2^2) + (m_1 - 2 m_2) m_\phi^2 \right] \sin(\phi)
}{
128 m_1^2 \sqrt{m_1 - m_2} \left(m_1^2 - m_2^2 + m_\phi^2\right)^3 \pi^2
}\text{ for } g_A^{(2)}=0,\\
\lambda_2^4 H\frac{
m_2 \left( -2 m_1^4 + m_2^4 - m_2^2 m_\phi^2 + m_1^2 \left(m_2^2 + 2 m_\phi^2\right) \right) \sin(\phi)
}{
32 m_1^2 \sqrt{(m_1 - m_2)(m_1 + m_2)} \left(m_1^2 - m_2^2 + m_\phi^2\right)^3 \pi^2
}\text{ for } g_A^{(2)}=1.
    \end{cases}
\end{equation}

\begin{tcolorbox}
    There is again an enhancement in the low-momentum limit.
\end{tcolorbox}

\subsubsection{Tree-level \(1\rightarrow 3\) decay process}\label{4-fermi}
The four-Fermi interaction, also known as the Fermi contact interaction, is an effective theory that describes weak interactions at low energies via a local, point-like coupling between four fermionic fields. Originally proposed by Enrico Fermi in 1933 to account for beta decay, the interaction is represented by a dimension-6 operator in the low-energy limit of the Standard Model. In modern physics, four-fermion interactions serve as effective descriptions of more complex interactions at low energies, such as those mediated by heavy gauge bosons.(\cite{PhysRev.109.193},\cite{Sudarshan:1984qq})

The interaction is similar to the contact interaction in the previous case, but with all 4 fermions distinct from each other. The interacting action term for the 4-Fermi interaction in $3 + 1$ Minkowski spacetime is given by is 
\begin{equation}
    S_{\text{int}} = \int d^3 \vec{x} dt \; G_F |J|^2.
\end{equation}
Extending this Lagrangian to a general coordinate system, we get
\begin{align}
    S_{\text{int}}= \int d^3 \vec{x}dt \sqrt{-g}\,\mathcal L_{\text{int}} = \int d^3 \vec{x}dt \sqrt{-g}\,G_F|J|^2,
\end{align}
where \( G_F \) is the effective Fermi constant and \( J^{\mu} \) is the Dirac current. For an interaction that completely violates parity, the form of the Dirac current is given by
\begin{align}
    J^{\mu} = \sum_a \overline{\psi}_a\gamma^\mu\frac{(1-\gamma^5)}{2}\psi_{b_a},
\end{align}
where the form of the current indicates that the spinor of type $a$ is coupled to the spinor of type $b_a$.

Let us now study the decay process $ \ket{\psi_1} \to \ket{\psi_2 \psi_3 \psi_4}$. We assume that $\psi_1$ is coupled to $\psi_2$ and $\psi_3$ is coupled to $\psi_4$. Therefore, the net relevant interacting Lagrangian term for the process $ \ket{\psi_1} \to \ket{\psi_2 \psi_3 \psi_4}$ is given by
\begin{equation}
    \mathcal{L}_{\text{int}} = G_F(-\tau H)^{-4}\left(\bar{\psi}_{2}\gamma^\mu\frac{(1-\gamma^5)}{2}\psi_{1}\right)\left(\bar{\psi}_4\gamma_\mu\frac{(1-\gamma^5)}{2}\psi_{3}\right).
\end{equation}
Rescaling the field $\psi_j \to (-\tau H)^{3 /2}\psi_j$, we get
\begin{equation}\label{eq: lagrangian 4-fermi}
    \mathcal{L}_{\text{int}} =  (-\tau H)^{2} G_F \left(\bar{\psi}_{2}\gamma^\mu\frac{(1-\gamma^5)}{2}\psi_{1}\right)\left(\bar{\psi}_4\gamma_\mu\frac{(1-\gamma^5)}{2}\psi_{3}\right) .
\end{equation}
In order to compute the S-matrix element, we use the LSZ reduction derived in Appendix \ref{LSZ derivation}. The (spin-label suppressed) matrix element is given by

\begin{align}
    \mathcal{M} = \frac{G_F}{4}\int dt e^{-3 Ht}  \; &\bar{u}_{2,\vec{p_2}}(t)\gamma^\mu {(1-\gamma^5)} u_{1,\vec{p_1}}(t) \times \bar{u}_{4,\vec{p_4}}(t) \gamma_\mu {(1-\gamma^5)} v_{3,\vec{p_3}}(t).
\end{align}
Using some simple properties of Dirac matrices, we get
\begin{align}
    |\mathcal{M}|^2 = \frac{G_F^2}{16} \iint dt dt' e^{-3 H(t+t')}&\bar{v}_{3,\vec{p_3}}(t')(1+\gamma^5)\gamma_\mu u_{4,\vec{p_4}}(t')\bar{u}_{1,\vec{p_1}}(t')(1+\gamma^5)\gamma^{\mu}u_{2,\vec{p_2}}(t') \notag \\ 
    \times &\bar{u}_{2,\vec{p_2}}(t)\gamma^\nu (1-\gamma^5) u_{1,\vec{p_1}}(t) \bar{u}_{4,\vec{p_4}}(t) \gamma_\nu (1-\gamma^5)v_{3,\vec{p_3}}(t).
\end{align}

We consider two processes that are relevant to collider physics. The first is when all the particles are unpolarized, which means that we have to sum over all the spins for all the particles. The second is when we have control over the spin of the incoming particle, which we will call polarized decay. 
\paragraph{Unpolarized Beam Decay}
\mbox{}\\[1ex]
Spin-averaging the matrix element and using the properties of Dirac matrices, we get
\begin{align}
|\mathcal{M}|^2 &= \iint dt\, dt'\; \frac{G_F^2}{32}\, e^{-3H(t + t')} \notag \\
&\quad \times \mathrm{Tr} \left[
    \sum_{s_1, s_2} \gamma^\mu (1 - \gamma^5)\, u_{2, \vec{p}_2}(t')\,
    \bar{u}_{2, \vec{p}_2}(t)\, \gamma^\nu (1 - \gamma^5)\,
    u_{1, \vec{p}_1}(t)\, \bar{u}_{1, \vec{p}_1}(t')
\right] \notag \\
&\quad \times \mathrm{Tr} \left[
    \sum_{s_3, s_4} \gamma_\mu (1 - \gamma^5)\, u_{4, \vec{p}_4}(t')\,
    \bar{u}_{4, \vec{p}_4}(t)\, \gamma_\nu (1 - \gamma^5)\,
    v_{3, \vec{p}_3}(t)\, \bar{v}_{3, \vec{p}_3}(t')
\right].
\label{eq:unpolarized_M2_fermi}
\end{align}

The matrix element is simply
\begin{equation}
    |\mathcal{M}|^2 = \frac{32\pi^2 G_F^2}{\Pi_j (2\omega_j)}(p_1\cdot p_3)(p_2 \cdot p_4) \delta^2(\Delta \omega).
\end{equation}
\begin{tcolorbox}
    There is no first-order correction to decay probability for the unpolarized case.
\end{tcolorbox}
\paragraph{Polarized Beam Decay}
\mbox{}\\[1ex]
Considering the case where we can differentiate between the polarization states of incoming particles, the matrix element is given by 
\begin{align}
|\mathcal{M}_{s_1}|^2 &= \iint dt\, dt'\; \frac{G_F^2}{32}\, e^{-3H(t + t')} \notag \\
&\quad \times \mathrm{Tr} \left[
    \sum_{s_2} \gamma^\mu (1 - \gamma^5)\, u_{2, \vec{p}_2}(t')\,
    \bar{u}_{2, \vec{p}_2}(t)\, \gamma^\nu (1 - \gamma^5)\,
    u_{1, \vec{p}_1}(t)\, \bar{u}_{1, \vec{p}_1}(t')
\right] \notag \\
&\quad \times \mathrm{Tr} \left[
    \sum_{s_3, s_4} \gamma_\mu (1 - \gamma^5)\, u_{4, \vec{p}_4}(t')\,
    \bar{u}_{4, \vec{p}_4}(t)\, \gamma_\nu (1 - \gamma^5)\,
    v_{3, \vec{p}_3}(t)\, \bar{v}_{3, \vec{p}_3}(t')
\right].
\label{eq:polarized_M2_fermi}
\end{align}

\begin{tcolorbox}
    This time we have corrections for order $O(H)$ and the matrix element is given by
\begin{align}
    |\mathcal{M}_{s_1}|^2 = \frac{32 \pi^2 G_F^2}{\prod_j (2\omega_j)} (p_2 \cdot p_4)  \Bigg( &\left(p_1 \cdot p_3 + \vec{\hat{j}} \cdot \left[\vec{p_1} \frac{\vec{p_1}\cdot \vec{p_3}}{m_1 + \omega_1} + m_1 \vec{p_3} - |\vec{p_3}|\vec{p_1}\right] \right)\delta^2(\Delta \omega) \notag \\ 
     & - H m_1 \vec{\hat{j}} \cdot
     \frac{\vec{p_1}\times\vec{p_3}}{2\omega_1^2}(\delta^2(\Delta \omega) - 2\delta(\Delta \omega)\delta'(\Delta \omega)\omega_1) 
     \Bigg). 
\end{align}
\end{tcolorbox}

\section{The emiT Experiment and Lower Bounds on T-Violations}
The correction term due to the presence of the Hubble expansion is of the form $\vec{\hat{j}}\cdot(\vec{p_1}\times\vec{p_2})$, which is odd under time reversal. A similar kind of term was considered by Jackson et al in \cite{jackson} by adding fundamental interactions to the effective Lagrangian with complex coefficients, which explicitly violate T-symmetry. Their  differential decay rate for polarized beta decay has the following form (we have shown only those terms relevant for the Fermi theory; \cite{jackson} includes a more general form of terms motivated by EFT)
\begin{align} \label{jackson}
        \frac{d \Gamma}{d E_e d \Omega_e d \Omega_\nu } \propto  \bigg(1 + \vec{\hat{j}}\cdot \left[ B\frac{\vec{p_\nu}}{E_\nu} + D\frac{\vec{p_e}\times\vec{p_\nu}}{E_eE_\nu}\right]+\cdots\bigg),
\end{align}
where the term with dimensionless coefficient $D$ is the term that is odd under time reversal ($\vec{\hat{j}}$ is the unit vector directed along the neutron polarization). The goal of the emiT experiment was to measure this T-violation in beta decay, and the current upper bound on $D$ is of the order of $10^{-5}$. The coefficient $B$ in our case is unity. Our findings in $O(H^0)$ agree with \cite{jackson}.

\begin{figure}[H]
    \centering    \includegraphics[width=0.35\linewidth]{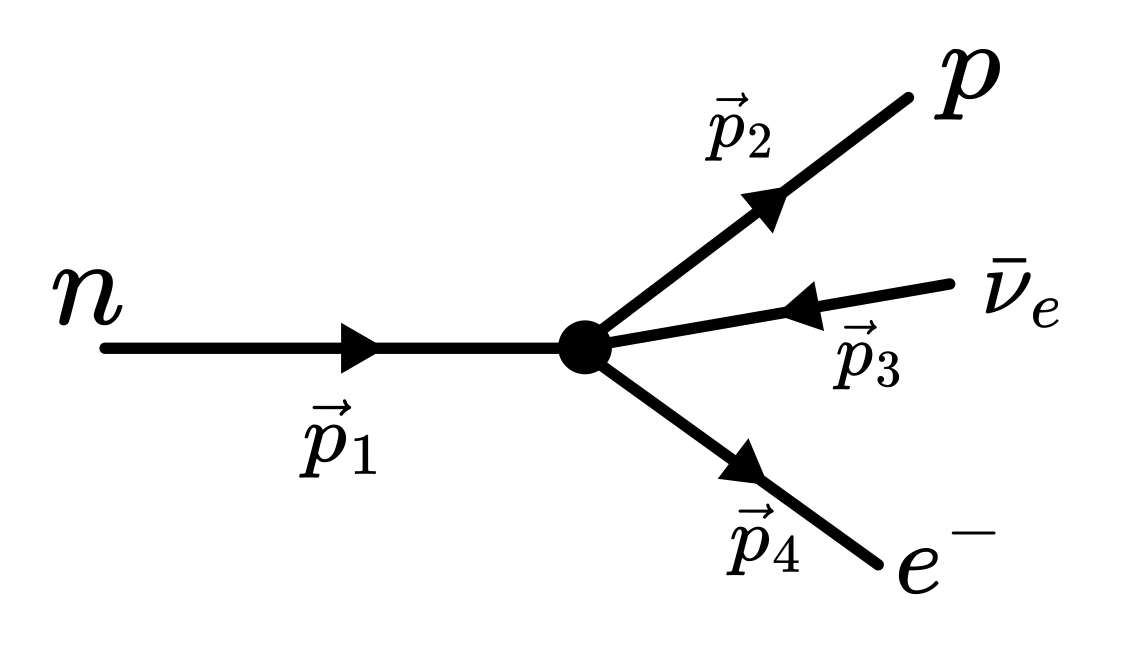}
    \caption{Beta decay}
    \label{fig:enter-label}
\end{figure}

As such terms are also present in the $|\mathcal{M}|^2$ in de Sitter, we could extract a similar kind of $D$ here as well. In our case, we will find that the analogous $D$-terms are somewhat different. One major difference in terms of experimental setup is that in the analysis that goes in \cite{jackson, emit}, the neutron is at rest, but in our case, it has to be moving with respect to the co-moving frame in order to show up any corrections. As boost is no longer a symmetry, we have inequivalent frames\footnote{It should be possible to constrain the matrix elements as a consequence of boost symmetry; we leave this as a future direction to explore.}. We use the analysis done in section (\ref{4-fermi}) and eq. (\ref{eq:polarized_M2_fermi}) to write down the matrix the expression for the $|\mathcal{M}|^2$ for polarized beta decay, which is given by 
\begin{align}
    |\mathcal{M}|^2_{\vec{\hat{j}}} \propto \frac{G_F^2}{\prod_l (\omega_{l})} (p_2 \cdot p_4)  \Bigg( &\left(p_1 \cdot p_3 + \vec{\hat{j}} \cdot \left[\vec{p_1} \frac{\vec{p_1}\cdot \vec{p_3}}{m_n + \omega_1} + m_n \vec{p_3} - |\vec{p_3}|\vec{p_1}\right] \right)\delta^2(\Delta \omega) \notag \\ 
     & - H m_n \, \vec{\hat{j}} \cdot
     \frac{\vec{p_1}\times\vec{p_3}}{2\omega_1^2}(\delta^2(\Delta \omega) - 2\delta(\Delta \omega)\delta'(\Delta \omega)\omega_1) 
     \Bigg).
\end{align}
Here $\Delta\omega=\omega_1-\omega_2-\omega_3-\omega_4$.
We can see that both energy-conserving and violating terms will contribute to the value of $D$. To get the differential cross section, we have to integrate over the phase space. Integration over $\vec{p_2}$ can be done using three momentum-conserving delta function, which implies $\vec{p_2} = \vec{p_1} - \vec{p_3}-\vec{p_4}$. Therefore, 
\begin{align}
    \frac{d\Gamma}{d\Omega_3 d\Omega_4 d|\vec{p_3}|} \propto \frac{G_F^2}{T} \int  
    \frac{(p_2 \cdot p_4)}{\omega_2 \omega_4}  \Bigg( &\left(
    \frac{p_1 \cdot p_3}{\omega_1 \omega_3} + \vec{\hat{j}} \cdot \left[\frac{\vec{p_1}}{m_n + \omega_1} \frac{\vec{p_1}\cdot \vec{p_3}}{\omega_1 \omega_3} +\frac{m_n \vec{p_3}}{\omega_1 \omega_3} - \frac{\vec{p_1}}{\omega_1} \right] \right)\delta^2(\Delta \omega) \notag \\ 
     & - H \frac{m_n}{2\omega_1^2}\, \vec{\hat{j}} \cdot
     \frac{\vec{p_1}\times\vec{p_3}}{\omega_1 \omega_3 }(\delta^2(\Delta \omega) - 2\delta(\Delta \omega)\delta'(\Delta \omega)\omega_1) 
     \Bigg) |\vec{p_3}|^2|\vec{p_4}|^2 d|\vec{p_4}|.
\end{align}



For the contribution from the energy-violating term, we use integration by parts to convert the $\delta \delta'$ to $\delta^2$, which is 
    \begin{equation}
        \frac{d \Gamma_{\text{violating}}}{d \Omega_3 d\Omega_4 d|\vec{p_3}| } 
        \propto \frac{G_F^2}{T} \int\frac{H m_n}{\omega_1}\frac{p_2\cdot p_4}{\omega_2 \omega_4}\frac{\vec{p_1}\times\vec{p_3}}{\omega_1 \omega_3} \delta(\Delta \omega)\delta'(\Delta \omega) |\vec{p_3}|^2|\vec{p_4}|^2 d |\vec{p_4}|\,.
    \end{equation}
We change the integration variable from $|\vec{p_4}|$ to $\Delta \omega$. Therefore, the above equation becomes\footnote{We have to be careful here as  $\Delta \omega$ is not always a strictly monotonic function of $|\vec{p_4}|$.} 
\begin{equation}
    \frac{d \Gamma_{\text{violating}}}{d \Omega_3 d\Omega_4 d|\vec{p_3}| } \propto \frac{G_F^2}{T}\int\frac{H m_n}{\omega_1}\frac{p_2\cdot p_4}{\omega_2 \omega_4}\frac{\vec{p_1}\times\vec{p_3}}{\omega_1 \omega_3} |\vec{p_3}|^2|\vec{p_4}|^2 \delta(\Delta \omega)\delta'(\Delta \omega)\frac{d|\vec{p_4}|}{d(\Delta \omega)} d (\Delta \omega). 
\end{equation}
Let $\displaystyle \frac{H m_n}{\omega_1}\frac{p_2\cdot p_4}{\omega_2 \omega_4}\frac{\vec{p_1}\times\vec{p_3}}{\omega_1 \omega_3} |\vec{p_3}|^2 |\vec{p_4}|^2\frac{d|\vec{p_4}|}{d(\Delta \omega)}  = g(\Delta \omega)$. Now notice that
\begin{align}
    \int g(\Delta \omega) \delta(\Delta \omega )\delta'(\Delta \omega ) d(\Delta \omega) &= \frac{1}{2}\int g(\Delta \omega) \frac{d \delta^2(\Delta \omega)}{d (\Delta \omega)} d(\Delta \omega) \notag \\
    &=  \frac{1}{2}g(\Delta \omega) \delta^2(\Delta \omega)\Big|_{|\vec{p_4}| = 0}^{|\vec{p_4}| = \infty} - \frac{1}{2}\int g'(\Delta \omega) \delta^2(\Delta \omega) d(\Delta \omega).
\end{align}
At $|\vec{p_4}| = 0 $, $g(\Delta \omega) = 0$. For large $|\vec{p_4}|$, $\vec{p_2} \approx -\vec{p_4}$, $\omega_2 \approx |\vec{p_4}|$, $\omega_4 \approx |\vec{p_4}|$ and $\Delta \omega \approx -2 |\vec{p_3}|$, therefore $g(\Delta \omega) \propto |\vec{p_4}|^2$. As the relevant representation of $\delta$ here is given by $\delta(\Delta \omega) = \frac{\sin(\Delta \omega T/2 )}{\pi \Delta \omega}$, hence for large $|\vec{p_4}|$, $\delta(\Delta \omega) \approx \frac{\sin(|\vec{p_4}| T)}{2 \pi |\vec{p_4}|}$, therefore $g(\Delta \omega) \delta^2(\Delta \omega) \propto \frac{\sin^2(|\vec{p_4}| T)}{4 \pi^2}$. For large $T$, it contributes to the decay rate $ \propto \frac{\sin^2(|\vec{p_4}| T)}{T} \to 0$. We change the integration variable to $|\vec{p_4}|$ in the last term of the above integration and rewrite $g'(\Delta \omega) d(\Delta \omega) \text{ as } \frac{d g(|\vec{p_4}|)}{ d|\vec{p_4}|} d|\vec{p_4}|$. This leaves us with 
\begin{align}
    \frac{d\Gamma}{d\Omega_3 d\Omega_4 d|\vec{p_3}|} \propto &\frac{G_F^2}{T}  \int  
     \Bigg( 
    \frac{p_1 \cdot p_3}{\omega_1 \omega_3}\frac{p_2 \cdot p_4}{\omega_2 \omega_4}  + \vec{\hat{j}}  \cdot \left[\frac{\vec{p_1}}{m_n + \omega_1} \frac{\vec{p_1}\cdot \vec{p_3}}{\omega_1 \omega_3} +\frac{m_n \vec{p_3}}{\omega_1 \omega_3} - \frac{\vec{p_1}}{\omega_1} \right]\frac{p_2 \cdot p_4}{\omega_2 \omega_4}   \notag \\ 
     &\!\!\!\!\!\!\!\! - H \frac{m_n}{2\omega_1^2} \, \vec{\hat{j}}  \cdot
     \frac{\vec{p_1}\times\vec{p_3}}{\omega_1 \omega_3 }\left(\frac{p_2 \cdot p_4}{\omega_2 \omega_4} + \frac{\omega_1}{|\vec{p_4}|^2}\frac{d}{d|\vec{p_4}|}\left(\frac{p_2\cdot p_4}{\omega_2 \omega_4}|\vec{p_4}|^2 \frac{d|\vec{p_4}|}{d(\Delta \omega)} \right)\right) 
     \Bigg) \delta^2(\Delta \omega) |\vec{p_3}|^2|\vec{p_4}|^2 d|\vec{p_4}|.
\end{align}
Integrating out $|\vec{p_4}|$ we get\footnote{See Appendix \ref{delta delta prime} for details on the integration of $\delta^2$.}
\begin{align}
    \frac{d\Gamma}{d\Omega_3 d\Omega_4 d|\vec{p_3}|} \propto \, & G_F^2 \,\omega_4 |\vec{p_4}||\vec{p_3}|^2 
     \frac{p_2 \cdot p_4}{\omega_2 \omega_4} \Bigg( 
    \frac{p_1 \cdot p_3}{\omega_1 \omega_3}  + \vec{\hat{j}}  \cdot \left[\frac{\vec{p_1}}{m_n + \omega_1} \frac{\vec{p_1}\cdot \vec{p_3}}{\omega_1 \omega_3} +\frac{m_n \vec{p_3}}{\omega_1 \omega_3} - \frac{\vec{p_1}}{\omega_1} \right]  \notag \\ 
     &\!\!\!\!\!\!\!\!- H \frac{m_n}{2\omega_1^2} \, \vec{\hat{j}}  \cdot
     \frac{\vec{p_1}\times\vec{p_3}}{\omega_1 \omega_3 }\left(  1 + \frac{\omega_2 \omega_4}{p_2 \cdot p_4} \frac{\omega_1}{|\vec{p_4}|^2}\frac{d}{d|\vec{p_4}|}\left(\frac{p_2\cdot p_4}{\omega_2 \omega_4}|\vec{p_4}|^2 \frac{d|\vec{p_4}|}{d(\Delta \omega)} \right)\right) 
     \Bigg).
\end{align}
where $|\vec{p_4}|$ is such that $\Delta \omega = 0$. We now take the large proton mass limit as mentioned in \cite{beta_decay} which is that we neglect the mass of electron with respect to neutron and proton and assume them to be non-relativistic. Under this approximation, the above takes the form given in (\ref{jackson}) and is exactly the same when $H =0$ and $D = 0$.  Comparing the above equation with (\ref{jackson}) we can define an effective $D$-parameter given by:
\begin{equation}
    \hat D^H= D_{\rm{cons}}^H+D_{\rm{violating}}^H\,,
\end{equation} where $D_{\rm{cons}}$ comes from the energy-conserving part and is given by 
\begin{equation}
    D_{\rm{cons}}^H= -\frac{H m_n}{2\omega_1^2}\,,
\end{equation}
whose order of magnitude in the present universe is $10^{-42}$, probably impossible to measure. Nevertheless, this already sets a lower limit on the T-violation. The energy-nonconserving piece gives $D_{\rm{violating}}^H$
given by
\begin{align}
    D^H_{\text{violating}} = -\frac{1}{2}\frac{H m_n}{\omega_1 |\vec{p_4}|^2}\frac{\omega_2 \omega_4}{p_2 \cdot p_4}\frac{d}{d|\vec{p_4}|}\left(\frac{p_2\cdot p_4}{\omega_2 \omega_4}|\vec{p_4}|^2 \frac{d|\vec{p_4}|}{d(\Delta \omega)} \right).
\end{align}
There are two important points to be aware of. The $D$-parameter in \cite{jackson} did not depend on the momenta, but the $\hat D^H$-parameter obtained above depends on the momenta. In particular, when the electron 3-momentum, $\vec{p_4}$, changes, the $\hat D^H$ value also changes. 
The second important point is that when $\vec{p_4}\rightarrow 0$, there could be huge enhancements to the effective $D$-parameter due to the $|\vec{p_4}|$ in the denominator. The energy violating term led to an integration by parts, which was responsible for the $1/|\vec{p_4}|^2$ factor outside. Without this integration by parts, this factor would have been canceled, as happened in the energy-conserving piece. To reach such configurations which lead to an effective enhancement of the $D$-parameter, we first solve $\Delta\omega=0$ subject to $|\vec{p_4}|=0$. Let us write $\vec{p}_i=p_i(\sin\theta_i\cos\phi_i, \sin\theta_i \sin\phi_i,\cos\theta_i)$ for $i=3,4$ and $\vec{p_1}=p_1(1,0,0)$. This gives the locus:
\begin{equation}
    \sqrt{m_1^2+p_1^2}=m_4+\sqrt{m_3^2+p_3^2}+\sqrt{m_2^2+p_1^2+p_3^2-2 p_1 p_3 \cos\phi_3\sin\theta_3}\,.
\end{equation}
Any kinematics respecting this constraint will lead to an effective enhancement of the $D$-parameter. One also needs to ensure that rest of the factors do not give a zero at the corresponding values. It is clear that the maximum value of $D_{\rm violating}$ near the above locus is given by
\begin{align}
    D_{\text{violating}}^{H,max} = -\frac{H m_n}{\omega_1 |\vec{p_4}|} \frac{d|\vec{p_4}|}{d(\Delta \omega)}\bigg|_{\Delta\omega=0} .
\end{align}

 To make $D^{H,max}_{\rm{violating}}\sim O(1)$, we need to offset the smallness of $H$ by fine-tuning the pole to $1/H$ precision. 
Of course, these highly tuned values of the momenta do not appear to be of experimental significance, as resolving such precision seems beyond reach. This also explains why such violations have not been observed. Thus, we can summarize our findings as

\begin{tcolorbox}
The intrinsic T-violation in polarized beta decay at $O(H)$ has an energy-conserving piece whose contribution is $O(H_0/m_n)$, and an energy-non-conserving piece whose contribution can be $O(1)$ for very fine-tuned kinematical configurations. 
\end{tcolorbox}

\section{Discussion}

In this paper, we examined cosmological correlators and de Sitter S-matrices involving fermions, primarily focusing on the latter. We developed approximations that allowed us to study the leading-order $O(H)$ correction to various processes. Even at tree-level, the calculations, while being conceptually straightforward, were technically involved. 

One of the main findings in our paper pertains to the Fermi theory calculation for polarized beta decay which leads to intrinsic T-violation which starts at $O(H)$. This had two contributions, one that was energy-conserving and another that was non-conserving. While the first piece contributed $H_0/m_n$ to the parity even triple correlation \cite{jackson,emit}, the second could lead to $O(1)$ contributions to the effective $D$-parameter for very finely tuned kinematical configurations. While plausible, it is not completely clear to us if we should think about these fine-tuned configurations as being of measure zero\footnote{Unlike Breit-Wigner resonances of finite width, which show up clearly in scattering cross-sections, these fine-tuned configurations will have close to zero width.}. We emphasize that while the effect in \cite{jackson, emit} could be considered in the rest frame of the neutron, in our case, it was important that the neutron was in motion. 

We list some potential future directions:

\begin{itemize}

\item
It will interesting to search for other observables that can give T-violation due to the presence of the Hubble constant. 

\item Our focus was the $O(H)$ contribution. Several steps of the calculations simplified due to this. For instance, the UdW$_T$ S-matrix and BD S-matrix were the same which led to some technical simplifications. Further, we did not have to calculate $O(H)$ corrections to the Fermi constant or to the phase-space integral. These simplifications will not hold at the next order $O(H^2)$. It will be interesting to set up the systematics in a manner that is ``non-perturbative" in $H$. 

\item
We considered the example of Fermi theory and showed that the in-in correlator and the in-out correlator as defined in \cite{pajer} are the same. 
In the future, it will be interesting to further develop the correspondence between the $BD_{-\infty}^{\infty}$ S-matrix and in-in correlators \cite{pajer} in the presence of fermions, especially at higher orders in perturbation theory. This would be relevant to simplify the calculations for in-in fermionic correlators in the primordial universe. 

\item In \cite{falkowski} a very general analysis of beta decay within the standard model was presented. Our analysis focused on the Fermi theory analysis. It will be useful to extend the results in this paper in the more general framework in \cite{falkowski}. 

\item Although we tried to motivate the relevance of our findings for terrestrial experiments, our crude arguments need to be refined to see if the tiny effects in our calculations have any hope of being detected on earth. 

\item Unlike flat space, where in some cases (for example, gapped theories) rigorous results are known for the analyticity of 2-2 scattering, very little is known about the structure of scattering amplitudes in de Sitter space. In recent times, considerable effort has been put into expanding our knowledge base on these results--see, for example, \cite{shota1}. In the same spirit, it will be useful to document particle physics processes to enlarge our dictionary of amplitudes in de Sitter space.

\item Finally, one can ask whether, using AdS/CFT techniques, one can analytically continue the AdS results to the dS results. For example, in \cite{sleight}, Mellin space techniques were used to calculate inflationary correlators. It will be interesting to figure out what the various notions of S-matrices, examined in the present paper, correspond to in AdS space and what the relevant analytic continuations are.

\end{itemize}

\section*{Acknowledgments}
We thank Biplob Bhattacharjee and Ranjan Laha for useful discussions. We also thank Ayngaran Thavanesan for useful comments and feedback.
 AS acknowledges support from the SERB core grant CRG/2021/000873 and a Quantum Horizons Alberta chair professorship. 

\begin{appendix}

\section{Canonical quantization of spin-0 and spin-$1/2$ fields}

In this appendix, we give a summary of the quantizing scheme for free real scalar and Dirac spinor fields in the expanding Poincaré patch of de Sitter spacetime using canonical quantization. For general aspects of quantum field theory in de Sitter space we refer to~\cite{dSDirac:1935zz,dSNachtmann1967,dSThirring:1967dd,dSGursey:1963ir,dSAkhmedov:2024npw,dSEpstein:2012zz}.
For our purpose, we shall quantize the quantum fields in the Bunch-Davies vacuum~\cite{Schaub:2023scu} of the Poincaré patch.

\subsection{Canonical quantization of real scalar fields}
The action for a minimally coupled scalar field in any curved spacetime is
\begin{equation}
    S[\phi] = \frac{1}{2}\int d^d\mathbf x\int d\tau \sqrt{-g}\left(g^{\mu\nu}\partial_\mu\phi\partial_\nu\phi - m^2\phi^2\right).
\end{equation}
In the case of de Sitter, we have \(\sqrt{-g}=(-\tau H)^{-d-1}\) and \(g^{\mu\nu}=(-\tau H)^2 \eta^{\mu\nu}\). Varying the field variable, we get the equation of motion for scalar fields as
\begin{equation}
    \tau^2\partial_\tau^2\phi - \tau (d-1)\partial_\tau \phi +  k^2 \tau^2\phi + \frac{m^2}{H^2}\phi = 0,
\end{equation}
\noindent in the momentum space. Rescaling the field $\phi \rightarrow (-\tau H)^{d/2}\phi$, allows us to write down the equation of motion as a Bessel differential equation with imaginary order;
\begin{equation}\label{scalar eom}
    \tau^2\partial_\tau^2\phi+\tau\partial_\tau\phi + (k^2\tau^2+\mu^2)\phi = 0, \quad \text{where}\quad 
    \mu^2 = \frac{m^2}{H^2} - \frac{d^2}{4}.
\end{equation}
Now we promote the field \(\phi\) to a quantum field operator $\hat{\phi}$, given by
\begin{equation}
    \hat{\phi}(\mathbf x,\tau) = \int \frac{d^d\mathbf k}{(2\pi)^d} \left(f_{\mathbf k}(\tau)e^{i\mathbf k\cdot\mathbf x} \hat{a}_{\mathbf k} + \bar f_{\mathbf k}(\tau)e^{-i\mathbf k\cdot\mathbf x} \hat{a}^\dagger_{\mathbf k}\right),
\end{equation}
where \(f_{\mathbf k}(\tau)\) satisfy the differential equation (\ref{scalar eom}). The choice of specific \(f_{\mathbf k}\)'s, of course, depends on the boundary condition. Unlike flat space, a single choice of mode functions does not diagonalize the Hamiltonian at all \(\tau\). At a certain time \(\tau = \tau_*\), the true particle states should diagonalize the Hamiltonian at that time, and this requirement translates to the following boundary condition:
\begin{equation}\label{eq: boundary condition scalar mode-function}
    (\partial_\tau + i \omega_k (\tau)) f_{\mathbf k}(\tau)|_{\tau = \tau_*} = 0\quad\text{ where }\quad \omega_k (\tau) = \sqrt{\frac{\mu^2}{\tau^2}+|\mathbf k|^2}.
\end{equation}
The `true vacuum' at \(\tau_*\), which we denote by \(\ket{0,\tau_*}\), is annihilated by the corresponding \(\hat{a}_{\mathbf k}\)'s. The Fock space built by applying creation operators on this `true vacuum' contains our physical particle states.\\

\noindent The next important step is to specify the normalization of the mode functions, where we need to use the canonical quantization condition for $\hat{\phi}$. The action of the rescaled \(\phi\) is,
\begin{equation}\label{eq: action of rescaled free scalar field}
    S[\phi] = \frac{1}{2} \int d^d\mathbf x\int d\tau\, (-\tau H)\left[(\partial_\tau\phi)
    ^2+\frac{d}{\tau}\phi\,\partial_\tau\phi-(\nabla\phi)^2 - \left(\frac{m^2}{H^2}-\frac{d^2}{4}\right)\frac{\phi^2}{\tau^2}\right]\,.
\end{equation}
Therefore, the canonical momentum is given by
\begin{equation}
    \Pi_\phi  =\frac{1}{2}(-\tau H)\left(2\partial_\tau\phi+\frac{d}{2}\phi\right)\,.
\end{equation}
The canonical quantization condition is
\begin{equation}
    [\phi(\mathbf x,\tau),\Pi_\phi(\mathbf x',\tau)] = i \delta(\mathbf x-\mathbf x')\,.
\end{equation}
Choosing the normalization \([a_{\mathbf k},a^\dagger_{\mathbf k'}]=(2\pi)^d \delta(\mathbf k-\mathbf k')\), we get the following condition,
\begin{equation}
\label{scalar mf normalisation}
    (-\tau H) f_{\mathbf k}(\tau)\,\overleftrightarrow{\partial_\tau} \bar f_{\mathbf k}(\tau) = i\,.
\end{equation}
Now, solving the equation of motion with the boundary condition Eq.(\ref{eq: boundary condition scalar mode-function}) at \(\tau*\rightarrow-\infty\) (the Bunch-Davies boundary condition) and the above normalization, we have the required mode function,
\begin{equation}\label{eq: scalar mode functions }
    f_{\mathbf k}(\tau) = \sqrt{\frac{\pi}{4 H}}e^{-\frac{\pi\mu}{2}+\frac{i\pi}{4}} H^{(1)}_{i\mu}(-k\tau) \, , \quad
    \bar f_{\mathbf k}(\tau) = \sqrt{\frac{\pi}{4 H}}e^{+\frac{\pi\mu}{2}-\frac{i\pi}{4}} H^{(2)}_{i\mu}(-k\tau)\,.
\end{equation}
\noindent The normalization of the mode function is guaranteed by the Wronskian of the Hankel functions,
\begin{equation}\label{wronskian hankel}
    -H^{(1)}_{i\mu}(-k\tau)\tau \overleftrightarrow{\partial_\tau} H^{(2)}_{i\mu}(-k\tau) = \frac{4i}{\pi}\,.
\end{equation}
This completes the canonical quantization of the free scalar field in the expanding Poincaré patch of de Sitter with Bunch-Davies boundary condition\footnote{Choosing a different boundary condition will yield a mode function which can be related to the Bunch-Davies mode functions via Bogoliubov transformation. For \(\tau_* \rightarrow 0\), the mode function becomes,\(f_{\mathbf k}(\tau) = \sqrt{\frac{\pi}{2 H \sinh (\pi \mu)}}J_{i\mu}(-k\tau)\). The case for general \(\tau_*\) is discussed in section 2.2 for both scalars and fermions.}. Since the Minkowski spacetime is the \(H\rightarrow0\) limit of the de-Sitter spacetime\footnote{In this limit \(\mu\rightarrow \infty\) (for \(m\neq 0\)) and \(\tau \rightarrow -1/H + t \)}, taking this limit, we get back the appropriate flat-space mode functions (the sanity check we mentioned). Thus, we have
\begin{equation}
    f_{\mathbf k} \rightarrow \frac{e^{-i\omega_k t}}{\sqrt{2\omega_k}}\,,\quad\text{where,}\quad
    \omega_k^2 = k^2 + m^2.
\end{equation}
For massless particle, the mode functions in de-Sitter is same as that in flat spacetime because of conformal symmetry. So, the flat space expansion works fine.\\

\subsection{Canonical quantization of Dirac fields}\label{dirac field quantization}

Canonical quantization of spinor fields in the de-Sitter space-time has been studied in \cite{poincare_quantization_spinor,Schaub:2023scu}. A study of Schwinger effect in de-Sitter spacetime has been performed in \cite{Stahl_2016}. In this appendix, we provide a general overview of the quantization. Extending the flat-space fermionic field action to general curved spacetime, we get
\begin{equation}\label{general coordinate spin action}
    S = \int dt d^\dimd x \sqrt{|g|}\;\Bar{\psi} ( i \slashed{\nabla} - m ) \psi
\end{equation}
\noindent where we have changed the integration measure and the gradient to be generally covariant.\\

\noindent Here \(\Bar{\psi} = \psi^\dagger \gamma^0\) with the usual flat space Dirac matrix.  The quantity \(\slashed{\nabla}= \Tilde{\gamma}^\mu \nabla_\mu\) is related to spinor covariant derivative and gamma matrices in curved spacetime. We have the following sets of Dirac matrices, 
\begin{equation}
    \{\gamma^a,\gamma^b\} = 2\eta^{ab}\,, \quad \Tilde{\gamma}^\mu = \tns{e}{^\mu_a}\gamma^a \Rightarrow \{\Tilde{\gamma}^\mu,\Tilde{\gamma}^\nu\} = 2g^{\mu\nu},
\end{equation}
\noindent where, \(e^{\mu}_{\text{\hspace{4 pt}}a}\) are the frame coefficients that take vectors to the spinor's tangent space. The gradient in the curved spacetime is
\begin{equation}
    \label{curved_gradient}
    \nabla_\mu = \del_\mu + \frac{1}{4}\omega_{\mu ab}\gamma^{ab}\,,\quad\text{where,}\quad\tns{\omega}{_\mu^a_b} = \tns{e}{_\nu^a}\tns{e}{^\lambda_b}\Gamma^\nu_{\mu\lambda} - \tns{e}{^\lambda_b}\del_\mu\tns{e}{_\lambda^a}.
\end{equation}
\noindent Here, \(\omega_{\mu ab}\) are called spin connections. Using the anticommutation relations for the two sets of Dirac matrices, and \(e^{\mu}_{\hspace{4 pt}a}e^{a}_{\hspace{4 pt}\nu}=\delta^{\mu}_{\hspace{4 pt}\nu}\) (and other similar orthogonality and inverse relations), we get \(e^{\mu}_{\hspace{4 pt}a}=(-\tau H)\delta^{\mu}_{\hspace{4 pt}a}\). Using this, we can now find \(\slashed{\nabla}\), which turns out to be \(\slashed{\nabla}=-\tau H \gamma^a \tns{\delta}{^\mu_a}\del_\mu +  \frac{\dimd H}{2}\gamma^0\). The equation of motion for the field $\psi$ is then, 
\begin{equation}
    (i\slashed{\nabla}-m)\psi = 0\,,
\end{equation}
\noindent which after making the necessary substitutions becomes,
\begin{equation}
\left(\tau H\gamma^a\tns{\delta}{^\mu_a}\del_\mu - \frac{\dimd H}{2}\gamma^0 - im \right)\psi = 0\,.
\end{equation}
Rescaling the field $\psi \rightarrow (-\tau H)^{\frac{d}{2}}\psi$\footnote{It can be checked that under this rescaling, the action has the same form as the flat space action. Thus, the conformal symmetry of the de Sitter spacetime is verified as a sanity check. For a similar treatment for scalar mode functions and consequent normalization the reader is requested to refer to \cite{Mukhanov_Winitzki_2007}}, and squaring the Dirac equation to get Klein-Gordon equation for Dirac fields gives us the equation of motion of a harmonic oscillator with time-dependent frequency, similar to the scalar case,
\begin{equation}
    \left( \del^2_\tau + k^2 + \frac{1}{\tau^2}\left(\left(\frac{m}{H} \mp \frac{i}{2}\right)^2 +\frac{1}{4}\right)\right)\Psi_\pm = 0\,.
\end{equation}
Here, \(\gamma^0\) mixes the positive and negative energy modes. To overcome this, the solution is split into eigenspaces of \(\gamma^0\), with the order of the Bessel equation (obtained after rescaling the field again by a factor of \(\sqrt{-\tau}\)) being \(\nu_\pm = \frac{m}{H} \mp \frac{i}{2}\); where \(\pm\) corresponds to \(\psi_\pm\), with \(\gamma^0\psi_\pm=\pm\psi_\pm\). To see how the positive and negative energy modes are mixed by \(\gamma^0\), we apply the projection operators \(\mc P_\pm=\frac{1\pm\gamma^0}{2}\) to the Dirac equation after rescaling of the fields. One finds that
\begin{equation} \label{eom of Psi}
    \left(\del_\tau \mp \frac{im}{\tau H}\right)\Psi_\pm = \mp \vec{ \gamma}\cdot\mathbf k\Psi_\mp\,.
\end{equation}
\noindent Taking inspiration from the scalar case, the two solutions for each eigenspace of \(\gamma^0\), if we use B.D. boundary conditions, are,
\begin{equation} \label{modefunc definition}
\begin{aligned}
\small
    g_{i{\nu_\pm}}(k, \tau) = \sqrt{-\tau\,\pi}\frac{e^{\textstyle \frac{-\pi\nu_\pm}{2} +i\frac{\pi}{4}}}{2}H^{(1)}_{i \nu_\pm}(-k\tau)\,,\quad
    f_{i \nu_\pm}(k, \tau) = \sqrt{-\tau\,\pi}\frac{e^{\textstyle \frac{\pi\nu_\pm}{2} -i\frac{\pi}{4}}}{2}H^{(2)}_{i \nu_\pm}(-k\tau),\\
    \text{where, }\nu_\pm = \frac{m}{H} \pm \frac{i}{2}.
    \end{aligned}
\end{equation} 
$g$ and $f$ correspond to positive and negative energy modes in the flat spacetime limit. Defining the spinors \(\xi_r\) and \(\chi_r\) such that they span the eigenspaces of \(\gamma^0\) with eigenvalues 1 and -1 respectively, with the normalization \(\xi^\dagger_r\xi_{r'} =\delta_{rr'}\) and \(\chi^\dagger_r\chi_{r'} =\delta_{rr'}\). Therefore, the general solution to the Dirac equation is given by. 
\begin{equation} \label{psi expansion}
\begin{aligned}
    \psi_+(\tau,\vec{x}) &= \intm{k}\left(\alpha_{+,\vec{k}}^r \;\expp{k}{x} g_{i\nu_+} + \beta_{+,\vec{k}}^r\; \expn{k}{x}f_{i\nu_+}\right)\xi_r\quad \text{and}    \\
    \psi_-(\tau,\vec{x}) &= \intm{k}\left(\alpha_{-,\vec{k}}^r \;\expp{k}{x} g_{i\nu_-} + \beta_{-,\vec{k}}^r\; \expn{k}{x}f_{i\nu_-}\right)\chi_r
\end{aligned}
\end{equation}
where the summation on index $r$ is understood. Using Eq.(\ref{eom of Psi}) we get the expression of $\psi$ which is
\begin{align}\label{eq: classical psi solution}
    \psi(\tau,\vec{x})=\intm{k}\left(\alpha_{r,\vec{k}}u^r_{\vec{k}}(\tau)\expp{k}{x}+\beta_{r,\vec{k}}v^r_{\vec{k}}(\tau)\expn{k}{x}\right).
\end{align}
where 
\begin{equation}
    u^r_{\mathbf k}(\tau) \sim \left(g_{i\nu_+}(\tau) - \frac{\mathbf \gamma\cdot\mathbf k}{k}g_{i\nu_-}(\tau) \right)\xi^r,\quad v^r_{\mathbf k}(\tau)\sim \left(f_{i\nu_-}(\tau) + \frac{\mathbf \gamma\cdot\mathbf k}{k}f_{i\nu_+}(\tau) \right)\chi^r.
\end{equation}
For quantizing the dirac field, we lift the integration constants into creation and annihilation operator, for fermions and anti-fermions. We have
\begin{equation}\label{spinor field solution}
    \hat{\psi}(\mathbf x,\tau) = \int \frac{d^d\mathbf k}{(2\pi)^d} \left(u^r_{\mathbf k}(\tau) e^{i\mathbf k \cdot \mathbf x} \hat{b}^r_{\mathbf k} + v^r_{\mathbf k}(\tau) e^{-i\mathbf k \cdot \mathbf x} \hat{d}^{r\dagger}_{\mathbf k}\right),
\end{equation}
where, \(\hat{b}^r_{\mathbf k}\) is the annihilation operator of the particle and \(\hat{d}^{r\dagger}_{\mathbf k}\) is the creation operator of the anti-particle, with anti-commutation relations,
\begin{equation}
    \{b^r_{\mathbf k},b^{s\dagger}_{\mathbf k'}\}  = (2\pi)^d \delta^{rs}\delta(\mathbf k-\mathbf k')\,,\quad\{d^r_{\mathbf k},d^{s\dagger}_{\mathbf k'}\}  = (2\pi)^d \delta^{rs}\delta(\mathbf k-\mathbf k').
\end{equation}
Now, to fix the normalizations of \(u^r_{\mathbf k}(\tau)\) and \(v^r_{\mathbf k}(\tau)\), we need to use the canonical anti-commutation relation of the Dirac field. The canonical momentum conjugate to the field \(\psi\) is (can be derived from the action), \(\Pi_\psi = i \psi^\dagger\), for which the canonical anti-commutation relation for the field variable becomes,
\begin{equation}
    \{\psi_\alpha(\mathbf x,\tau),i\psi_\beta^\dagger (\mathbf x',\tau)\} = i \delta_{\alpha\beta}\delta(\mathbf x-\mathbf x').
\end{equation}
Here \(\alpha,\beta\) are spinor indices. Substituting the mode decomposition of the field, we will have
\begin{equation}
   \sum_r u^r_{\mathbf k}(\tau)u^{r\dagger}_{\mathbf k}(\tau) + v^r_{-\mathbf k}(\tau)v^{r\dagger}_{-\mathbf k}(\tau) = 1.
\end{equation}
This tells that \(u^r_{\mathbf k}(\tau),v^r_{-\mathbf k}(\tau)\) forms complete orthonormal basis in the spinor space. Now demanding the condition of orthogonality, it can be checked that \(v^{r\dagger}_{-\mathbf k}(\tau)u^s_{\mathbf k}(\tau)=0\) is satisfied, and this can be used to get,
\begin{equation}
\label{eq: spinor mode function orthogonality}
    u^{r\dagger}_{\mathbf k}(\tau)u^s_{\mathbf k}(\tau)=\delta^{rs},\quad v^{r\dagger}_{-\mathbf k}(\tau)v^s_{-\mathbf k}(\tau)= \delta^{rs}.
\end{equation}
This fixes the normalization, and finally, we have,
\begin{equation}
    \begin{aligned}
        u^r_{\mathbf k}(\tau) =\sqrt{k} \left(g_{i\nu_+}(\tau) - \frac{\mathbf \gamma\cdot\mathbf k}{k}g_{i\nu_-}(\tau) \right)\xi^r_{\mathbf k}\,,\quad
        v^r_{\mathbf k}(\tau)&=\sqrt{k} \left(f_{i\nu_-}(\tau) + \frac{\mathbf \gamma\cdot\mathbf k}{k}f_{i\nu_+}(\tau) \right)\chi^r_{\mathbf k}\\
        &= \sqrt{k} \left(\bar g_{i\nu_+}(\tau) + \frac{\mathbf \gamma\cdot\mathbf k}{k}\bar g_{i\nu_-}(\tau) \right)\chi^r_{\mathbf k}.
    \end{aligned}
\end{equation}
In the intermediate step we have to use the relation,
\begin{equation}
    |g_{i\nu_+}(\tau)|^2 + |g_{i\nu_-}(\tau)|^2 = \frac{1}{k},
\end{equation}
which follows from Wronskian relations of the Hankel function. These mode functions also satisfy the flat space limit, under which they become,
\begin{equation}
    \begin{aligned}
        u^r_{\mathbf k} (t) &= \frac{e^{-i\omega_k t}}{\sqrt{2\omega_k}} (\omega_k + m)^{-1/2} \left((\omega_k + m)-\mathbf \gamma\cdot \mathbf k \right)\xi^r_{\mathbf k}\,,\\
        v^r_{\mathbf k} (t) &= \frac{e^{+i\omega_k t}}{\sqrt{2\omega_k}} (\omega_k + m)^{-1/2} \left((\omega_k + m)+\mathbf \gamma\cdot \mathbf k \right)\chi^r_{\mathbf k}.
    \end{aligned}
\end{equation}

\paragraph{Choosing a different basis:} The above solution corresponds to choosing the Bunch-Davies boundary condition, where,
\begin{equation}\label{eq: dirc eqn dS}
    g_{i\nu_{\pm}}(\tau) \sim e^{-i k\tau}\quad\text{and}\quad f_{i\nu_{\pm}}(\tau) \sim e^{+i k\tau}\quad \text{ for }\quad \tau \rightarrow -\infty.
\end{equation}
The key idea is that these mode functions diagonalize the Dirac Hamiltonian at \(\tau \rightarrow -\infty\). More generally, we are interested in finding the mode functions (\(g'_{i\nu_{\pm}},(\tau),f'_{i\nu_{\pm}}(\tau)=\bar g'_{i\nu_{\pm}},(\tau)\)) that diagonalize the Hamiltonian at some intermediate \(\tau=\tau_*\). It turns out that the relevant condition is
\begin{equation}\label{eq: boundary condition spinor}
    (\partial_\tau + i\omega_k (\tau))g'_{i\nu_{\pm}}(\tau)|_{\tau=\tau_*} = 0;\quad
    (\partial_\tau - i\omega_k (\tau))f'_{i\nu_{\pm}}(\tau)|_{\tau=\tau_*} = 0,
\end{equation}
where \(\omega_k(\tau) = \sqrt{m^2/(-\tau H)^2+ k^2}\). First, we expand
\begin{equation}\label{eq: new spinor mode functions}
     g'_{i\nu_{+}}(\tau) = \alpha_1 g_{i\nu_{+}}(\tau) + \beta_1 f_{i\nu_{+}}(\tau),\quad
     g'_{i\nu_{-}}(\tau) = \alpha_2 g_{i\nu_{-}}(\tau) + \beta_2 f_{i\nu_{-}}(\tau).
\end{equation}
Equations of motion demand,
\begin{equation}
        \left(\partial_\tau \mp \frac{i m}{\tau H}\right) g'_{i\nu_\pm}(\tau) = - i k g'_{i\nu_\mp}(\tau),\quad
        \left(\partial_\tau \mp \frac{i m}{\tau H}\right) f'_{i\nu_\pm}(\tau) = + i k f'_{i\nu_\mp}(\tau),
\end{equation}
which are already true for the BD mode functions. Then the above condition tells us that
\begin{equation}\label{eq: relation between spinor bogoliubovs}
   \alpha := \alpha_1 = \alpha_2,\quad \beta:=\beta_1 = -\beta_2.
\end{equation}
This is consistent with the solution of the boundary conditions Eq.(\ref{eq: boundary condition spinor}),
\begin{equation}\label{eq: beta by alpha spinors}
    r_{\pm}(\tau_*) = - \frac{\left(\pm \frac{i m}{\tau H} g_{i\nu_\pm}(\tau_*) + i \omega_k (\tau_*) g_{i\nu_\pm}(\tau_*) - i k  g_{i\nu_\mp}(\tau_*) \right)}{\left(\pm \frac{i m}{\tau H} f_{i\nu_\pm}(\tau_*) + i \omega_k (\tau_*) f_{i\nu_\pm}(\tau_*) + i k  f_{i\nu_\mp}(\tau_*) \right)}, \quad r_+ = \frac{\beta_1}{\alpha_1},\,\, r_- = \frac{\beta_2}{\alpha_2},
\end{equation}
as can be checked that \(r_+ = - r_-\). Now applying canonical normalization we find that
\begin{equation}\label{eq: spinor alpha beta normalization}
    |\alpha|^2 + |\beta|^2 = 1.
\end{equation}

\paragraph{Form of the spinors \(\xi\) and \(\chi\)}
The \(r\)-label denotes degeneracy indices in the spinor space for positive/negative energy solutions. For \(d=3\) case (in which we are interested in) \(r\) can take 2 values, can be associated with the eigenvalue of spin-projection along some unit vector \(\mathbf n\), that is, eigenvalue of \(\mathbf S\cdot\mathbf n\), where the components of \(\mathbf S\) are given by,
\begin{equation}
    S_j = \frac{1}{2}\epsilon_{jkl} J^{kl} =\frac{i}{8}\epsilon_{jkl}[\gamma^k,\gamma^l],
\end{equation}
where \(J^{kl}\) are the generators of \(SO(3)\).\\

\noindent In the polar parameterization of \(\mathbf n = (\sin\theta\cos\phi,\sin\theta\sin\phi,\cos\theta)\), for up-spin along \(\mathbf n\), we have,
\begin{equation}
    \xi^\uparrow = \begin{pmatrix}
       1\\
        0
    \end{pmatrix} \otimes
    \begin{pmatrix}
        \cos(\frac{\theta}{2})e^{-i\phi}\\
        \sin(\frac{\theta}{2})
    \end{pmatrix},
\end{equation}
Similarly, for down-spin,
\begin{equation}
    \xi^\downarrow = \begin{pmatrix}
       1\\
        0
    \end{pmatrix} \otimes
    \begin{pmatrix}
        -\sin(\frac{\theta}{2})\\
        \cos(\frac{\theta}{2})e^{i\phi}
    \end{pmatrix}.
\end{equation}
For antiparticle spinors, we have~\cite{Peskin:1995ev}
\begin{equation}
\begin{aligned}
    &\chi^\uparrow = (i\gamma^2\gamma^0) \xi^{\uparrow *} =( \sigma_1 \otimes(-i\sigma_2))\,\xi^{\uparrow *} = (\sigma_1 \otimes I) \,\,\xi^\downarrow = \begin{pmatrix}
       0\\
        1
    \end{pmatrix} \otimes
    \begin{pmatrix}
        -\sin(\frac{\theta}{2})\\
        \cos(\frac{\theta}{2})e^{i\phi}
    \end{pmatrix},\\
    &\chi^\downarrow =(i\gamma^2\gamma^0) \xi^{\downarrow *}= ( \sigma_1 \otimes(-i\sigma_2))\,\xi^{\downarrow *} =-(\sigma_1 \otimes I) \,\, \xi^\uparrow = - \begin{pmatrix}
       0\\
        1
    \end{pmatrix} \otimes
    \begin{pmatrix}
        \cos(\frac{\theta}{2})e^{-i\phi}\\
        \sin(\frac{\theta}{2})
    \end{pmatrix}.
\end{aligned}
\end{equation}
It follows that
\begin{equation}
\begin{aligned}
    &\xi^\uparrow = (-i\gamma^2\gamma^0)\chi^{\uparrow*} = -C \chi^{\uparrow*},\quad
   \xi^\downarrow = (-i\gamma^2\gamma^0)\chi^{\downarrow*}=-C \chi^{\downarrow*},\quad\\
   &\chi^\uparrow = (i\gamma^2\gamma^0)\xi^{\uparrow*} = C \xi^{\uparrow*},\quad
   \chi^\downarrow = (i\gamma^2\gamma^0)\xi^{\downarrow*}=C \xi^{\downarrow*},
\end{aligned}
\end{equation}
where we have defined the charge-conjugation matrix for the Dirac spinors,
\begin{equation}\label{eq: charge conjugation}
    C = i\gamma^2\gamma^0.
\end{equation}
This has the property
\begin{equation}\label{eq: charge conugation on gamma}
    C (\gamma^{\mu})^T C^{-1} = -\gamma^\mu.
\end{equation}
Related to this, the mode-function spinors have the properties
\begin{equation}\label{charge conjugation on uv}
    u_{\mathbf k}^r(\tau) = C\, (\bar v_{\mathbf k}^r (\tau))^T,\quad v_{\mathbf k}^r(\tau) = C\, (\bar u_{\mathbf k}^r (\tau))^T.
\end{equation}
These properties will be useful while discussing charge-conjugation symmetry of the Dirac fields.

\paragraph{Helicity basis} The helicity operator for a particle with momentum \(\mathbf k\) is,
\begin{equation}\label{eq: helicity operator}
    \hat h_{\mathbf k} = \frac{1}{2} \frac{\mathbf \sigma\cdot\mathbf k}{k} =  \frac{\mathbf S\cdot\mathbf k}{k} = \mathbf S\cdot \hat{\mathbf k}\,.
\end{equation}
This operator has eigenvalues \(\pm 1/2\), known as positive (negative) helicity. For 2-spinor, the helicity eigenstates are denoted by \(\ket{\pm \hat{\mathbf k} }\) with,
\begin{equation}
    \mathbf S\cdot \hat{\mathbf k} \ket{\pm \hat{\mathbf k} } = \pm \frac{1}{2} \ket{\pm \hat{\mathbf k} }\,.
\end{equation}
Therefore, the relevant 4-spinors are
\begin{equation}\label{eq: helicity spinors}
    \begin{aligned}
        \xi_{\mathbf k}^{\pm} &= \begin{pmatrix}
            1\\
            0
        \end{pmatrix}\otimes \ket{\pm \hat{\mathbf k} }\quad \text{and}\\
        \chi_{\mathbf k}^{\pm} &= \pm \begin{pmatrix}
            0\\
            1
        \end{pmatrix}\otimes \ket{\mp \hat{\mathbf k} }\,.
    \end{aligned}
\end{equation}

\paragraph{Spinor Propagator}
In Section \ref{small h} we have documented the mode functions at sub-leading order in Hubble's constant. Taking inspiration from it, we will find the spinor propagator at the sub-leading order. In momentum space the propagator is defined as
\begin{equation}
    G^F_{\mathbf{k}}(t,t') = \int \! \mathrm{d}^3 \mathbf{x} \, e^{-i \mathbf{k} \cdot \mathbf{x}}\big[
    \bra{0} \psi(\mathbf{x}, t) \, \bar{\psi}(\mathbf{0}, t') \ket{0} \, \Theta(t - t') - \bra{0} \bar{\psi}(\mathbf{0}, t') \, \psi(\mathbf{x}, t) \ket{0} \, \Theta(t' - t)\big],
\end{equation}
where, \(\psi\) is given by \ref{spinor field solution}. For the Heaviside-step function we have,
\begin{equation}
    \Theta(t - t')\, e^{-i \omega_k (t - t')} = \int \! \frac{\mathrm{d}k^0}{2\pi} \, \frac{i}{k^0 - \omega_k + i \epsilon} \, e^{-i k^0 (t - t')}, \quad \epsilon \to 0^+.
\end{equation}
Making use of this and Eq.(\ref{uu vv first order expansion}), we get the spinor propagator as (spin indices suppressed on the mode functions),
\begin{equation}
\begin{aligned}
    G^F_{\mathbf{k}}(t, t') &= \sum_s u(t, \mathbf{k}) \, \bar{u}(t', \mathbf{k}) \, \Theta(t - t') 
    - \sum_s v(t, -\mathbf{k}) \, \bar{v}(t', -\mathbf{k}) \, \Theta(t' - t) \\
    &= \int \! \frac{\mathrm{d}k^0}{2\pi} \, e^{-i k^0 (t - t')} \, 
    \frac{i}{(k^0)^2 - \omega_k^2 + i \epsilon} \, \left( \slashed{k} + m + \Delta(H, t, t') \right) \\
    &= G_{\mathbf{k}}^{F(0)}(t, t') + G_{\mathbf{k}}^{F(1)}(t, t'),
\end{aligned}
\label{spinor-prop-form}
\end{equation}
where $\Delta$ is given by,
\begin{equation}
\Delta(H, t, t') = 
H \begin{pmatrix}
    \displaystyle -\frac{N_1}{\omega_k} 
    & \displaystyle (\boldsymbol{\sigma} \cdot \mathbf{k}) \, \frac{N_2}{\omega_k^2} \\[10pt]
    \displaystyle -(\boldsymbol{\sigma} \cdot \mathbf{k}) \, \frac{N_3}{\omega_k^2} 
    & \displaystyle -\frac{N_4}{\omega_k}
\end{pmatrix},
\label{spinor-prop-correction}
\end{equation}
where
\begin{align*}
N_1 &=  (t + t')\big[m^3 - i k^2 m (t - t') k^0 - (m + i k^2 (t - t')) \omega_k^2\big] \\
N_2 &= m (i + m (t + t')) k^0 - (t - t')(m + i k^2 (t + t')) \omega_k^2 \\
N_3 &= m (i - m (t + t')) k^0 + (t - t')(m - i k^2 (t + t')) \omega_k^2 \\
N_4 &=  (t + t')\big[m^3 - i k^2 m (t - t') k^0 - (m - i k^2 (t - t')) \omega_k^2\big].
\end{align*}

\section{LSZ reduction of Bunch-Davies \(S\)-matrix scalars and Dirac spinors}\label{LSZ derivation}

\subsection{LSZ reduction prescription for real scalar fields}
The normalisation of the scalar mode functions, in \ref{scalar mf normalisation}, tells us that the mode functions are orthogonal under the Wronskian defined. Using this, we can find the creation/annihilation operator in terms of the fields and mode-functions. We can thus write,
\begin{equation}
\begin{aligned}
    {}_{BD}\bra{0}(-i \,  \bar f_{\mathbf k} (\tau) \overleftrightarrow{\tau\partial_\tau} \phi_0 (\mathbf k,\tau)) &= {}_{BD}\bra{\mathbf k}\quad\text{and} \\
    i\,  f_{\mathbf k} (\tau) \overleftrightarrow{\tau\partial_\tau} \phi_{0} (-\mathbf k,\tau) \ket{0}_{BD} &= \ket{\mathbf k}_{BD}
\end{aligned}
\end{equation}
(note that \(\phi (-\mathbf k,\tau) =\phi^\dagger (\mathbf k,\tau) \)).\\

\noindent For an interacting theory, we need to employ the adiabatic approximation (discussed later), which says that the interactions turn off sufficiently quickly. This is a necessary condition in order to write the ``in" and ``out" particle states as,
\begin{equation}\label{eq: Adiabatic approx for scalars}
    \ket{\mathbf k}_{\text{in}} = \lim_{\tau\rightarrow -\infty} i\,  f_{\mathbf k} (\tau) \overleftrightarrow{\tau\partial_\tau} \phi (-\mathbf k,\tau) \ket{0}_{\text{in}}\quad\text{and}\quad {}_{\text{out}}\bra{\mathbf k} = \lim_{\tau\rightarrow 0}{}_{\text{out}}\bra{0} (-i\,  \bar f_{\mathbf k} (\tau) \overleftrightarrow{\tau\partial_\tau} \phi (\mathbf k,\tau)).
\end{equation}
The adiabatic approximation ensures that the creation/annihilation operators, create/annihilate one particle states.\\

\noindent Now consider the \(S\)-matrix,
\begin{equation}
    S_{\text{in}\rightarrow \text{out}} = {}_{\text{out}}\bra{n'} T (\cdots) \ket{\mathbf k,n}_{\text{in}}
\end{equation}
where \(\ket{\mathbf k,n}_{\text{in}}\) is a \((n+1)\) particle state with one scalar particle with momentum \(\mathbf k\) and some other \(n\) particles. Then we have
\begin{subequations}
    \begin{align}
        S_{\text{in}\rightarrow \text{out}} &= {}_{\text{out}}\bra{n'} T (\cdots) \ket{\mathbf k,n}_{\text{in}}\\
        &= i\,\lim_{\tau\rightarrow -\infty} \, {}_{\text{out}}\bra{n'} T (\cdots) \,f_{\mathbf k} (\tau) \overleftrightarrow{\tau\partial_\tau} \phi (-\mathbf k,\tau) \ket{n}_{\text{in}}\\
        &= - i \int_{-\infty}^0 \frac{d\tau}{\tau}\, f_{\mathbf k}(\tau)\,\mathcal E_{\mathbf k}[\tau]\,{}_{\text{out}}\bra{n'} T (\cdots \, \phi (-\mathbf k,\tau) )\ket{n}_{\text{in}}\\
        &= -i \int d^d\mathbf x\int_{-\infty}^0 \frac{d\tau}{\tau}\, f_{\mathbf k}(\mathbf x,\tau)\,\mathcal E_{\mathbf x}[\tau]\,{}_{\text{out}}\bra{n'} T (\cdots \, \phi (\mathbf x,\tau) )\ket{n}_{\text{in}}.
    \end{align}
\end{subequations}
Similarly, we have
\begin{equation}
    \begin{aligned}
       {}_{\text{out}}\bra{n',\mathbf k'} T (\cdots) \ket{n}_{\text{in}}= -i \int d^d \mathbf x \int_{-\infty}^{0} \frac{d\tau}{\tau}\,\bar f_{\mathbf k'}(\mathbf x,\tau) \mathcal E_{\mathbf x}[\tau] {}_{\text{out}}\bra{n'} T (\phi(\mathbf x,\tau)\cdots) \ket{n}_{\text{in}}.
    \end{aligned}
\end{equation}
In this way, for each asymptotic state, there will be a field insertion in the correlator and action of the equation of motion, and then smearing over spacetime with particular mode functions carrying the information of the asymptotic states. Therefore, for an \(S\)-matrix element describing the scattering of \(n\) particles into \(n'\) particles in the Bunch-Davies basis will have a final form in terms of ``in-out cosmological correlators",
\begin{align}
    S_{n\to n'} &= {}_{\text{out}}\!\braket{\{\mathbf k_p'\} | \{\mathbf k_q\}}_{\text{in}} \notag\\
    &= (-i)^{n+n'} \int \Bigg[ \prod_{p=1}^{n'} \frac{d^d\mathbf x_p'\, d\tau_p'}{\tau_p'}\, \bar f_{\mathbf k_p'}\, \mathcal E_p' \Bigg]
       \Bigg[ \prod_{q=1}^{n} \frac{d^d\mathbf x_q\, d\tau_q}{\tau_q}\, f_{\mathbf k_q}\, \mathcal E_q \Bigg] \notag\\
    &\quad \times {}_{\text{out}}\!\bra{0} T\bigg( \prod_{p=1}^{n'}\phi(\mathbf x_p',\tau_p') 
       \prod_{q=1}^{n}\phi(\mathbf x_q,\tau_q) \bigg) \ket{0}_{\text{in}}.
\end{align}

The in-out correlator appearing in the LSZ formula can be evaluated using interaction picture perturbation theory. Using the above formula, we can deduce the Feynman rules for external scalar legs. This goes as follows. For an incoming scalar, we have the following.
\begin{equation}
    (-i) \int \frac{d^d\mathbf x\,d\tau }{\tau} f_{\mathbf k} (\mathbf x,\tau)\, \mathcal E_{\mathbf x}[\tau]\,G_0(\mathbf x,\tau;\mathbf x_j,\tau_j)\times (\cdots)= f_{\mathbf k}(\tau_j)\,e^{i\mathbf k\cdot \mathbf x_j}\times (\cdots)
\end{equation}
where \(G_0(\mathbf x,\tau;\mathbf x_j,\tau_j)= \bra{0} T (\phi_0 (\mathbf x,\tau)\phi_0 (\mathbf x_j,\tau_j))\ket{0}\), with $(\vec{x_j}, \tau_j)$ are the coordinates of an internal vertex. For outgoing mode functions, it gets replaced by the complex conjugate mode function, \(f_{\mathbf k}(\tau_j)\,e^{-i\mathbf k\cdot \mathbf x_j}\).

\subsection{LSZ reduction prescription for Dirac fields}
The spinor mode functions form a complete orthonormal basis under Wronskian relation of the Hankel function as per \ref{eq: spinor mode function orthogonality}. As already seen for the scalars, these orthogonality relations let us write the creation/annihilation operators for fermions and anti-fermions in the free theory.
\begin{subequations}
    \begin{align}
        &b^r_{\mathbf k} = u^{r\dagger}_{\mathbf k}(\tau) \psi(\mathbf k,\tau)\,,\quad
        b^{r\dagger}_{\mathbf k} = \psi(\mathbf k,\tau)^\dagger  u^{r}_{\mathbf k}(\tau)=\psi^\dagger(-\mathbf k,\tau)  u^{r}_{\mathbf k}(\tau)\,;\\
        &d^r_{\mathbf k} = \psi(-\mathbf k,\tau)^\dagger v^{r}_{\mathbf k}(\tau) = \psi^\dagger(\mathbf k,\tau) v^{r}_{\mathbf k}(\tau)\,,\quad
        d^{r\dagger}_{\mathbf k} = v^{r\dagger}_{\mathbf k}(\tau) \psi(-\mathbf k,\tau)\,.
    \end{align}
\end{subequations}
As we did for the scalars, the adiabatic approximation can be used to create single particle states in the interacting theory, at asymptotic times. Thus we have
\begin{subequations}\label{eq: Adiabatic approx for spinors}
    \begin{align}
        &\ket{\mathbf k,r}_{\text{in}} = \lim_{\tau\rightarrow-\infty} \psi^\dagger(-\mathbf k,\tau) u^{r}_{\mathbf k}(\tau)  \ket{0}_{\text{in}}\,,\quad
        \ket{\bar{\mathbf k},r}_{\text{in}} = \lim_{\tau\rightarrow-\infty}v^{r\dagger}_{\mathbf k}(\tau) \psi(-\mathbf k,\tau)\ket{0}_{\text{in}}\,;
        \\ &{}_{\text{out}} \bra{\mathbf k,r} = \lim_{\tau\rightarrow0}{}_{\text{out}}\bra{0} u^{r\dagger}_{\mathbf k}(\tau) \psi(\mathbf k,\tau)\,,\quad
      {}_{\text{out}}\bra{\bar{\mathbf k},r} = \lim_{\tau\rightarrow0}{}_{\text{out}}\bra{0} \psi^\dagger(\mathbf k,\tau) v^{r}_{\mathbf k}(\tau)
    \end{align}
\end{subequations}
where a `bar' above momentum indicates an anti-particle state. Let us now discuss the LSZ prescription for each kind of particle.

\paragraph{Ingoing fermion} Consider some \(S\)-matrix containing an ingoing fermion among other things,
\begin{subequations}
\begin{align}
    S_{\text{in} \to \text{out}} 
    &= {}_{\text{out}}\!\bra{n'}\, T(\cdots)\, \ket{\mathbf{k}, r; n}_{\text{in}} \\
    &= \lim_{\tau \to -\infty} {}_{\text{out}}\!\bra{n'}\, T\big(\cdots\, \psi^\dagger(-\mathbf{k}, \tau)\, u^r_{\mathbf{k}}(\tau)\big) \ket{n}_{\text{in}} \\
    &= - \int_{-\infty}^{0} d\tau\, \partial_\tau \Big[ {}_{\text{out}}\!\bra{n'}\, T\big(\cdots\, \bar\psi(-\mathbf{k}, \tau)\, \gamma^0\, u^r_{\mathbf{k}}(\tau)\big) \ket{n}_{\text{in}} \Big] \\
    &= - \int_{-\infty}^{0} d\tau\, {}_{\text{out}}\!\bra{n'}\, T\big(\cdots\, \big[ \partial_\tau \bar\psi(-\mathbf{k}, \tau)\, \gamma^0\, u^r_{\mathbf{k}}(\tau) 
       + \bar\psi(-\mathbf{k}, \tau)\, \gamma^0\, \partial_\tau u^r_{\mathbf{k}}(\tau) \big] \big) \ket{n}_{\text{in}} \\
    &= - \int_{-\infty}^{0} d\tau\, {}_{\text{out}}\!\bra{n'}\, T\big(\cdots\, \bar\psi(-\mathbf{k}, \tau)\, 
       \big( \overleftarrow{\partial}_\tau \gamma^0 - i \boldsymbol{\gamma} \cdot \mathbf{k} + \tfrac{i m}{\tau H} \big) u^r_{\mathbf{k}}(\tau) \big) \ket{n}_{\text{in}} \\
    &= -i \int d^d\mathbf{x} \int_{-\infty}^{0} d\tau\, {}_{\text{out}}\!\bra{n'}\, T\big(\cdots\, \bar\psi(\mathbf{x}, \tau) \big) \ket{n}_{\text{in}}\,
       \overleftarrow{\bar{\mathcal{E}}}_{\mathbf{x}}[\tau]\, e^{i \mathbf{k} \cdot \mathbf{x}}\, u^r_{\mathbf{k}}(\tau)\,.
\end{align}
\end{subequations}

This expresses the \(S\)-matrix in a form where an incoming fermion is replaced by \(\bar\psi\) with the equation of motion acting on it and smearing over spacetime with the appropriate mode-function. What it effectively does is amputate the external legs of various diagrams, contributing to computing the correlator. Suppose a diagram in the interaction picture perturbation has an external leg connected to some vertex \((\mathbf x_j,\tau_j)\),
\begin{equation}
    (\cdots) \times (-i)\int d^d\mathbf x \,d\tau \,G(\mathbf x_j,\tau_j;\mathbf x,\tau)\,\overleftarrow{\bar{ \mathcal E}}_{\mathbf x}[\tau]\,e^{i\mathbf k\cdot\mathbf x}\,u^{r}_{\mathbf k}(\tau)= (\cdots) \times u_{\mathbf k}^r(\tau_j)\,e^{i\mathbf k\cdot\mathbf x_j}
\end{equation}
where, \(G(\mathbf x_j,\tau_j;\mathbf x,\tau) = \bra{0}T(\psi(\mathbf x_j,\tau_j)\bar\psi(\mathbf x,\tau))\ket{0}\). Therefore, the external propagator attached to that internal vertex is simply replaced by the mode function attached to that vertex.

\paragraph{Outgoing anti-fermion} For the \(S\)-matrix involving outgoing anti-fermions, the calculations will follow steps similar to those of the incoming fermion with a few changes.
\begin{equation}
    \begin{aligned}
        S_{in\rightarrow out} &= {}_{\text{out}}\bra{\bar{\mathbf k},r;n'}T(\cdots)\ket{n}_{\text{in}}\\
        &= \lim_{\tau\rightarrow0}{}_{\text{out}}\bra{n'} \psi^\dagger(\mathbf k,\tau) v^{r}_{\mathbf k}(\tau) T (\cdots)\ket{n}_{\text{in}}\\
        &= + i \int d^d\mathbf x \int_{-\infty}^0 d\tau\, {}_{\text{out}}\bra{n'} T ( \bar\psi(\mathbf x,\tau)\cdots)  \ket{n}_{\text{in}} \overleftarrow{\bar{ \mathcal E}}_{\mathbf x}[\tau]\,e^{-i\mathbf k\cdot\mathbf x}\,v^{r}_{\mathbf k}(\tau),
    \end{aligned}
\end{equation}
which means that the external leg of a diagram contributing to the correlator will be replaced by \( v_{\mathbf k}^r(\tau_j)\,e^{-i\mathbf k\cdot\mathbf x_j}\).

\paragraph{Outgoing fermion} Now we focus on \(S\)-matrix with an outgoing fermion. It has the following form,
\begin{subequations}
\begin{align}
    S_{\text{in} \to \text{out}} 
    &= {}_{\text{out}}\!\bra{\mathbf{k}, r; n'}\, T(\cdots)\, \ket{n}_{\text{in}} \\
    &= \lim_{\tau \to 0} {}_{\text{out}}\!\bra{n'}\, u^{r\dagger}_{\mathbf{k}}(\tau)\, \psi(\mathbf{k}, \tau)\, T(\cdots)\, \ket{n}_{\text{in}} \\
    &= \int_{-\infty}^{0} d\tau\, \partial_\tau \Big[ {}_{\text{out}}\!\bra{n'}\, T\big( u^{r\dagger}_{\mathbf{k}}(\tau)\, \psi(\mathbf{k}, \tau)\, \cdots \big) \ket{n}_{\text{in}} \Big] \\
    &= \int_{-\infty}^{0} d\tau\, {}_{\text{out}}\!\bra{n'}\, T\Big( 
       \big[ \partial_\tau u^{r\dagger}_{\mathbf{k}}(\tau)\, \psi(\mathbf{k}, \tau) + u^{r\dagger}_{\mathbf{k}}(\tau)\, \partial_\tau \psi(\mathbf{k}, \tau) \big]\, \cdots \Big) \ket{n}_{\text{in}} \\
    &= \int_{-\infty}^{0} d\tau\, {}_{\text{out}}\!\bra{n'}\, T\Big( 
       \bar u^r_{\mathbf{k}}(\tau)\, \big( i\boldsymbol{\gamma} \cdot \mathbf{k} - \tfrac{i m}{\tau H} + \gamma^0 \partial_\tau \big)\, \psi(\mathbf{k}, \tau)\, \cdots \Big) \ket{n}_{\text{in}} \\
    &= -i \int d^d \mathbf{x} \int_{-\infty}^{0} d\tau\, 
       \bar u^r_{\mathbf{k}}(\tau)\, e^{-i \mathbf{k} \cdot \mathbf{x}}\, 
       \mathcal{E}_{\mathbf{x}}[\tau]\, {}_{\text{out}}\!\bra{n'}\, T\big( \psi(\mathbf{x}, \tau)\, \cdots \big)\, \ket{n}_{\text{in}}.
\end{align}
\end{subequations}

Here, we see that the equation of motion is acting on the correlator. If in a particular diagrammatic contribution, there is an external leg connecting to an internal vertex \((\mathbf x_j,\tau_j)\), then the equation of motion will act on the free propagator \(G(\mathbf x,\tau;\mathbf x_j,\tau_j)\). Therefore, the external free propagator will be replaced by \(\bar u^r_{\mathbf k}(\tau_j)e^{-i\mathbf k\cdot\mathbf x_j}\) for outgoing fermions.

\paragraph{Incoming antifermion} The calculation for incoming antifermion is similar to that of outgoing fermion with a few changes. The \(S\)-matrix element can be written as
\begin{equation}
    \begin{aligned}
        S_{in\rightarrow out} &= {}_{\text{out}}\bra{n'}T (\cdots)\ket{\bar{\mathbf k}, r;n}_{\text{in}}\\
        &= \lim_{\tau\rightarrow-\infty}{}_{\text{out}}\bra{n'}T (\cdots)v^{r\dagger}_{\mathbf k}(\tau) \psi(-\mathbf k,\tau)\ket{n}_{\text{in}}\\
        &= i \int d^d\mathbf x\,\int_{-\infty}^0 d\tau\,\bar v^r_{\mathbf k}(\tau)e^{i\mathbf k\cdot\mathbf x}\mathcal E_{\mathbf x}[\tau]\,{}_{\text{out}}\bra{n'} T\left(\cdots\psi(\mathbf x,\tau)\right)\ket{n}_{\text{in}}.
    \end{aligned}
\end{equation}

\noindent Therefore, in the case of an incoming antifermion, external free propagators of Feynman diagrams will be replaced by \(\bar v^r_{\mathbf k}(\tau_j)\,e^{i\mathbf k\cdot\mathbf x_j}\).

\subsection{Adiabatic Approximation for LSZ prescription}
In the proof of LSZ, that we have shown for fermions\footnote{Detailed discussions on LSZ for scalars and adiabatic approximation for the same can be found in \cite{pimentel1}.}, we have assumed that the creation and annihilation operators shown creates single particle states at far past/future, as is shown in \ref{eq: Adiabatic approx for scalars} and \ref{eq: Adiabatic approx for spinors}. However, the mode functions used here are the free theory mode functions, which are the solutions for the homogeneous equation of motion. Thus, for interacting theory, the creation operator can create multi-particle states at finite times. For the validity of our method, the probability for extra particle creation should approach 0 as we increase the time interval of the experiment.

For the UdW\(_T\) S matrix that we use, both the states are defined with respect to the vacuum in the far past. In far past, the equation of motion is the same as its Minkowski counterpart for both scalars and spinors\footnote{The discussion that will follow later is applicable for weak limit of the adiabatic approximation. For a more detailed discussion of this, see \cite{collins2019newapproachlszreduction}}. Thus, the subtleties of the adiabatic approximation for this is the same as that in Minkowski space-time. The time-dependent effective mass term goes to 0, which leads to possible IR divergence. We will be concerned only with scattering of hard modes, \(\mathbf{k}\neq0\), which will get rid of these possible IR divergences. The adiabatic approximation in a flat space has been treated quite rigourously in (\cite{Hepp1965}\cite{PhysRev.112.669}\cite{Lehmann1955}\cite{collins2019newapproachlszreduction}).

Let us work out what happens if we try to create particles on the vacuum at finite times, using the creation operator. We have
\begin{subequations}
    \begin{align}
        i\,  f_{\mathbf k} (\tau) \overleftrightarrow{\tau\partial_\tau} \phi (-\mathbf k,\tau) \ket{0,-\infty}_{\text{out}} &=\ket{\mathbf{k},-\infty}_{\text{out}}+\sum_{n=2}^{\infty}c^{S}_{\text{out}}\left(k\tau;n\right)\ket{n,-\infty}_{\text{out}}\,,\\
        \psi^\dagger(-\mathbf k,\tau)  u^{r}_{\mathbf k}(\tau)\ket{0}_{\text{out}} &= \ket{\mathbf k,r}_{\text{out}} + \sum_{n=2}^{\infty}c^{F}_{\text{out}}\left(k\tau;n\right)\ket{n,r}_{\text{out}}\quad \text{and}\\
        v^{r\dagger}_{\mathbf k}(\tau) \psi(-\mathbf k,\tau)\ket{0}_{\text{out}} &= \ket{\bar{\mathbf k},r}_{\text{out}} + \sum_{n=2}^{\infty}c^{\bar{F}}_{\text{out}}\left(k\tau;n\right)\ket{\bar{n},r}_{\text{out}}
    \end{align}
\end{subequations}
where the bar on F denotes the constant for antiparticle production. There is a corresponding set of coefficients for the in states. The weak limit for the adiabatic approximation means that, taking \(\tau\) asymptotic times, these coefficients of extra particle/antiparticle production must vanish. Thus,
\begin{equation}
    \lim_{ \tau \rightarrow 0 } c^{ \alpha }_{ \text{out} } = 0 \quad \text{and}\quad \lim_{ \tau \rightarrow -\infty } c^{ \alpha }_{ \text{in} } = 0 \quad \text{where} \quad \alpha \in \{ S,\, F,\, \bar{F}\}\,.
\end{equation}
For weak limits in the far past, the discussion is the same as that of Minkowski spacetime, as mentioned before. For late times, in the case of scalars in the principal series\footnote{For subtleties regarding scalars in the complementary series and detailed calculations for scalars in principal series, refer to \cite{pimentel1}}, \(c^{S}_{\text{out}}\) approaches zero, if \(\sum_{b=1}^n|\Im\mu_b|>\frac{d}{2}\left(n-2\right)\), for any n-point interaction\footnote{\(\mu_b\) is the order of the scalar fields involved, as given in eq. \ref{scalar eom}}. For spinors, in the Heisenberg picture, the field evolves as follows:
\begin{equation}
    \mc E[k\tau]\psi(\tau,\mathbf k)=\frac{\delta S_{\text{int}}}{\delta \bar{\psi}}\,,
\end{equation}
And we can write the general solution as
\begin{equation}\label{general solution spinor}
    \psi(\tau,\mathbf k)=\hat{b}^r_{\mathbf k}u^r_{\mathbf k}+\hat{d}^{r\,\dagger}_{\mathbf k}v^r_{\mathbf k}+\int_\tau^0 d\tau \, \mc G^{\text{ret}}(k\tau,k\tau')\frac{\delta S_{\text{int}}}{\delta\bar{\psi}}
\end{equation}
where \(S_{\text{int}}\) is the non-linear part of the action, and \(\mc G^{\text{ret}}(k\tau,k\tau')\) is the retarded propagator built from the mode functions. We can verify if the weak limit holds for an \(S_{\text{int}}\). Let us take the 4-Fermi interaction as an example, the non-linear part of the action is given by,
\begin{equation}
    S_{\text{int}} = \int d^dx\sqrt{-g}\, G_F\left(\overline{\psi}_{2}\gamma^\mu\frac{(1-\gamma^5)}{2}\psi_{1}\right)\left(\overline{\psi}_4\gamma_\mu\frac{(1-\gamma^5)}{2}\psi_{3}\right)\,.
\end{equation}
The retarded propagator is \(u^r_{\mathbf k}\bar{u}^r_{\mathbf k}\). On substituting this in \ref{general solution spinor}, and using that to create particles at late times (we need to choose a field, let us choose \(\psi_2\)), it turns out there will be some non-trivial contribution to the state \(\ket{\mathbf k_1 \mathbf k_2 \bar{\mathbf k}_3}\). The spin indices are suppressed as it will not be important to consider. As it turns out,
\begin{equation}
    c^{F}_{\text{out}}(k\tau;\mathbf k_1\tau,\mathbf k_2\tau,\bar{\mathbf k}_3\tau)=\int_\tau^0d\tau(-\tau)^{d-1}(\bar{u}_3P_L\gamma_\mu v_4)(\bar{u}_{1}P_L\gamma_\mu u_2),
\end{equation}
can be verified to vanish as \(\tau\rightarrow0\). We have suppressed a momentum-conserving delta function \((2\pi)^d\delta^d(\mathbf k_1+\mathbf k_3+\mathbf k_4-\mathbf k_2)\). In general, for any n-point contact interaction of spinors, the integrand behaves like \(\sim\tau^{\alpha-1}\), where \(\alpha=\frac{d}{2}(n-2)\). Thus, for \(n>2\) the integration will vanish for \(\tau\rightarrow0\). A more general proof for the validity of LSZ for scalars/spinors in line with \cite{collins2019newapproachlszreduction} is left for future exploration.

\begin{tcolorbox}
    Unlike scalars, spinors do not have a complementary series. This is because the order of the fermionic mode functions has a mass-independent imaginary part, as given in \ref{modefunc definition}.
\end{tcolorbox}

\section{Scalar example further details}\label{scalar}
\paragraph{Two-derivative contact interaction} Consider the following two-derivative contact interaction,
\begin{equation}
    \mathcal L_{EFT} \supset \frac{\lambda_2}{(2!)^2} (\phi_1\partial_\mu\phi_2) (\phi_1 \partial^\mu \phi_2) = \frac{\lambda_2}{(2!)^2}  g^{\mu\nu} (\phi_1\partial_\mu\phi_2) (\phi_1 \partial_\nu \phi_2)\,.
\end{equation}
We note that \(g^{\mu\nu} = (-\tau H)^2 \eta^{\mu\nu}\) and also we need to rescale \(\phi_2 \rightarrow (-\tau H)^{d/2}\phi_2\) and same for \(\phi_1\). Whereas \(\nabla\phi_2 \rightarrow (-\tau H)^{d/2}\nabla\phi_2\), the time-derivative of \(\phi_2\) has some structure. Let us see this in detail.
\begin{equation}
\begin{aligned}
    \partial_\tau \phi_2 &\rightarrow  \partial_\tau( (-\tau H)^{d/2}\phi_2)\\
    \implies (\partial_\tau \phi_2)^2 &\rightarrow (e^{-Ht})^d \left[(\partial_t\phi_2)^2 + 2 H \left(t\partial_t\phi_2 -\frac{d}{2}\phi_2\right)\partial_t \phi_2\right]\,.
\end{aligned}  
\end{equation}
Therefore,
\begin{equation}
    \mathcal L_{EFT} \supset \frac{\lambda_2}{(2!)^2} (e^{- Ht})^{2d+2} \phi_1^2 \left[(\partial_t\phi_2)^2 + 2 H \left(t\partial_t\phi_2 -\frac{d}{2}\phi_2\right)\partial_t \phi_2 - \nabla\phi_2\cdot \nabla\phi_2\right]\,.
\end{equation}
The matrix element coming from the above interaction term is
\begin{equation}
    \begin{aligned}
        \mathcal M_2 &= \lambda_2 \int dt \, (e^{- Ht})^{d+2} f_{1,\mathbf p_1}(t)\bar f_{1,\mathbf p_3} (t)\bigg (-\mathbf p_2\cdot\mathbf p_4\,f_{2,\mathbf p_2}(t)\bar f_{2,\mathbf p_4} (t) + \partial_t f_{2,\mathbf p_2}(t)\partial_t\bar f_{2,\mathbf p_4} (t)\\
        &- \frac{Hd}{2}(\partial_t f_{2,\mathbf p_2}(t)\bar f_{2,\mathbf p_4} (t)+f_{2,\mathbf p_2}(t)\partial_t \bar f_{2,\mathbf p_4} (t)) + 2 H t\, \partial_t f_{2,\mathbf p_2}(t)\partial_t\bar f_{2,\mathbf p_4} (t)
        \bigg)\,.
    \end{aligned}
\end{equation}
The zeroth-order term is
\begin{equation}
    \mathcal M_2^{(0)} = \frac{\lambda_2}{\Pi_i \sqrt{2\omega_i}} (\omega_2\omega_4 - \mathbf p_2 \cdot \mathbf p_4) 2\pi \delta(\Delta\omega)\,.
\end{equation}
This is real. \\

\noindent First order corrections coming from the explicit \(H\) dependent term is,
\begin{equation}
    \mathcal M_2^{(1)} \supset \frac{\lambda_2 H}{\Pi_i \sqrt{2\omega_i}} \int dt\, \left(2t\, \omega_2\omega_4 + i\frac{d}{2} (\omega_2-\omega_4)\right) e^{-i\Delta\omega\,t}\,.
\end{equation}
Evidently, this contribution is imaginary. Next, we focus on contributions from the derivatives of the mode functions. We note that,
\begin{equation}
    \partial_t f_{\mathbf k}(t) = \frac{e^{-i\omega_k t}}{\sqrt{2\omega_k}} (-i\omega_k -i\omega_k(i c_0^B H + c_1^B Ht + i c_2^B Ht^2) + c_1^B H + 2 i c_2^B H t)\,.
\end{equation}
Here we have explicitly shown the \(O(H)\) corrections in the derivative of the mode functions. Now, by direct calculation, it can be shown that the corrections originating from this in the matrix element are also imaginary. Therefore, in this case also, there will be no \(O(H)\) correction in the mode function.

\paragraph{Higher derivative contact interactions and a general argument}

By explicit calculations, we have seen that for non-derivative contact interactions as well as two-derivative contact interactions, the \(O(H)\) correction to the scalar scattering amplitude is imaginary. However, this conclusion is true even for higher derivative contact interactions. Even though it becomes more cumbersome to verify this, we have explicitly checked for the following four derivative contact interaction
\begin{equation}
    \mathcal L_{EFT} \supset \frac{\lambda_3}{(2!)^2} (\phi_1 \nabla_\mu\nabla_\nu \phi_2)(\phi_1 \nabla^\mu\nabla^\nu \phi_2)\,.
\end{equation}
In the covariant derivatives there will be now explicit dependence of the Christoffel symbols. In this case also the corrections are again imaginary.

Now we will give a general argument so as to why the \(O(H)\) correction to the scalar scattering amplitude will be always imaginary for any possible contact interaction. Suppose the full amplitude has the following contribution,
\begin{equation}
    \mathcal M = \int dt\, \Pi_i \mathcal O_i (t)\,.
\end{equation}
For \(H \rightarrow 0\), \(\mathcal O_i (t)\rightarrow \mathcal O_i^{(0)}\), and the zeroth order amplitude is,
\begin{equation}
    \mathcal M^{(0)} = \int dt\, \Pi_i \mathcal O_i^{(0)}(t)\,.
\end{equation}
Now we introduce the concept of ``reality \(\mathcal R\)" of a term. \(\mathcal R\) can take values \(\pm1\). A function \(\mathcal F(t)\) has \(\mathcal R = +1\) if \(\int dt\,\mathcal F(t)\) is real. \(\mathcal{R}=-1\) if \(\int dt\,e^{-i\Delta\omega\,t}\mathcal F(t)\) is purely imaginary. Since \(\mathcal M^{(0)}\) is always real, we have,
\begin{equation}
    \mathcal R \left(\Pi_i \mathcal O_i^{(0)}\right) = \Pi_i \mathcal R(\mathcal O_i^{(0)}) =  +1\,.
\end{equation}
The component terms \( \mathcal O_i^{(0)}\) can have whatever reality, but the above condition has to be satisfied. We also note that
\begin{equation}
    \begin{aligned}
        \mathcal R (t^n\,e^{-i\omega\,t}) &= +1\,\, \text{ for }\,\, n \in 2\mathbb Z \quad \text{and}\\
        \mathcal R (t^n\,e^{-i\omega\,t}) &= -1\,\, \text{ for }\,\, n \in 2\mathbb Z + 1\,.
    \end{aligned}
\end{equation}
Now the first order correction to \(\mathcal M\) is,
\begin{equation}
    \begin{aligned}
        \delta\mathcal M
        &= \mathcal M - \mathcal M^{(0)}\\
        &= \sum_i \int dt\, (\mathcal O_1^{(0)}\cdots \delta\mathcal O_{i}\cdots)
    \end{aligned}
\end{equation}
where \(\delta\mathcal O_{i} = \mathcal O_{i}-\mathcal O_{i}^{(0)}\). In the following we will show that \(\mathcal R (\delta\mathcal O_i)= -\mathcal R (\mathcal O_i^{(0)}) \). It will imply that \(\mathcal{R}(\mathcal O_1^{(0)}\cdots \delta\mathcal O_{i}\cdots)=-1\) and therefore,
\begin{equation}
    (\delta \mathcal M)^* = - \delta \mathcal M\,.
\end{equation}
The first-order correction term is imaginary.\\

\noindent The proof of our claim \(\mathcal R (\delta\mathcal O_i)= -\mathcal R (\mathcal O_i^{(0)})\) is as follows. The operators entering the integrand are always scalar-mode functions or their multiple (possibly covariant) derivatives. If \(\mathcal O_i\) is a pure mode function \(f_{\mathbf k}(t)\), then
\begin{equation}
    \delta\mathcal O_i = \frac{e^{-i\omega_k t}}{\sqrt{2\omega_k}} (i c_0^B H + c_1^B H t + i c_2^B H t^2)\,.
\end{equation}
Clearly, \(\mathcal R (\delta\mathcal O_i)=-1 =  \mathcal R (\mathcal O_i^{(0)})\). Also, by simple time derivative,s this conclusion is not going to change because,
\begin{equation}
    \mathcal R (\partial_t \mathcal O(t)) = - \mathcal R (\mathcal O(t))\,.
\end{equation}
Therfore, under time derivative the reality of both \(\delta\mathcal O_i\)  and \(\mathcal O_i^{(0)}\) will flip maintaining their opposite reality. 

Now we consider the extra first-order stuff that arises while doing a time derivative due to the \((-\tau H)^{d/2}\) rescaling of the fields. Now recall that under this rescaling,
\begin{equation}
    \partial_\tau \rightarrow \partial_t + H \left(t\partial_t - \frac{d}{2}\right)\,.
\end{equation}
Now \(\partial_t\) flips reality, but the first-order correction does not when it acts on some function. Therefore, in this case also the first order correction will have opposite reality with respect to the zeroth order term.\\

\noindent Another possibility of extra terms comes from the Christoffel symbols. However, the non-zero Christoffel symbols are proportional to
\begin{equation}
    \Gamma^{\mu}_{\rho\sigma} \propto \frac{-1}{\tau} \sim H \,,
\end{equation}
which has a positive reality. These Christoffel symbols appear in the following scenario (schematically),
\begin{equation}
   \nabla_\mu = \partial_\mu + \Gamma^{\rho}_{\mu\sigma}\,.
\end{equation}
Since the zeroth order term (which is a derivative) is reality flipping, but the first order correction is not, again we will have, \(\mathcal R (\delta\mathcal O_i) =  -\mathcal R (\mathcal O_i^{(0)})\).

 This concludes the proof that for all possible contact interactions also, the first-order correction to the matrix element will always be imaginary for scattering amplitude involving scalar fields.

\paragraph{Derivative exchange interaction} Now we consider the following interaction Lagrangian,
\begin{equation}
    \mathcal L_{int} = \lambda\, \phi_1 \partial_\mu\phi_1 \partial^\mu\phi_2 = \lambda\, g^{\mu\nu} \phi_1 \partial_\mu\phi_1 \partial_\nu\phi_2\,.
\end{equation}
We need to analyse this term carefully before moving on to calculate the \(S\)-matrix element. First, recall that \(g^{\mu\nu} = (-\tau H)^2 \eta^{\mu\nu}\). We also need to do rescale the fields by \((-\tau H)^{d/2}\), and in the presence of derivatives there will some additional terms appearing. The following relation will be useful.
\begin{equation}
\begin{aligned}
    &\partial_\tau ((-\tau H)^{d/2} \phi)\\
    &= (-\tau H)^{d/2} (\partial_t \phi + H (t\partial_t \phi - (d/2) \phi))\,.
\end{aligned}
\end{equation}
Therefore, the interaction term becomes the following.
\begin{equation}
    \mathcal L_{int} = \lambda\, (e^{-H t})^{3d/2+2} \phi_1\left(\partial_t\phi_1\partial_t \phi_2 - \nabla\phi_1\cdot\nabla\phi_2 + H (t\partial_t \phi_1 - \frac{d}{2} \phi_1)\partial_t \phi_2 + H \partial_t \phi_1(t\partial_t \phi_2 - \frac{d}{2} \phi_2) \right)\,.
\end{equation}
Now, let us calculate the matrix element of the \(s\) channel for the process \(\phi_1\phi_1 \rightarrow \phi_1\phi_1\). The zeroth-order term is
\begin{equation}
    \begin{aligned}
        \mathcal M_{(s)}^{(0)} = \frac{2\pi \lambda^2 \delta(\Delta\omega)}{\Pi_i \sqrt{2\omega_i}} \frac{((\omega_1+\omega_2)^2-\mathbf p^2)((\omega_3+\omega_4)^2-\mathbf p^2)}{\omega_p^2 - (\omega_1 + \omega_2)^2}\,.
    \end{aligned}
\end{equation}
Next, we see the effect of the term \(\lambda H \phi_1 \left((t\partial_t \phi_1 - \frac{d}{2} \phi_1)\partial_t \phi_2 +  \partial_t \phi_1(t\partial_t \phi_2 - \frac{d}{2} \phi_2) \right)\) in the matrix element. When it contributes to one of the vertices, then the correction is
\begin{equation}
    \begin{aligned}
        \mathcal M_{(s)}^{(1)} &\supset \frac{\lambda^2 H}{\Pi_i \sqrt{2\omega_i}} \frac{(\mathbf p^2 - (\omega_3+\omega_4)^2)}{\omega_p^2 - (\omega_3+\omega_4)^2} (i(\omega_3+\omega_4)) \int dt_1\, (-it_1 (\omega_1+\omega_2)-d/2) e^{-i\Delta \omega \,t_1} \\
        &+ \frac{\lambda^2 H}{\Pi_i \sqrt{2\omega_i}} \frac{(\mathbf p^2 - (\omega_3+\omega_4)^2)}{\omega_p^2 - (\omega_3+\omega_4)^2} (-i(\omega_1+\omega_2))\int dt_1 (i t_1 (\omega_3 + \omega_4)-d/2) e^{-i\Delta \omega \,t_1}\,.
    \end{aligned}
\end{equation} 
Therefore, the extra \(O(H)\) term in the Lagrangian adds only imaginary contribution to the matrix element.\\

\noindent Lets analyze the contribution of \(\lambda\phi_1\left(\partial_t\phi_1\partial_t \phi_2 - \nabla\phi_1\cdot\nabla\phi_2\right)\) (which will contain the zeroth order term also),

\begin{equation}
    \begin{aligned}
        \mathcal M_{(s)} &\supset {i\lambda^2} \int \int dt_1\,dt_2 (e^{-H (t_1+t_2)})^{d/2+2} (\left(f_{\mathbf p_2}(t_1) \partial_{t_1} f_{\mathbf p_1}(t_1) + f_{\mathbf p_1}(t_1) \partial_{t_1} f_{\mathbf p_2}(t_1)\right) \overrightarrow{\partial_{t_1}} - \mathbf p^2 f_{\mathbf p_2}(t_1) f_{\mathbf p_1}(t_1))\\
        &G_{\mathbf p} (t_1,t_2) (\overleftarrow{\partial_{t_2}}\left(\bar f_{\mathbf p_3}(t_2) \partial_{t_2} \bar f_{\mathbf p_4}(t_2) + \bar f_{\mathbf p_4}(t_1) \partial_{t_2} \bar f_{\mathbf p_3}(t_2)\right)  - \mathbf p^2 \bar f_{\mathbf p_4}(t_1) \bar f_{\mathbf p_3}(t_2))\,.
    \end{aligned}
\end{equation}
We need to care about the corrections coming from the external mode-functions as well as the internal propagator. Since we are working in \(O(H)\), we can analyze each correction independently. Correction coming from \(f_{\mathbf p_1}(t_1)\) (without time-derivative) will be similar to what was analyzed in the non-derivative case, that is, it will be proportional to,
\begin{equation}
    \int dt_1 e^{-i\Delta\omega\,t_1} f_{\mathbf p_1}^{(1)} (t_1)\,,
\end{equation}
which is imaginary. Next, we analyze the correction from \(\partial_{t_1}f_{\mathbf p_1}(t_1)\). We need to use,
\begin{equation}
    \partial_{t_1}f_{\mathbf p_1}(t_1) = \frac{- i \omega_1}{\sqrt{2\omega_1}}\, e^{-i\omega_1 t_1} + \frac{1}{\sqrt{2\omega_1}}\, e^{-i\omega_1 t_1} (-i\omega_1  f_{\mathbf p_1}^{(1)} (t_1) + \partial_{t_1} f_{\mathbf p_1}^{(1)} (t_1))
\end{equation}
where the second term is the first order correction. Now the contribution of this to the matrix element is
\begin{equation}
    \begin{aligned}
        \mathcal M_{(s)}^{(1)} &\supset \frac{\lambda^2}{\Pi_i \sqrt{2\omega_i}} \frac{((\omega_3+\omega_4)^2-\mathbf p^2)}{\omega_p^2 - (\omega_3+\omega_4)^2} i (\omega_3+\omega_4) \int dt_1 \, (-i\omega_1  f_{\mathbf p_1}^{(1)} (t_1) + \partial_{t_1} f_{\mathbf p_1}^{(1)} (t_1)) e^{-i\Delta \omega \,t_1}\\
        &\supset \frac{\lambda^2}{\Pi_i \sqrt{2\omega_i}} \frac{((\omega_3+\omega_4)^2-\mathbf p^2)}{\omega_p^2 - (\omega_3+\omega_4)^2} (\omega_3+\omega_4) \int dt_1 \, (\omega_1  f_{\mathbf p_1}^{(1)} (t_1) + i \partial_{t_1} f_{\mathbf p_1}^{(1)} (t_1)) e^{-i\Delta \omega \,t_1}\,.
    \end{aligned}
\end{equation}
Again the form of \(f_{\mathbf p_1}^{(1)}\) implies that this contribution is also purely imaginary. Now we can focus on the correction coming from the derivative of the internal propagator. There can be both double derivatives and single derivative on the internal propagator. The corrections have the following form,
\begin{equation}
\begin{aligned}
    \partial_{t_1} G_{\mathbf p}^{(1)} (t_1,t_2) &\supset  \int \frac{d p^0}{2\pi } e^{- i p^0 (t_1-t_2)} \frac{i}{(p^0)^2 - \omega_k^2 + i \epsilon} \left(-i p^0 c_1^B H t_1 + \frac{ (p^0)^2}{\omega_p} c_2^B H t_1^2 + c_1^B H + \frac{2 i p^0}{\omega_p} c_2^B H  t_1\right)\\
    &=: \int \frac{d p^0}{2\pi } e^{- i p^0 (t_1-t_2)} \frac{i}{(p^0)^2 - \omega_k^2 + i \epsilon} F_1 (t_1,p^0)\,,\\
    \partial_{t_2}\partial_{t_1} G_{\mathbf p}^{(1)} (t_1,t_2) &\supset \int \frac{d p^0}{2\pi } e^{- i p^0 (t_1-t_2)} \frac{i}{(p^0)^2 - \omega_k^2 + i \epsilon} (i p^0) F_1 (t_1,p^0)\,.
\end{aligned}
\end{equation}
We have shown some representative terms, there are similar looking terms involving \(t_2\). Let us see some first-order correction coming from the single derivative of the internal propagator,
\begin{equation}
    \begin{aligned}
        \mathcal M_{(s)}^{(1)} &\supset \frac{\lambda^2}{\Pi_i \sqrt{2\omega_i}} \frac{(-i (\omega_1+\omega_2)) (-\mathbf p^2)}{\omega_p^2 - (\omega_3+\omega_4)^2} \int dt_1 F_1 (t_1,\omega_3+\omega_4) e^{-i\Delta\omega\,t_1}\,.
    \end{aligned}
\end{equation}
This term is imaginary because the integral is real as \(\bar F_1 (-t_1,p^0) = F_1 (t_1,p^0)\). Now a representative correction term coming from the double derivative of the internal propagator is
\begin{equation}
    \begin{aligned}
        \mathcal M_{(s)}^{(1)} &\supset \frac{\lambda^2}{\Pi_i \sqrt{2\omega_i}} \frac{(\omega_1+\omega_2)(\omega_3+\omega_4)^2}{\omega_p^2 - (\omega_3+\omega_4)^2} (-i)\int dt_1\, F_1 (t_1,\omega_3+\omega_4) e^{-i\Delta\omega\,t_1}\,.
    \end{aligned}
\end{equation}
Evidently, this is also purely imaginary. Therefore, we conclude that even in this case of tree-level exchange diagram with derivative interactions, all the \(O(H)\) corrections are purely imaginary. Therefore, there is no correction to the amplitude squared or the scattering cross-section.

\section{Dealing with Dirac Delta function and its derivative} \label{delta delta prime}

In the main text, for computing the first-order correction in \(H\) we have to deal with the product of the delta function and derivative of the delta function. Here we will analyze this aspect in detail. The finite-size version that naturally appears in the amplitude calculation is
\begin{equation}
    \delta_T(u) =\frac{1}{2\pi} \int_{-\frac{T}{2}}^{+\frac{T}{2}}\,dt \, e^{-i u t} = \frac{\sin\left(\frac{u T}{2}\right)}{\pi u }\,.
\end{equation}
and the derivative of delta function (with respect to \(u\)),
\begin{equation}
    \delta_T'(u) =\frac{1}{2\pi} \int_{-\frac{T}{2}}^{+\frac{T}{2}}\,dt \, (- i t)\,e^{-i u t} = \frac{T \cos\left(\frac{u T}{2}\right)}{2 \pi u } - \frac{\sin\left(\frac{u T}{2}\right)}{\pi u^2 }\,.
\end{equation}

\paragraph{Dealing with \( \delta_T^2 (u)\)--Revisited} In usual flat-space QFT, some ``transition probability \(P\)" probability is proportional to \(\delta_T^2 (u)\). The origin of this is the following integral,
\begin{equation}
    P \propto \int dt \int dt' e^{-i u (t-t')}\cdots \propto \delta_T^2 (u)
\end{equation}
where \(\cdots\) indicate time-independent part of the integrand and \(u\) is the energy difference of the initial and final states.

The usual of dealing with this is the replacement \(\delta_T^2 (u) = \delta_T (0)\delta_T (u) = T \delta_T (u)/2\pi\). Then the \(\delta_T (u)\) is done with proper jacobian factor for change of variable to some other variable \(x\) which is an argument of \(u = u(x)\), and everywhere \(x\) is replaced with the solution of \(u(x)=0\). Let's see this more clearly. We have some transition probability \(P\) as,
\begin{equation}
    P = \int d x\,\, F(x) \,\delta_T^2 (u (x)) = \int du \frac{F(x)}{|u'(x)|}\delta_T^2 (u (x)) = \int d u\,\, \Tilde F(x (u)) \,\delta_T^2 (u)\,.
\end{equation}
Now, \(\delta_T^2 (u)\) has a peak at \(u\) which is sharper for larger \(T\), and the value of  \(\delta_T^2 (u)\) at this location is simply, \((T/2\pi)^2\). Therefore, the integral is approximated as
\begin{equation}
    P \approx \Delta u \times \Tilde F(x_0) \left(\frac{T}{2\pi}\right)^2
\end{equation}
(\(x_0\) is the physically allowed solution for \(u(x)=0\)).


Since \(u\) is Fourier conjugate of \(t\), \(\Delta u = 2\pi/T\), we have the following `transition rate \(R\)',
\begin{equation}
    R = \frac{P}{T} = \frac{\Tilde F(x_0)}{2\pi}\,.
\end{equation}
This is essentially the prescription that we follow in usual flat space QFT. 

\paragraph{Dealing with \( \delta_T (u)\delta_T' (u)\):} This type of term arises while calculating first-order de Sitter corrections. We have correction terms proportional to
\begin{equation}
    P \propto \int dt\int dt' \, (-i t) e^{-i u(t-t')}\cdots \propto \delta_T (u)\delta_T' (u).
\end{equation}
Being schematic, we have to deal with the following integral.
\begin{subequations}
    \begin{align}
        P &= \int dx\,\,F(x)\, \delta_T (u(x))\delta_T' (u(x))
        = \int du \frac{F(x)}{|u'(x)|} \, \delta_T (u)\delta_T' (u)\\
        &= \int du \,\,\Tilde F(x(u)) \, \delta_T (u)\delta_T' (u)
        \equiv - \frac{1}{2} \int du \,\,\left(\frac{\partial}{\partial u}\Tilde F(x(u))\right) \, \delta_T^2 (u)\\
        &= - \frac{1}{2} \int du \,\,\Tilde F'(x(u)) \frac{1}{u'(x)}\, \delta_T^2 (u)\,.
    \end{align}
\end{subequations}
In the second line we have used that the boundary term, while doing the integration by parts, is negligible compared to the bulk term. In particular, the bulk-term is proportional to \(T\), where the boundary term is independent of \(T\). Therefore, in the rate \(R=P/T\), the contribution of the boundary term is negligible for large enough \(T\).

Now using the similar technique as in the \(\delta^2\)-case, we find,
\begin{equation}
    R = \frac{P}{T} = -\frac{1}{2\pi} \frac{\tilde F'(x_0)}{2u'(x_0)}\,.
\end{equation}
\begin{tcolorbox}
The important point to note is that $T$ drops out of the final answers. This holds as long as $H T\ll 1$.
\end{tcolorbox}
\end{appendix}

\printbibliography

@article{shota1,
    author = "Di Pietro, Lorenzo and Gorbenko, Victor and Komatsu, Shota",
    title = "{Analyticity and unitarity for cosmological correlators}",
    eprint = "2108.01695",
    archivePrefix = "arXiv",
    primaryClass = "hep-th",
    reportNumber = "CERN-TH-2021-118",
    doi = "10.1007/JHEP03(2022)023",
    journal = "JHEP",
    volume = "03",
    pages = "023",
    year = "2022"
}

@misc{pdg,
  title = {Particle Data Group},
  year = {2024},
  url = {https://pdg.lbl.gov/},
  note = {Reference to the Particle Data Group's data and information.}
}

@article{snowmass,
    author = "Baumann, Daniel and Green, Daniel and Joyce, Austin and Pajer, Enrico and Pimentel, Guilherme L. and Sleight, Charlotte and Taronna, Massimo",
    title = "{Snowmass White Paper: The Cosmological Bootstrap}",
    eprint = "2203.08121",
    archivePrefix = "arXiv",
    primaryClass = "hep-th",
    doi = "10.21468/SciPostPhysCommRep.1",
    journal = "SciPost Phys. Comm. Rep.",
    volume = "2024",
    pages = "1",
    year = "2024"
}

@article{appl1,
    author = "Chowdhury, Chandramouli and Lipstein, Arthur and Marshall, Joe and Mei, Jiajie and Sachs, Ivo",
    title = "{Cosmological Dressing Rules}",
    eprint = "2503.10598",
    archivePrefix = "arXiv",
    primaryClass = "hep-th",
    reportNumber = "LMU-ASC 03/25",
    month = "3",
    year = "2025"
}

@article{appl2,
    author = "Ghosh, Diptimoy and Ullah, Farman",
    title = "{Cosmological cutting rules for Bogoliubov initial states: any mass and spin}",
    eprint = "2502.05630",
    archivePrefix = "arXiv",
    primaryClass = "hep-th",
    month = "2",
    year = "2025"
}

@article{appl3,
    author = "Pueyo, Carlos Duaso and Goodhew, Harry and McCulloch, Ciaran and Pajer, Enrico",
    title = "{Perturbative unitarity bounds from momentum-space entanglement}",
    eprint = "2410.23709",
    archivePrefix = "arXiv",
    primaryClass = "hep-th",
    month = "10",
    year = "2024"
}

@article{arkanimalda,
    author = "Arkani-Hamed, Nima and Maldacena, Juan",
    title = "{Cosmological Collider Physics}",
    eprint = "1503.08043",
    archivePrefix = "arXiv",
    primaryClass = "hep-th",
    month = "3",
    year = "2015"
}

@article{arkani1,
    author = "Arkani-Hamed, Nima and Baumann, Daniel and Lee, Hayden and Pimentel, Guilherme L.",
    title = "{The Cosmological Bootstrap: Inflationary Correlators from Symmetries and Singularities}",
    eprint = "1811.00024",
    archivePrefix = "arXiv",
    primaryClass = "hep-th",
    doi = "10.1007/JHEP04(2020)105",
    journal = "JHEP",
    volume = "04",
    pages = "105",
    year = "2020"
}

@article{arkani2,
    author = "Arkani-Hamed, Nima and Baumann, Daniel and Hillman, Aaron and Joyce, Austin and Lee, Hayden and Pimentel, Guilherme L.",
    title = "{Differential Equations for Cosmological Correlators}",
    eprint = "2312.05303",
    archivePrefix = "arXiv",
    primaryClass = "hep-th",
    month = "12",
    year = "2023"
}

@article{bau2,
    author = "Baumann, Daniel and Goodhew, Harry and Lee, Hayden",
    title = "{Kinematic Flow for Cosmological Loop Integrands}",
    eprint = "2410.17994",
    archivePrefix = "arXiv",
    primaryClass = "hep-th",
    month = "10",
    year = "2024"
}

@article{sleight,
    author = "Sleight, Charlotte and Taronna, Massimo",
    title = "{Bootstrapping Inflationary Correlators in Mellin Space}",
    eprint = "1907.01143",
    archivePrefix = "arXiv",
    primaryClass = "hep-th",
    reportNumber = "PUPT-2590",
    doi = "10.1007/JHEP02(2020)098",
    journal = "JHEP",
    volume = "02",
    pages = "098",
    year = "2020"
}

@article{suvrat,
    author = "Chakraborty, Tuneer and Chakravarty, Joydeep and Godet, Victor and Paul, Priyadarshi and Raju, Suvrat",
    title = "{Holography of information in de Sitter space}",
    eprint = "2303.16316",
    archivePrefix = "arXiv",
    primaryClass = "hep-th",
    doi = "10.1007/JHEP12(2023)120",
    journal = "JHEP",
    volume = "12",
    pages = "120",
    year = "2023"
}

@article{falkowski,
    author = "Falkowski, Adam and Gonz\'alez-Alonso, Mart\'\i{}n and Naviliat-Cuncic, Oscar",
    title = "{Comprehensive analysis of beta decays within and beyond the Standard Model}",
    eprint = "2010.13797",
    archivePrefix = "arXiv",
    primaryClass = "hep-ph",
    reportNumber = "IFIC/20-49, FTUV/20-1027",
    doi = "10.1007/JHEP04(2021)126",
    journal = "JHEP",
    volume = "04",
    pages = "126",
    year = "2021"
}

@article{loga,
    author = "Loganayagam, R. and Shetye, Omkar",
    title = "{Influence phase of a dS observer. Part I. Scalar exchange}",
    eprint = "2309.07290",
    archivePrefix = "arXiv",
    primaryClass = "hep-th",
    doi = "10.1007/JHEP01(2024)138",
    journal = "JHEP",
    volume = "01",
    pages = "138",
    year = "2024"
}

@article{marolf1,
    author = "Marolf, Donald and Morrison, Ian A. and Srednicki, Mark",
    title = "{Perturbative S-matrix for massive scalar fields in global de Sitter space}",
    eprint = "1209.6039",
    archivePrefix = "arXiv",
    primaryClass = "hep-th",
    doi = "10.1088/0264-9381/30/15/155023",
    journal = "Class. Quant. Grav.",
    volume = "30",
    pages = "155023",
    year = "2013"
}

@article{marolf2,
    author = "Marolf, Donald and Morrison, Ian A.",
    title = "{The IR stability of de Sitter QFT: Physical initial conditions}",
    eprint = "1104.4343",
    archivePrefix = "arXiv",
    primaryClass = "gr-qc",
    doi = "10.1007/s10714-011-1233-3",
    journal = "Gen. Rel. Grav.",
    volume = "43",
    pages = "3497--3530",
    year = "2011"
}

@article{bau1,
    author = "Baumann, Daniel and Duaso Pueyo, Carlos and Joyce, Austin and Lee, Hayden and Pimentel, Guilherme L.",
    title = "{The cosmological bootstrap: weight-shifting operators and scalar seeds}",
    eprint = "1910.14051",
    archivePrefix = "arXiv",
    primaryClass = "hep-th",
    doi = "10.1007/JHEP12(2020)204",
    journal = "JHEP",
    volume = "12",
    pages = "204",
    year = "2020"
}

@article{fermi1,
    author = "Chowdhury, Chandramouli and Chowdhury, Pratyusha and Moga, Radu N. and Singh, Kajal",
    title = "{Loops, recursions, and soft limits for fermionic correlators in (A)dS}",
    eprint = "2408.00074",
    archivePrefix = "arXiv",
    primaryClass = "hep-th",
    doi = "10.1007/JHEP10(2024)202",
    journal = "JHEP",
    volume = "10",
    pages = "202",
    year = "2024"
}

@article{penedones,
    author = "Hogervorst, Matthijs and Penedones, Jo\~ao and Vaziri, Kamran Salehi",
    title = "{Towards the non-perturbative cosmological bootstrap}",
    eprint = "2107.13871",
    archivePrefix = "arXiv",
    primaryClass = "hep-th",
    doi = "10.1007/JHEP02(2023)162",
    journal = "JHEP",
    volume = "02",
    pages = "162",
    year = "2023"
}

@article{altflat,
    author = "Cespedes, Sebastian and Jazayeri, Sadra",
    title = "{The Massive Flat Space Limit of Cosmological Correlators}",
    eprint = "2501.02119",
    archivePrefix = "arXiv",
    primaryClass = "hep-th",
    month = "1",
    year = "2025"
}

@article{pajer,
    author = "Donath, Yaniv and Pajer, Enrico",
    title = "{The in-out formalism for in-in correlators}",
    eprint = "2402.05999",
    archivePrefix = "arXiv",
    primaryClass = "hep-th",
    doi = "10.1007/JHEP07(2024)064",
    journal = "JHEP",
    volume = "07",
    pages = "064",
    year = "2024"
}

@inproceedings{smatrixmarathon,
    author = "Lee, Mang Hei Gordon and Pajer, Enrico and Giroux, Mathieu and Hannesdottir, Holmfridur S. and Mizera, Sebastian and Pasiecznik, Celina",
    title = "{Records from the S-Matrix Marathon: A Timeless History of Time}",
    eprint = "2410.00227",
    archivePrefix = "arXiv",
    primaryClass = "hep-th",
    month = "9",
    year = "2024"
}

@article{jackson,
    author = "Jackson, J. D. and Treiman, S. B. and Wyld, H. W.",
    title = "{Possible tests of time reversal invariance in Beta decay}",
    doi = "10.1103/PhysRev.106.517",
    journal = "Phys. Rev.",
    volume = "106",
    pages = "517--521",
    year = "1957"
}

@article{emit,
    author = "Mumm, H. P. and others",
    title = "{A New Limit on Time-Reversal Violation in Beta Decay}",
    eprint = "1104.2778",
    archivePrefix = "arXiv",
    primaryClass = "nucl-ex",
    doi = "10.1103/PhysRevLett.107.102301",
    journal = "Phys. Rev. Lett.",
    volume = "107",
    pages = "102301",
    year = "2011"
}

@article{pimentel1,
  title = {de Sitter $S$ matrix for the masses},
  author = {Melville, Scott and Pimentel, Guilherme L.},
  journal = {Phys. Rev. D},
  volume = {110},
  issue = {10},
  pages = {103530},
  numpages = {19},
  year = {2024},
  month = {Nov},
  publisher = {American Physical Society},
  doi = {10.1103/PhysRevD.110.103530},
  url = {https://link.aps.org/doi/10.1103/PhysRevD.110.103530}
}

@article{pimentel2,
    author = "Melville, Scott and Pimentel, Guilherme L.",
    title = "{A de Sitter S-matrix from amputated cosmological correlators}",
    eprint = "2404.05712",
    archivePrefix = "arXiv",
    primaryClass = "hep-th",
    doi = "10.1007/JHEP08(2024)211",
    journal = "JHEP",
    volume = "08",
    pages = "211",
    year = "2024"
}

@article{Schaub:2023scu,
    author = "Schaub, Vladimir",
    title = "{Spinors in (Anti-)de Sitter Space}",
    eprint = "2302.08535",
    archivePrefix = "arXiv",
    primaryClass = "hep-th",
    doi = "10.1007/JHEP09(2023)142",
    journal = "JHEP",
    volume = "09",
    pages = "142",
    year = "2023"
}

@book{bateman_bateman_manuscript_project_1953, 
title={Higher Transcendental Functions [Volumes I-III]}, abstractNote={The work of which this book is the first volume might be described as an up-to-date version of Part II. The Transcendental Functions of Whittaker and Watson's celebrated "Modern Analysis". Bateman (who was a pupil of E. T. Whittaker) planned his "Guide to the Functions" on a gigantic scale. In addition to a detailed account of the properties of the most important functions, the work was to include the historic origin and definition of, the basic formulas relating to, and a bibliography for all special functions ever invented or investigated. These functions were to be catalogued and classified under twelve different headings according to their definition by power series, generating functions, infinite products, repeated differentiations, indefinite integrals, definite integrals, differential equations, difference equations, functional equations, trigonometric series, series of orthogonal functions, or integral equations. Tables of definite integrals representing each function and numerical tables of a few new functions were to form part of the "Guide". An extensive table of definite integrals and a Guide to numerical tables of special functions were planned as companion works.}, note={Funding by Office of Naval Research (ONR)}, publisher={McGraw-Hill Book Company}, author={Bateman, Harry and Bateman Manuscript Project}, year={1953} }

@article{PhysRevD.110.030001,
  title = {Review of Particle Physics},
  author = {Navas, S. and Amsler, C. and Gutsche, T. and Hanhart, C. and Hern\'andez-Rey, J. J. and Louren\ifmmode \mbox{\c{c}}\else \c{c}\fi{}o, C. and Masoni, A. and Mikhasenko, M. and Mitchell, R. E. and Patrignani, C. and Schwanda, C. and Spanier, S. and Venanzoni, G. and Yuan, C. Z. and Agashe, K. and Aielli, G. and Allanach, B. C. and Alvarez-Mu\~niz, J. and Antonelli, M. and Aschenauer, E. C. and Asner, D. M. and Assamagan, K. and Baer, H. and Banerjee, Sw. and Barnett, R. M. and Baudis, L. and Bauer, C. W. and Beatty, J. J. and Beringer, J. and Bettini, A. and Biebel, O. and Black, K. M. and Blucher, E. and Bonventre, R. and Briere, R. A. and Buckley, A. and Burkert, V. D. and Bychkov, M. A. and Cahn, R. N. and Cao, Z. and Carena, M. and Casarosa, G. and Ceccucci, A. and Cerri, A. and Chivukula, R. S. and Cowan, G. and Cranmer, K. and Crede, V. and Cremonesi, O. and D'Ambrosio, G. and Damour, T. and de Florian, D. and de Gouv\^ea, A. and DeGrand, T. and Demers, S. and Demiragli, Z. and Dobrescu, B. A. and D'Onofrio, M. and Doser, M. and Dreiner, H. K. and Eerola, P. and Egede, U. and Eidelman, S. and El-Khadra, A. X. and Ellis, J. and Eno, S. C. and Erler, J. and Ezhela, V. V. and Fava, A. and Fetscher, W. and Fields, B. D. and Freitas, A. and Gallagher, H. and Gershon, T. and Gershtein, Y. and Gherghetta, T. and Gonzalez-Garcia, M. C. and Goodman, M. and Grab, C. and Gritsan, A. V. and Grojean, C. and Groom, D. E. and Gr\"unewald, M. and Gurtu, A. and Haber, H. E. and Hamel, M. and Hashimoto, S. and Hayato, Y. and Hebecker, A. and Heinemeyer, S. and Hikasa, K. and Hisano, J. and H\"ocker, A. and Holder, J. and Hsu, L. and Huston, J. and Hyodo, T. and Ianni, Al. and Kado, M. and Karliner, M. and Katz, U. F. and Kenzie, M. and Khoze, V. A. and Klein, S. R. and Krauss, F. and Kreps, M. and Kri\ifmmode \check{z}\else \v{z}\fi{}an, P. and Krusche, B. and Kwon, Y. and Lahav, O. and Lellouch, L. P. and Lesgourgues, J. and Liddle, A. R. and Ligeti, Z. and Lin, C.-J. and Lippmann, C. and Liss, T. M. and Lister, A. and Littenberg, L. and Lugovsky, K. S. and Lugovsky, S. B. and Lusiani, A. and Makida, Y. and Maltoni, F. and Manohar, A. V. and Marciano, W. J. and Matthews, J. and Mei\ss{}ner, U.-G. and Melzer-Pellmann, I.-A. and Mertsch, P. and Miller, D. J. and Milstead, D. and M\"onig, K. and Molaro, P. and Moortgat, F. and Moskovic, M. and Nagata, N. and Nakamura, K. and Narain, M. and Nason, P. and Nelles, A. and Neubert, M. and Nir, Y. and O'Connell, H. B. and O'Hare, C. A. J. and Olive, K. A. and Peacock, J. A. and Pianori, E. and Pich, A. and Piepke, A. and Pietropaolo, F. and Pomarol, A. and Pordes, S. and Profumo, S. and Quadt, A. and Rabbertz, K. and Rademacker, J. and Raffelt, G. and Ramsey-Musolf, M. and Richardson, P. and Ringwald, A. and Robinson, D. J. and Roesler, S. and Rolli, S. and Romaniouk, A. and Rosenberg, L. J and Rosner, J. L. and Rybka, G. and Ryskin, M. G. and Ryutin, R. A. and Safdi, B. and Sakai, Y. and Sarkar, S. and Sauli, F. and Schneider, O. and Sch\"onert, S. and Scholberg, K. and Schwartz, A. J. and Schwiening, J. and Scott, D. and Sefkow, F. and Seljak, U. and Sharma, V. and Sharpe, S. R. and Shiltsev, V. and Signorelli, G. and Silari, M. and Simon, F. and Sj\"ostrand, T. and Skands, P. and Skwarnicki, T. and Smoot, G. F. and Soffer, A. and Sozzi, M. S. and Spiering, C. and Stahl, A. and Sumino, Y. and Takahashi, F. and Tanabashi, M. and Tanaka, J. and Ta\ifmmode \check{s}\else \v{s}\fi{}evsk\'y, M. and Terao, K. and Terashi, K. and Terning, J. and Thoma, U. and Thorne, R. S. and Tiator, L. and Titov, M. and Tovey, D. R. and Trabelsi, K. and Urquijo, P. and Valencia, G. and Van de Water, R. and Varelas, N. and Verde, L. and Vivarelli, I. and Vogel, P. and Vogelsang, W. and Vorobyev, V. and Wakely, S. P. and Walkowiak, W. and Walter, C. W. and Wands, D. and Weinberg, D. H. and Weinberg, E. J. and Wermes, N. and White, M. and Wiencke, L. R. and Willocq, S. and Woody, C. L. and Workman, R. L. and Yao, W.-M. and Yokoyama, M. and Yoshida, R. and Zanderighi, G. and Zeller, G. P. and Zhu, R.-Y. and Zhu, S.-L. and Zimmermann, F. and Zyla, P. A. and Anderson, J. and Kramer, M. and Schaffner, P. and Zheng, W.},
  collaboration = {Particle Data Group Collaboration},
  journal = {Phys. Rev. D},
  volume = {110},
  issue = {3},
  pages = {030001},
  numpages = {5},
  year = {2024},
  month = {Aug},
  publisher = {American Physical Society},
  doi = {10.1103/PhysRevD.110.030001},
  url = {https://link.aps.org/doi/10.1103/PhysRevD.110.030001}
}

@book{Schwartz:2014sze,
    author = "Schwartz, Matthew D.",
    title = "{Quantum Field Theory and the Standard Model}",
    isbn = "978-1-107-03473-0, 978-1-107-03473-0",
    publisher = "Cambridge University Press",
    month = "3",
    year = "2014"
}

@book{Peskin:1995ev,
    author = "Peskin, Michael E. and Schroeder, Daniel V.",
    title = "{An Introduction to quantum field theory}",
    doi = "10.1201/9780429503559",
    isbn = "978-0-201-50397-5, 978-0-429-50355-9, 978-0-429-49417-8",
    publisher = "Addison-Wesley",
    address = "Reading, USA",
    year = "1995"
}

@article{dSDirac:1935zz,
    author = "Dirac, P. A. M.",
    title = "{The Electron Wave Equation in De-Sitter Space}",
    doi = "10.2307/1968649",
    journal = "Annals Math.",
    volume = "36",
    pages = "657--669",
    year = "1935"
}

@article{dSNachtmann1967,
  author       = {Nachtmann, Otto},
  title        = {Quantum theory in de-Sitter space},
  journal      = {Communications in Mathematical Physics},
  year         = {1967},
  volume       = {6},
  number       = {1},
  pages        = {1--16},
  month        = mar,
  doi          = {10.1007/BF01646319},
  url          = {https://doi.org/10.1007/BF01646319},
  issn         = {1432-0916}
}

@article{dSGursey:1963ir,
    author = "Gursey, F. and Lee, T. D.",
    editor = "Feinberg, G.",
    title = "{Spin 1/2 Wave Equation in De Sitter Space}",
    doi = "10.1073/pnas.49.2.179",
    journal = "Proc. Nat. Acad. Sci.",
    volume = "49",
    pages = "179--186",
    year = "1963"
}

@article{dSThirring:1967dd,
    author = "Thirring, W.",
    editor = "Urban, P.",
    title = "{Quantum Field Theory in de Sitter Space}",
    doi = "10.1007/978-3-7091-5485-4_10",
    journal = "Acta Phys. Austriaca Suppl.",
    volume = "4",
    pages = "267--287",
    year = "1967"
}

@article{dSAkhmedov:2024npw,
    author = "Akhmedov, E. T. and Lapushkin, V. I. and Sadekov, D. I.",
    title = "{Light fields in various patches of de Sitter space-time}",
    eprint = "2411.11106",
    archivePrefix = "arXiv",
    primaryClass = "hep-th",
    month = "11",
    year = "2024"
}

@article{dSEpstein:2012zz,
    author = "Epstein, Henri",
    editor = "Alvarez-Gaume, Luis and Boudjema, Fawzi and Sorba, Paul",
    title = "{Remarks on quantum field theory on de Sitter and anti-de Sitter space-times}",
    doi = "10.1007/s12043-012-0312-7",
    journal = "Pramana",
    volume = "78",
    pages = "853--864",
    year = "2012"
}

@article{Goodhew:2020hob,
    author = "Goodhew, Harry and Jazayeri, Sadra and Pajer, Enrico",
    title = "{The Cosmological Optical Theorem}",
    eprint = "2009.02898",
    archivePrefix = "arXiv",
    primaryClass = "hep-th",
    doi = "10.1088/1475-7516/2021/04/021",
    journal = "JCAP",
    volume = "04",
    pages = "021",
    year = "2021"
}

@article{Jazayeri:2021fvk,
    author = "Jazayeri, Sadra and Pajer, Enrico and Stefanyszyn, David",
    title = "{From locality and unitarity to cosmological correlators}",
    eprint = "2103.08649",
    archivePrefix = "arXiv",
    primaryClass = "hep-th",
    doi = "10.1007/JHEP10(2021)065",
    journal = "JHEP",
    volume = "10",
    pages = "065",
    year = "2021"
}

@article{Melville:2021lst,
    author = "Melville, Scott and Pajer, Enrico",
    title = "{Cosmological Cutting Rules}",
    eprint = "2103.09832",
    archivePrefix = "arXiv",
    primaryClass = "hep-th",
    doi = "10.1007/JHEP05(2021)249",
    journal = "JHEP",
    volume = "05",
    pages = "249",
    year = "2021"
}

@article{Hillman:2021bnk,
    author = "Hillman, Aaron and Pajer, Enrico",
    title = "{A differential representation of cosmological wavefunctions}",
    eprint = "2112.01619",
    archivePrefix = "arXiv",
    primaryClass = "hep-th",
    doi = "10.1007/JHEP04(2022)012",
    journal = "JHEP",
    volume = "04",
    pages = "012",
    year = "2022"
}

@article{Lee:2023jby,
    author = "Lee, Mang Hei Gordon and McCulloch, Ciaran and Pajer, Enrico",
    title = "{Leading loops in cosmological correlators}",
    eprint = "2305.11228",
    archivePrefix = "arXiv",
    primaryClass = "hep-th",
    doi = "10.1007/JHEP11(2023)038",
    journal = "JHEP",
    volume = "11",
    pages = "038",
    year = "2023"
}

@misc{collins2019newapproachlszreduction,
      title={A new approach to the LSZ reduction formula}, 
      author={John Collins},
      year={2019},
      eprint={1904.10923},
      archivePrefix={arXiv},
      primaryClass={hep-ph},
      url={https://arxiv.org/abs/1904.10923}, 
}

@book{Mukhanov_Winitzki_2007, place={Cambridge}, title={Introduction to Quantum Effects in Gravity}, publisher={Cambridge University Press}, author={Mukhanov, Viatcheslav and Winitzki, Sergei}, year={2007}}

@article{PhysRev.109.193,
  title = {Theory of the Fermi Interaction},
  author = {Feynman, R. P. and Gell-Mann, M.},
  journal = {Phys. Rev.},
  volume = {109},
  issue = {1},
  pages = {193--198},
  numpages = {0},
  year = {1958},
  month = {Jan},
  publisher = {American Physical Society},
  doi = {10.1103/PhysRev.109.193},
  url = {https://link.aps.org/doi/10.1103/PhysRev.109.193}
}

@article{Sudarshan:1984qq,
    author = "Sudarshan, E. C. G. and Marshak, R. E.",
    editor = "Mann, Alfred K. and Cline, David B.",
    title = "{Origin of the Universal V-A theory}",
    reportNumber = "VPI-HEP-84/8, IMSc/PP/85-001",
    doi = "10.1063/1.45454",
    journal = "AIP Conf. Proc.",
    volume = "300",
    pages = "110--124",
    year = "1994"
}

@article{Lehmann1955,
    author = "Lehmann, H. and Symanzik, K. and Zimmermann, W.",
    title = "Zur Formulierung quantisierter Feldtheorien",
    doi = "10.1007/BF02731765",
    journal = "Il Nuovo Cimento",
    pages = "205-225",
    year = " 1955"
}

@article{PhysRev.112.669,
  title = {Quantum Field Theories with Composite Particles and Asymptotic Conditions},
  author = {Haag, R.},
  journal = {Phys. Rev.},
  volume = {112},
  issue = {2},
  pages = {669--673},
  numpages = {0},
  year = {1958},
  month = {Oct},
  publisher = {American Physical Society},
  doi = {10.1103/PhysRev.112.669},
  url = {https://link.aps.org/doi/10.1103/PhysRev.112.669}
}

@article{Hepp1965,
    author = "Hepp and Klaus",
    title = "On the connection between the LSZ and Wightman quantum field theory",
    doi = "10.1007/BF01646494",
    journal = "Communications in Mathematical Physics",
    pages = "95-111",
    year = " 1965"
}

@article{poincare_quantization_spinor,
author = {Cotaescu, Ion I.},
title = {Canonical quantization of the covariant fields on de Sitter space–times},
journal = {International Journal of Modern Physics A},
volume = {33},
number = {08},
pages = {1830007},
year = {2018},
doi = {10.1142/S0217751X18300077},

URL = { 
    
        https://doi.org/10.1142/S0217751X18300077
    
    

},
eprint = { 
    
        https://doi.org/10.1142/S0217751X18300077
    
    

}
}

@article{Anninos_2015,
doi = {10.1088/1475-7516/2015/11/048},
url = {https://dx.doi.org/10.1088/1475-7516/2015/11/048},
year = {2015},
month = {nov},
publisher = {},
volume = {2015},
number = {11},
pages = {048},
author = {Anninos, Dionysios and Anous, Tarek and Freedman, Daniel Z. and Konstantinidis, George},
title = {Late-time structure of the Bunch-Davies de Sitter wavefunction},
journal = {Journal of Cosmology and Astroparticle Physics}
}

@article{Stahl_2016,
   title={Fermionic current and Schwinger effect in de Sitter spacetime},
   volume={93},
   ISSN={2470-0029},
   url={http://dx.doi.org/10.1103/PhysRevD.93.025004},
   DOI={10.1103/physrevd.93.025004},
   number={2},
   journal={Physical Review D},
   publisher={American Physical Society (APS)},
   author={Stahl, Clément and Strobel, Eckhard and Xue, She-Sheng},
   year={2016},
   month=jan }

@article{beta_decay,
  title = {Neutron ${\ensuremath{\beta}}^{\mathbf{\ensuremath{-}}}$ decay as a laboratory for testing the standard model},
  author = {Ivanov, A. N. and Pitschmann, M. and Troitskaya, N. I.},
  journal = {Phys. Rev. D},
  volume = {88},
  issue = {7},
  pages = {073002},
  numpages = {25},
  year = {2013},
  month = {Oct},
  publisher = {American Physical Society},
  doi = {10.1103/PhysRevD.88.073002},
  url = {https://link.aps.org/doi/10.1103/PhysRevD.88.073002}
}

\end{document}